\documentclass[10pt, letter, onecolumn]{arxiv}
\usepackage{tcolorbox}
\tcbuselibrary{breakable} 
\usepackage{enumitem}
\usepackage{enumitem}
\usepackage{booktabs}
\usepackage{float}
\usepackage{tabularx}
\usepackage{ragged2e}
\usepackage{tcolorbox}
\usepackage{enumitem}
\usepackage{kantlipsum, lipsum}
\usepackage{placeins}
\usepackage{dm-colors}
\usepackage{amsmath}
\usepackage{microtype}
\usepackage{pstricks, pst-node}
\usepackage{verbatim}
\usepackage{multirow}
\usepackage{scalerel}
\usepackage{booktabs}
\usepackage{enumitem}
\usepackage{xspace}
\usepackage{bm}
\usepackage{bbm}
\usepackage{mathtools}
\usepackage{soul}
\usepackage{epsfig}
\usepackage{graphicx}
\usepackage{tcolorbox}
\usepackage{subcaption}
\usepackage{amssymb}
\usepackage{colortbl}
\usepackage{csquotes}
\usepackage{etex}
\usepackage{setspace}
\usepackage{svg}
\usepackage{colortbl}
\usepackage{tabularx,ragged2e}
\usepackage{placeins}
\usepackage[symbol]{footmisc}
\usepackage{tabularx}
\usepackage{booktabs}
\usepackage{tcolorbox}
\tcbuselibrary{breakable,skins}
\usepackage{caption} 

\usepackage{csvsimple,tcolorbox,ragged2e}
\tcbuselibrary{breakable,skins}
\newcommand{\preservebreaks}[1]{\begingroup\RaggedRight\parindent=0pt\parskip=0.5\baselineskip\obeylines #1\endgroup}
\tcbset{casecard/.style={enhanced,breakable,colback=white,colframe=black!65,boxrule=0.5pt,arc=2mm,
                         title filled,colbacktitle=black!75,coltitle=white,fonttitle=\bfseries,
                         left=2mm,right=2mm,top=1.5mm,bottom=1.5mm}}

\usepackage[bibstyle=nature,citestyle=numeric-comp,%
            natbib=true,backend=biber,maxbibnames=99,%
            giveninits=false,sorting=none]{biblatex}
\usepackage{nameref}
\usepackage{varioref}
\usepackage[pagebackref=false,breaklinks=false,%
            colorlinks=true,bookmarks=true,citecolor=ourdarkblue,%
            urlcolor=ourdarkblue,linkcolor=ourdarkblue]{hyperref}
\usepackage[noabbrev,capitalize]{cleveref}
\usepackage{etoc}

\title{Complementary Human-AI Clinical Reasoning in Ophthalmology}

\author[1,2,$\ddagger$]{Mertcan Sevgi}
\author[1,2,3,4,$\ddagger$]{Fares Antaki}
\author[5]{Abdullah Zafar Khan}
\author[1,6,7]{Ariel Yuhan Ong}
\author[1,2,8]{\\David Adrian Merle}
\author[1,2]{Kuang Hu}
\author[1,2]{Shafi Balal}
\author[9,10]{Sophie-Christin Kornelia Ernst}
\author[1,6,11]{\\Josef Huemer}
\author[4]{Gabriel T. Kaufmann}
\author[6]{Hagar Khalid}
\author[6]{Faye Levina}
\author[6]{Celeste Limoli}
\author[1,2]{\\Ana Paula Ribeiro Reis}
\author[12]{Samir Touma}
\author[13]{Anil Palepu}
\author[14]{Khaled Saab}
\author[14]{Ryutaro Tanno}
\author[14]{Valentin Liévin}
\author[14]{Tao Tu}
\author[14]{Yong Cheng}
\author[13]{Mike Schaekermann}
\author[14]{S. Sara Mahdavi}
\author[14]{Elahe Vedadi}
\author[14]{David Stutz}
\author[14]{Vivek Natarajan}
\author[14]{Alan Karthikesalingam}
\author[1,2,$\ddagger$]{Pearse A. Keane}
\author[14,$\ddagger$]{Wei-Hung Weng}
\affil[1]{UCL Institute of Ophthalmology, }
\affil[2]{NIHR Biomedical Research Centre at Moorfields Eye Hospital, }
\affil[3]{University of Montreal, }
\affil[4]{Cleveland Clinic Cole Eye Institute, }
\affil[5]{University College Dublin, }
\affil[6]{Moorfields Eye Hospital, }
\affil[7]{Oxford Eye Hospital, }
\affil[8]{University Hospital Tübingen, }
\affil[9]{Spross Research Institute, }
\affil[10]{University Hospital Basel, }
\affil[11]{Kepler University Hospital, }
\affil[12]{Massachusetts Eye and Ear, }
\affil[13]{Google Research, }
\affil[14]{Google DeepMind}

\renewcommand{\correspondingauthor}[1]{
$^{\ddagger}$~Corresponding authors: \texttt{m.sevgi@ucl.ac.uk}, \texttt{fares.antaki@umontreal.ca}, \texttt{p.keane@ucl.ac.uk},
\texttt{ckbjimmy@google.com}
}

\usepackage[utf8]{inputenc}
\usepackage{textcomp}
\DeclareUnicodeCharacter{2265}{\ensuremath{\geq}}
\begin{document}

\begin{refsection}

\begin{abstract}
Vision impairment and blindness are a major global health challenge where gaps in the ophthalmology workforce limit access to specialist care. We evaluate AMIE, a medically fine-tuned conversational system based on Gemini with integrated web search and self critique reasoning, using real world clinical vignettes that reflect scenarios a general ophthalmologist would be expected to manage. We conducted two complementary evaluations: (1) a human-AI interactive diagnostic reasoning study in which ophthalmologists recorded initial differentials and plans, then reviewed AMIE’s structured output and revised their answers; and (2) a masked preference and quality study comparing AMIE’s narrative outputs with case author reference answers using a predefined rubric. AMIE showed standalone diagnostic performance comparable to clinicians at baseline. Crucially, after reviewing AMIE's responses, ophthalmologists tended to rank the correct diagnosis higher, reached greater agreement with one another, and enriched their investigation and management plans. Improvements were observed even when AMIE’s top choice differed from or underperformed the clinician baseline, consistent with a complementary effect in which structured reasoning support helps clinicians re-rank rather than simply accept the model output. Preferences varied by clinical grade, suggesting opportunities to personalise responses by experience. Without ophthalmology specific fine-tuning, AMIE matched clinician baseline and augmented clinical reasoning at the point of need, motivating multi axis evaluation, domain adaptation, and prospective multimodal studies in real world settings.  
\end{abstract}

\maketitle


\clearpage
\section{Introduction}
\label{sec:introduction}

Vision loss is a major and inequitable global health challenge. At least one billion people live with untreated or preventable vision impairment, with the burden concentrated in low- and middle-income countries (LMICs) \cite{worldhealthorganizationWorldReportVision2019}. This unmet need is compounded by a projected global shortfall of millions of health workers by 2030 \cite{worldhealthorganizationHealthWorkforceRequirements2016} and stark disparities in specialist availability: the mean number of ophthalmologists is 3.7 per million in low income countries versus 76 per million in high income countries \cite{resnikoffEstimatedNumberOphthalmologists2020a}. Task shifting and task sharing have therefore been advocated to extend access through rational redistribution of clinical tasks across cadres \cite{worldhealthorganizationTaskShiftingRational2007}. To work safely at scale, eye care also needs tools that surface subspecialist knowledge at the point of need, support structured clinical reasoning, and standardise decision making across variable experience levels.

Large language models (LLMs) are emerging as a candidate technology for this purpose, but credible evaluation is as essential as capability. For example, Med-PaLM achieved a passing score on USMLE style multiple choice questions, although its long form answers still lagged behind physicians \cite{singhalLargeLanguageModels2023a}. Med-PaLM 2 advanced this line of work by using a stronger base model, medical fine-tuning, and strategies such as ensemble refinement and chain of retrieval. Crucially, it helped establish an evaluation approach that goes beyond knowledge tests to include structured, multi-axis human ratings of clinical helpfulness and safety alongside accuracy \cite{singhalExpertlevelMedicalQuestion2025}.

AMIE (Articulate Medical Intelligence Explorer) extends this agenda from single turn question answering to diagnostic dialogue. Based on the Gemini family and incorporating integrated web search with self critique reasoning, AMIE is optimised for history taking, differential diagnosis, and management planning \cite{tuConversationalDiagnosticArtificial2025}. In subspecialty evaluations including cardiology and oncology, AMIE performed comparably to residents and fellows. Clinicians improved after reviewing AMIE's output, showing the benefits of AI and human interaction \cite{osullivanDemocratizationSubspecialityMedical2024a,palepuExploringLargeLanguage2024}.

Within ophthalmology, interest in LLMs is accelerating but methods remain heterogeneous. A systematic review across clinical medicine identified surgery as the most commonly evaluated major specialty, with ophthalmology the most common surgical subspecialty \cite{shoolSystematicReviewLarge2025}. A scoping review in eye care found that studies have largely centred on exam style knowledge tests and patient education tasks, which limits synthesis and real world inference \cite{seeUseLargeLanguage2025}. Many studies emphasise diagnosis or short form answers rather than the broader clinical workflow of constructing and re-ordering a differential, justifying reasoning, and planning longitudinal management \cite{antakiPerformanceGPT5Frontier2025,haghighiEYELlamaIndomainLarge2025,mikhailPerformanceDeepSeekR1Ophthalmology2025}. Head-to-head evaluations suggest steady gains across model generations, with the latest systems outperforming earlier ones on ophthalmic benchmarks, while domain specific weaknesses persist \cite{antakiPerformanceGPT5Frontier2025,srinivasan2025-vllm-ophth}.

Here we address these needs by evaluating AMIE for diagnostic reasoning in ophthalmology using 100 real world clinical case vignettes spanning common and challenging scenarios for a general ophthalmologist. We compare AMIE generated differentials, justifications, and management recommendations with expert responses, and assess whether exposure to AMIE shifts clinicians' diagnostic ranking and planning. In doing so, we move beyond accuracy only benchmarks toward clinically relevant, multi-axis evaluation of both model quality and AI-human interaction.

\section{Methods}
\label{sec:methods}

\subsection{Study design}
Five ophthalmologist case authors (“case authors”) created 100 clinical vignettes and a free-text reference response for each scenario (\cref{vignette}). Each vignette was submitted to AMIE to generate a structured response (\cref{AMIE-generated responses}), which was then rewritten into an AMIE narrative response for masked side-by-side comparison. A panel of nine ophthalmologist raters (three residents, three fellows, three consultants) participated. For each vignette, three raters, one from each grade, were randomly assigned, yielding three independent ratings per vignette and 300 ratings in total. 

\begin{figure}[htbp!]
\centering
\includegraphics[width=\textwidth,keepaspectratio]{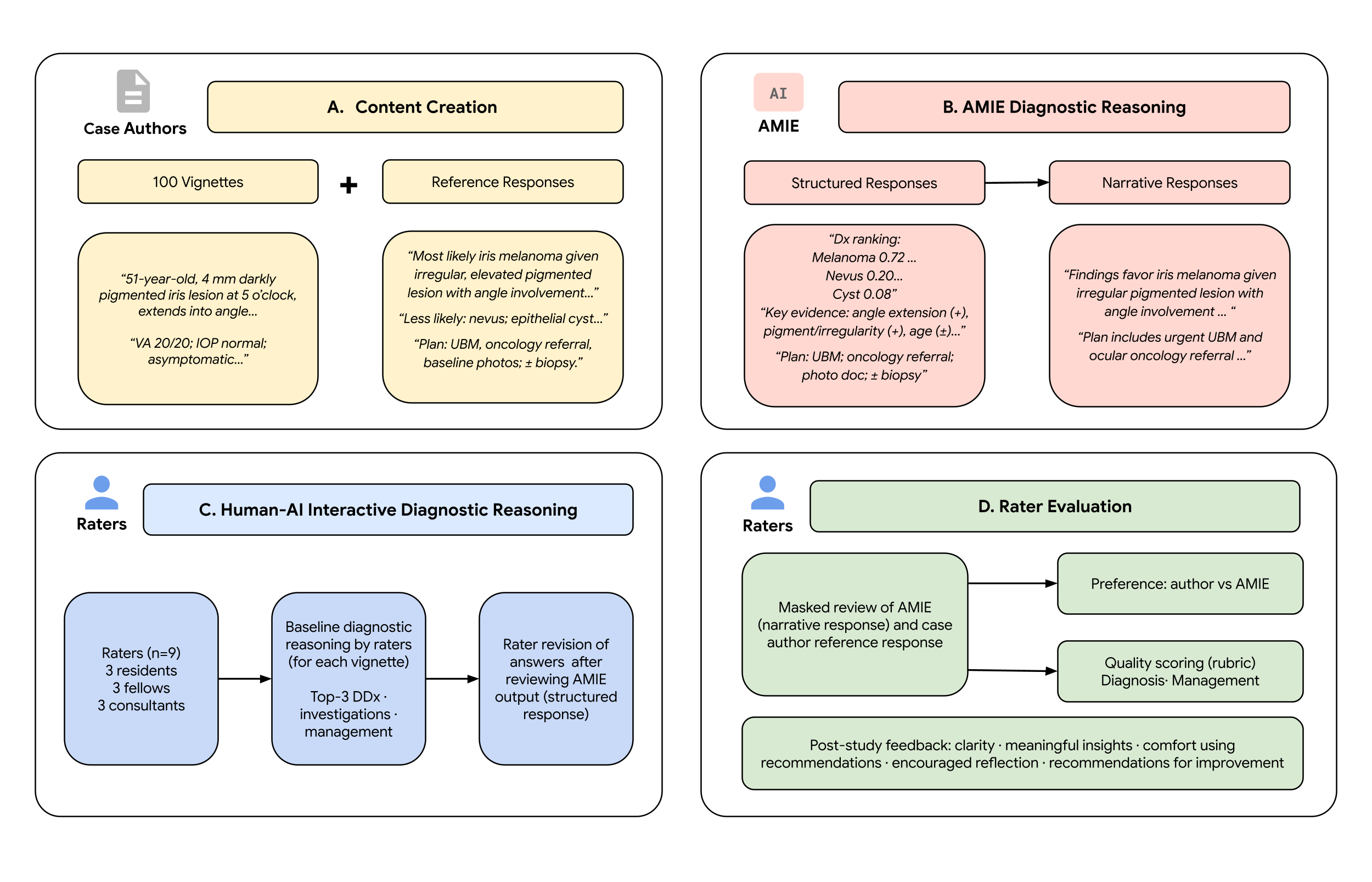}
\vspace{0.1cm}
\caption{\textbf{Study overview.} (A) Content creation by five ophthalmologist case authors. (B) AMIE responses used in the study. (C) Task 1---Human–AI interactive diagnostic reasoning: raters record baseline reasoning, then revise after viewing the AMIE structured response. (D) Task 2---Rater evaluation: masked comparison of the AMIE narrative response versus case author reference; rater preference and quality scoring; post-study feedback.}
\label{fig:overview}
\end{figure}

\clearpage
We conducted two tasks (\cref{fig:overview}):
\begin{itemize}
    \item \textbf{Task 1---Human-AI interactive diagnostic reasoning.} Raters provided a top-three differential, an initial investigation and management plan for each vignette. They then reviewed the AMIE structured response. Where indicated, they revised their responses to diagnosis, investigation and management. 
    \item \textbf{Task 2---Rater evaluation.} The same raters, masked to the authorship of given responses, performed a side-by-side comparison of AMIE narrative responses and case author reference responses, indicated their preference and scored each using a predefined rubric covering diagnosis and management domains (\cref{table:eval_rubrics}).
\end{itemize} 

After completing all cases, raters also provided brief post-study feedback on the AMIE responses, focusing on the AMIE structured responses (clarity, usefulness, clinical acceptability) and free-text comments on helpful aspects and suggested improvements (\cref{fig:post_survey_feedback}).

\begin{table}[htbp!]
\centering
\resizebox{1.0\textwidth}{!}{
\begin{tabular}{cll}
\toprule
\textbf{Category} & \multicolumn{1}{c}{\textbf{Criteria}} & \multicolumn{1}{c}{\textbf{Options}} \\
\midrule
Diagnosis & \begin{tabular}[c]{@{}l@{}}How APPROPRIATE was the doctor's\\ differential diagnosis (DDx) compared\\ to the answer key?\end{tabular} & \begin{tabular}[c]{@{}l@{}}1: Very Inappropriate\\ 2: Inappropriate\\ 3: Neither Appropriate Nor Inappropriate\\ 4: Appropriate\\ 5: Very Appropriate\end{tabular} \\ \cline{2-3} 
 & \begin{tabular}[c]{@{}l@{}}How close did the doctor's differential\\ diagnosis (DDx) come to including the\\ PROBABLE DIAGNOSIS from the\\ answer key?\end{tabular} & \begin{tabular}[c]{@{}l@{}}1: Nothing in the DDx is related to the \\ probable diagnosis.\\ 2: DDx contains something that is related, \\ but unlikely to be helpful in determining \\ the probable diagnosis.\\ 3: DDx contains something that is closely \\ related and might have been helpful in \\ determining the probable diagnosis.\\ 4: DDx contains something that is very \\ close, but not an exact match to the \\ probable diagnosis.\\ 5: DDx includes the probable diagnosis.\end{tabular} \\
\midrule
Management & \begin{tabular}[c]{@{}l@{}}Did the doctor SUGGEST appropriate\\ INVESTIGATIONS?\end{tabular} & \begin{tabular}[c]{@{}l@{}}1: No - The doctor did not recommend \\ investigations, but the correct action \\ would be to order investigations\\ 2: No - The doctor recommended \\ investigations but these were not \\ comprehensive (some were missing)\\ 3: Yes - The doctor recommended a \\ comprehensive and appropriate set \\ of investigations (including correctly \\ selecting zero investigations if this\\  was best for the case)\end{tabular} \\ \cline{2-3} 
 & \begin{tabular}[c]{@{}l@{}}Did the doctor SUGGEST appropriate \\ TREATMENTS?\end{tabular} & \begin{tabular}[c]{@{}l@{}}1: No - The doctor did not recommend \\ treatments, but the correct action \\ would be to recommend investigations\\ 2: No - The doctor recommended \\ treatments but these were not \\ comprehensive (some were missing)\\ 3: Yes - The doctor recommended a \\ comprehensive and appropriate set of \\ treatments (including correctly selecting \\ zero treatments if this was best for the \\ case or if further investigation should \\ precede treatment)\end{tabular} \\ \cline{2-3} 
 & \begin{tabular}[c]{@{}l@{}}To what extent was the doctor's MANAGEMENT \\ PLAN appropriate, including recommending \\ emergency or red-flag presentations to go to ED?\end{tabular} & \begin{tabular}[c]{@{}l@{}}1: Very Inappropriate\\ 2: Inappropriate\\ 3: Neither Appropriate Nor Inappropriate\\ 4: Appropriate\\ 5: Very Appropriate\end{tabular} \\
\bottomrule
\end{tabular}
}
\vspace{0.2cm}
\caption{\textbf{Rater evaluation rubric.} Raters scored responses across diagnosis and management using the listed criteria. DDx: differential diagnosis.}
\label{table:eval_rubrics}
\end{table}

\subsection{Case vignette development}
\label{vignette}

In this study, we generated 100 clinical vignettes, each centred on a single ophthalmic disorder that a general ophthalmologist would be expected to recognise and manage. We selected conditions through a structured mapping exercise aligning the American Board of Ophthalmology Written Qualifying Examination blueprint with high-yield tables from the AAO Basic and Clinical Science Course and content from the Wills Eye Manual \cite{americanboardofophthalmologyWQETestBlueprint2025}. This strategy prioritised scenarios of broad educational value rather than exhaustive coverage.

The five case authors were based in Canada, the United States, the United Kingdom and Ireland, and included both board-certified ophthalmologists and senior residents. Their expertise spanned retina, neuro-ophthalmology and general ophthalmology. Each vignette was created by an individual case author who sometimes wrote outside their subspecialty, they consulted authoritative sources in an “open book” format, with all material written to the level expected of a general ophthalmologist.

Each vignette was crafted to lead to a pre-specified principal diagnosis and reinforce a discrete learning objective. For example, recognising the diagnostic criteria of idiopathic intracranial hypertension, distinguishing typical from atypical optic neuritis or applying an algorithmic approach to anisocoria. Every vignette comprised (1) a detailed history (demographics, presenting complaint, relevant background), (2) examination findings, and (3) key investigations including imaging where appropriate (see example, \crefrange{fig:case_vignette_1}{fig:case_vignette_2}). The final corpus spanned Paediatrics and Strabismus (n=17), Neuro-ophthalmology (n=14), Cornea (n=14), Retina (n=12), Glaucoma (n=11), Oculoplastics (n=9), Uveitis (n=9), Lens and Cataract (n=5), Refractive Surgery (n=5), and Ocular Oncology (n=4).

For each vignette, the case author also produced a free text reference response, which was treated as the ground truth. Authors were asked to present their typical clinical reasoning, including diagnostic considerations and management recommendations, without prescriptive formatting. Responses also included relevant pathophysiological or epidemiological insights and, when appropriate, contextualised the case within broader clinical practice. All content underwent peer review by the senior case author, who is board certified and has demonstrated excellence in the AAO OKAP examination.

\subsection{AMIE inference strategy}
\label{model}

\begin{figure}[htbp!]
\centering
\includegraphics[width=\textwidth,keepaspectratio]{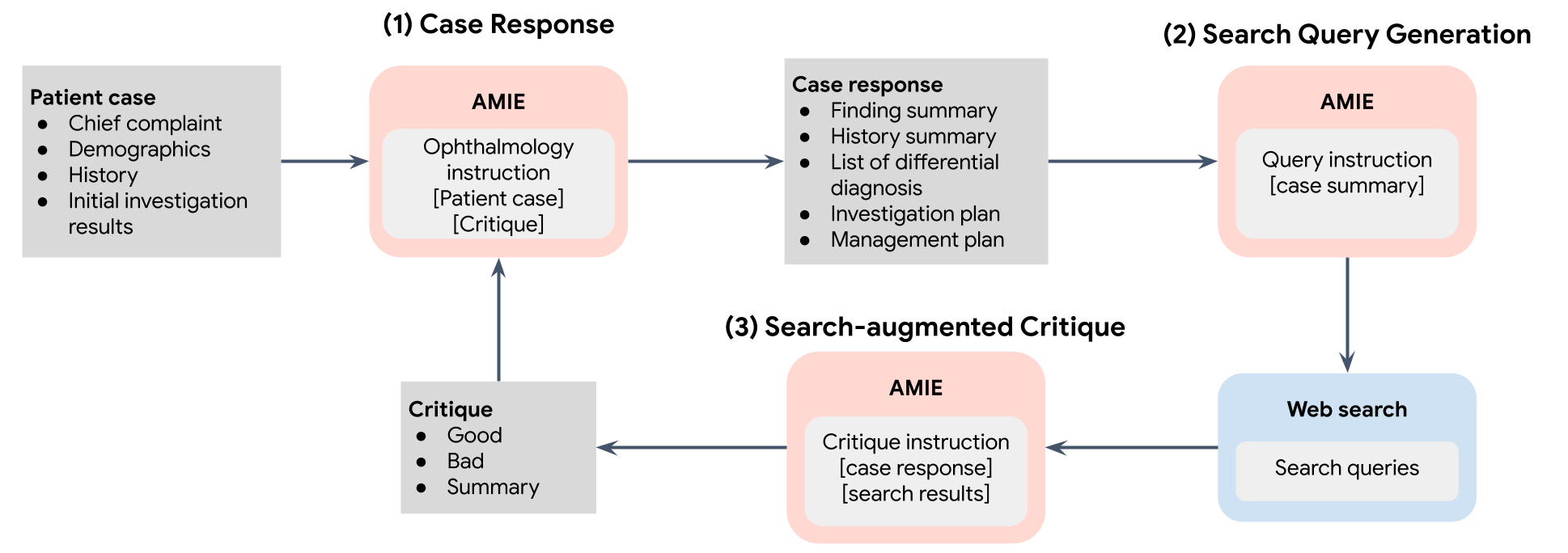}
\vspace{0.1cm}
\caption{\textbf{AMIE self-critique reasoning loop for inference.} AMIE first generates an initial case response, then formulates search queries to retrieve additional context. Using the retrieved information, AMIE critiques its draft and produces a revised response.}
\label{fig:inference}
\end{figure}

AMIE is a medically fine-tuned LLM based on the Gemini model developed by Google \cite{tuConversationalDiagnosticArtificial2025}. We used AMIE with an integrated web search and self-critique reasoning process to generate responses for the 100 case vignettes. Details regarding the base LLM can be found in \cite{team2023gemini,reid2024gemini}, and the fine-tuning approach is described in \cite{palepuConversationalAIDisease2025,tuConversationalDiagnosticArtificial2025}. We did not apply any additional fine-tuning for this ophthalmology task. Instead, we focused on adapting the inference strategy, in line with prior work, to make the best use of AMIE's capabilities. Following earlier studies \cite{osullivanDemocratizationSubspecialityMedical2024a,palepuExploringLargeLanguage2024}, our inference strategy involved search retrieval of relevant context and self-critique and revision of the initial response from AMIE. This was a multi-step process (\cref{fig:inference}): AMIE first generated a draft response, then identified key search queries for web search, used the retrieved context to critique its initial draft, and finally revised the response based on this critique.

For the initial zero-shot response, AMIE was prompted to include the following structured elements based on the case vignette:
\begin{itemize}
    \item Identification of all positive and negative clinical findings.
    \item Listing of all medications, medical history, social history, and family history.
    \item Generation of a list of the most probable differential diagnoses, each with supporting and opposing evidence drawn from the vignette.
    \item Based on the differential diagnosis list, provision of a prioritized list of recommended investigation plans and a list of management plans.
\end{itemize}

The complete prompt for generating the initial response is detailed in \cref{fig:inference_prompt}. Following the initial draft, relevant information regarding findings, diagnoses, and management plans was retrieved by AMIE using Google Search. AMIE then acted as a self-critique agent, generating critiques highlighting the strengths and weaknesses of its initial response based on the retrieved search results. Finally, AMIE revised the initial response incorporating the critique and search findings. The prompts for search, critique, and revision are described \cref{fig:search_prompts,fig:critique_prompt,fig:revision_prompt}.

\subsection{AMIE-generated responses}
\label{AMIE-generated responses}

Providing the vignette through the AMIE inference strategy prompts produced a structured response (\cref{fig:structured_response_1,fig:structured_response_2}), which presented a bullet-point case summary, confidence-weighted differential diagnoses with supporting and opposing arguments, the most likely diagnosis, and detailed next-step investigations and management.
This structured response was subsequently used as input for a narrative-generation prompt (\cref{fig:narrative_generation}), which produced a narrative response (\cref{fig:narrative_response}): a prose reformulation of the same content designed to emulate human clinical reasoning while omitting explicit AI identifiers, thereby enabling masked comparison with the case author reference response.

\subsection{Raters}

Nine practising ophthalmologists (henceforth termed “raters”) were recruited from four centres: Moorfields Eye Hospital (London, UK), University Eye Hospital Tübingen (Tübingen, Germany), University Hospital Basel (Basel, Switzerland) and Ambimed Basel (Basel, Switzerland). The panel comprised three consultants or attendings (mean clinical experience 14.6 years, range 8–20 years), three subspecialty fellows (mean 8.3 years, range 7–10 years) and three residents (mean 4.3 years, range 3.5-6 years). Subspecialty interests included cornea, cataract, glaucoma, uveitis and retina. The residents practised general ophthalmology.

\subsection{Data analysis}

We used a mixed-methods approach to quantify model performance and understand its impact on clinician reasoning.
\begin{itemize}
    \item \textbf{Quantitative outcomes:} We calculated top‑1, top‑2 and top‑3 accuracy for AMIE, for raters before viewing AMIE's structured response, and after viewing it. Subgroup analyses were performed by rater subspecialty and clinical grade. To capture ranking information, we applied a pre‑specified four‑level ordinal score for each rater (3 = correct diagnosis ranked first; 2 = second; 1 = third; 0 = not in the top three). Inter‑rater agreement was summarised as percent consensus among the three raters per vignette at each top‑N level. For all accuracy outcomes, 95\% confidence intervals (CI) were obtained by bootstrap resampling.
    \item \textbf{Hypothesis testing:} Paired Wilcoxon signed‑rank tests were used for within‑rater before/after comparisons of the 0-3 ordinal rank score, and for head‑to‑head comparisons between AMIE and the rater-only baseline using the same ordinal score. Chi‑squared tests or Fisher's exact tests were used for categorical outcomes (e.g., preference proportions, grade‑stratified revision rates) as appropriate. All tests were two‑sided with $p<0.05$ considered significant.
    \item \textbf{Qualitative outcomes involved thematic analyses of management plan changes and post-study feedback.} For management plan changes, we reviewed all rater-entered revisions to their original plans, inductively identified recurrent categories (e.g. “add imaging test”, “refer to subspecialist”), and then coded every revision according to this schema to quantify how AMIE's responses influence decision-making. For post-study feedback, raters' post-study feedback comments on helpful aspects and suggestions for improvement (\cref{fig:post_survey_feedback}) were analysed thematically. 
\end{itemize}

\section{Results}
\label{sec:results}

\subsection{Standalone diagnostic performance and subspecialty analysis}

\subsubsection{Overall}

\begin{figure}[htbp!]
\centering
\includegraphics[width=0.7\textwidth]{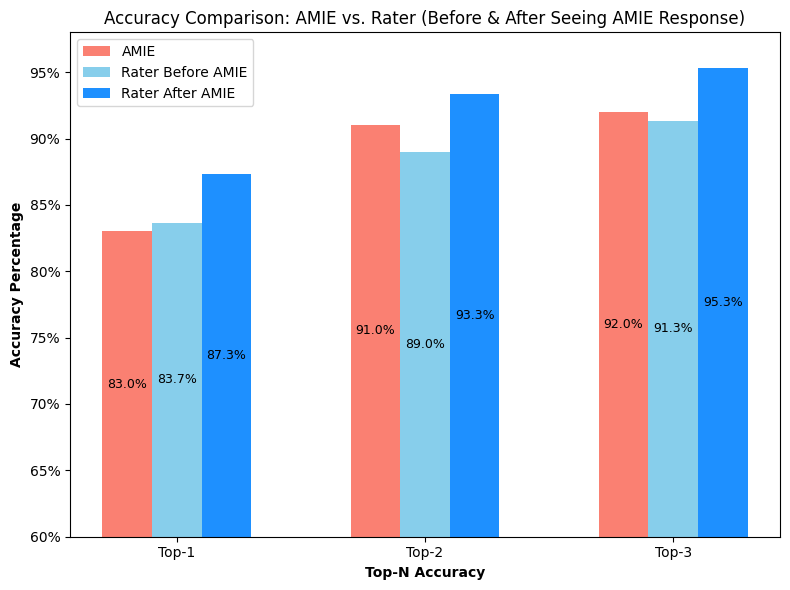}
\vspace{0.1cm}
\caption{\textbf{Accuracy comparison: AMIE versus raters (before/after seeing AMIE's response.)} Bar plot showing the Top-1, Top-2, and Top-3 diagnosis accuracy for AMIE,  raters before and after seeing AMIE's response.}
\label{fig:accuracy_comparison}
\end{figure}

Across the 100-case corpus, AMIE's standalone diagnostic accuracy was 83.0\% for top-1 (95\% CI 78.7–87.0\%), 91.0\% for top-2 (87.7–94.0\%) and 92.0\% for top-3 (88.7–95.0\%). Before viewing AMIE's structured response, raters achieved 83.7\% (95\% CI 79.3–87.7\%) for top-1, 89.0\% (85.3–92.3\%) for top-2, and 91.3\% (88.0–94.3\%) for top-3 (\cref{fig:accuracy_comparison}). Using the pre-specified 0-3 ordinal rank score, AMIE did not differ from the human baseline ($p=0.647$). By contrast, within-rater scores improved after viewing AMIE's structured response ($p=0.0014$), reflecting higher placement of the correct diagnosis in rater lists. Inter-rater agreement also increased after viewing AMIE's structured response: top-1 consensus rose from 66\% to 77\%, top-2 from 76\% to 86\%, and top-3 from 81\% to 91\%.

\subsubsection{Performance by Subspecialty}

\begin{figure}[htbp!]
\centering
\includegraphics[width=0.7\textwidth]{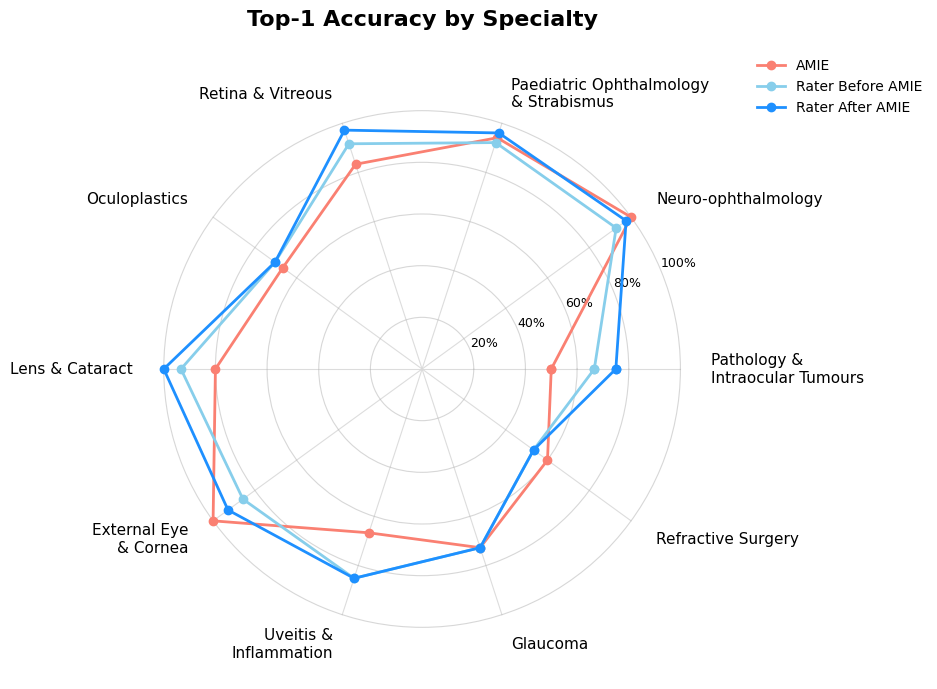}
\vspace{0.1cm}
\caption{\textbf{Top-1 accuracy by subspecialty (Rater Before AMIE, Rater After AMIE, AMIE)} Radar plot showing Top-1 diagnostic accuracy within each subspecialty for AMIE, raters before seeing AMIE, and the same raters after review. Radial axis: 0-100\%.}
\label{fig:radar_plot_top_1}
\end{figure}

Top-1 accuracy varied by subspecialty (\cref{fig:radar_plot_top_1}). Raters before AMIE ranged from 53.3\% in Refractive Surgery (95\% CI 26.7-80.0\%) to 93.3\% in Lens \& Cataract (95\% CI 80.0-100.0\%). AMIE reached 100\% Top-1 accuracy in External Eye Disease \& Cornea and Neuro-ophthalmology (95\% CI 100-100\%), with its lowest performance in Ophthalmic Pathology \& Intraocular Tumours (Top-1 50.0\%; 95\% CI 25.0–75.0\%). Patterns for Top-2 and Top-3 were similar (\cref{fig:radar_plot_top_2,fig:radar_plot_top_3}), and exact estimates with 95\% CIs for all metrics are provided in \cref{tab:diagnostic_accuracy}. Chi-squared tests indicated accuracy differed across subspecialties for AMIE, raters before, and raters after seeing AMIE (all $p<0.05$)

\subsubsection{Performance by Clinical Grade}

\cref{table:diag_by_grade} summarises top-1, top-2 and top-3 accuracy for AMIE, and for the raters stratified by clinical grade and by whether this was done before or after viewing AMIE's structured responses. Performance of human raters versus AMIE was not significantly different across subgroups: consultants ($p=0.4993$), fellows ($p=0.0557$) and residents ($p=0.1406$). Within grade, paired Wilcoxon tests on the 0–3 rank score showed significant improvement after viewing AMIE's structured response for consultants ($p=0.027$) and fellows ($p=0.022$) , but not for residents ($p=0.414$). Taken together, the accuracy summaries (\cref{table:diag_by_grade}) and the ordinal tests indicate that AMIE performed comparably to raters within each grade at baseline, while reviewing its structured response was associated with upward re-ordering of correct diagnoses for mid and senior level raters. 

\vspace{0.5cm}
\begin{table}[htbp!]
\centering
\small
\renewcommand{\arraystretch}{1.2}
\begin{tabular}{llccc}
\toprule
\textbf{Rater} & \textbf{Approach} & \textbf{Top-1 (\%)} & \textbf{Top-2 (\%)} & \textbf{Top-3 (\%)} \\
\midrule
AMIE     & N/A                & 83.0 (78.7--87.0) & 91.0 (87.7--94.0) & 92.0 (88.7--95.0) \\
\midrule
Resident       & Rater Before AMIE  & 90.0 (84.0--95.0) & 94.0 (89.0--98.0) & 96.0 (92.0--99.0) \\
               & Rater After AMIE   & 89.0 (83.0--94.0) & 96.0 (92.0--99.0) & 97.0 (94.0--100.0) \\
\midrule
Fellow         & Rater Before AMIE  & 78.0 (70.0--86.0) & 84.0 (77.0--91.0) & 88.0 (81.0--94.0) \\
               & Rater After AMIE   & 86.0 (79.0--93.0) & 91.0 (85.0--96.0) & 94.0 (89.0--98.0) \\
\midrule
Consultant     & Rater Before AMIE  & 83.0 (75.0--90.0) & 89.0 (83.0--95.0) & 90.0 (84.0--96.0) \\
               & Rater After AMIE   & 87.0 (80.0--93.0) & 93.0 (88.0--98.0) & 95.0 (90.0--99.0) \\
\bottomrule
\end{tabular}
\vspace{0.2cm}
\caption{\textbf{Diagnostic accuracy by rater's clinical grade and approach (Before/After AMIE).} Values are percentages with 95\% confidence intervals. “Approach” indicates whether clinician ratings were made before or after reviewing AMIE. Top-N = correct diagnosis appears within the top N choices.}
\label{table:diag_by_grade}
\end{table}

\clearpage
\subsection{Rater Revision}

\begin{figure}[htbp!]
\centering
\includegraphics[width=0.7\textwidth]{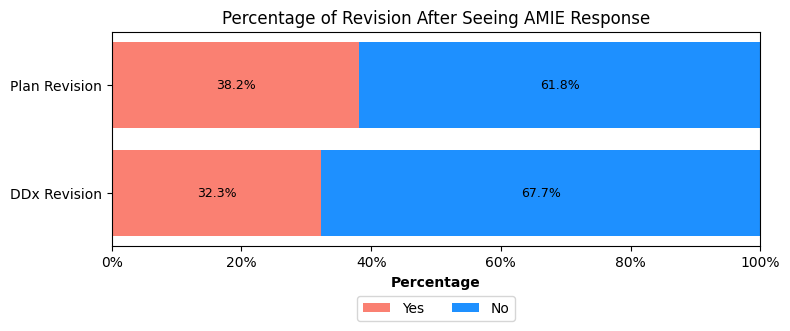}
\vspace{0.1cm}
\caption{\textbf{Percentage of revision after seeing AMIE's responses.} Horizontal stacked bar plot showing the percentage of raters who revised their Differential Diagnosis (DDx) list and Management Plan after reviewing AMIE's responses.}
\label{fig:revision_changes}
\end{figure}

Consulting AMIE's structured answer prompted raters to revise their differential diagnosis in 32.3\% of cases and the initial management plan in 38.2\% (\cref{fig:revision_changes}).

\vspace{-0.5cm}
\begin{itemize}
    \item \textbf{Differential diagnosis revision.} Revision behaviour varied by clinical grade: after viewing the structured response generated by AMIE, consultants amended their differential lists in 26\% of vignettes, residents in 28\%, and fellows in 43\%. A chi-squared test confirmed a significant difference across grades ($p=0.019$). Revision propensity also differed markedly between individual raters: across their 33-34 assigned cases, the number of changes in cases ranged from 2 to 18 (6\%–55\% of cases), with a median of 11, highlighting substantial heterogeneity both by experience level and by individual clinician.
    \item \textbf{Management plan revision.} Among the 102 encounters in which raters modified their management plan after reading AMIE's structured response, we coded every change against eight predefined themes (\cref{table:theme_revisions}). A single case could belong to more than one theme, so percentages refer to the proportion of revised plans that contained at least one change in that category rather than summing to 100\%.
\end{itemize}

\begin{table}[htbp!]
\centering
\small
\renewcommand{\arraystretch}{1.1}
\begin{tabular}{lcc}
\toprule
\textbf{Theme} & \textbf{Revised plans (n)} & \textbf{Share of revised plans (\%)} \\
\midrule
Clinical \& functional tests             & 63 & 61.8 \\
Diagnostic imaging                       & 57 & 55.9 \\
Laboratory investigations \& biopsy      & 52 & 51.0 \\
Medications                              & 44 & 43.1 \\
Surgical \& procedural interventions     & 39 & 38.2 \\
Referrals \& multidisciplinary care      & 39 & 38.2 \\
Patient education \& counselling         & 33 & 32.4 \\
Social \& support-services referrals     &  6 &  5.9 \\
\bottomrule
\end{tabular}
\vspace{0.2cm}
\caption{\textbf{Themes of management plan revision after seeing AMIE's responses}. Counts and percentage of revised plans containing at least one change in each theme among encounters with any plan revision (n=102). Themes are not mutually exclusive; percentages do not sum to 100\%}
\label{table:theme_revisions}
\end{table}

The most frequent changes involved clinical and functional tests (61.8\%), typically adding gonioscopy and pachymetry when angle-closure mechanisms were considered, orthoptic assessments for strabismus work-ups, or electrophysiology in suspected melanoma-associated retinopathy. Diagnostic imaging (55.9\%) changes often refined vague orders into specific protocols, for example expanding “orbital imaging” to MRI of brain and orbits with contrast and cavernous sinus sequences, adding CT angiography for carotid pathology, or specifying repeat imaging to gauge antibiotic response. Laboratory investigations and biopsy (51.0\%) commonly introduced targeted serology or tissue sampling, such as ACE testing for suspected sarcoidosis, anterior-chamber paracentesis with PCR for toxoplasma, or conjunctival biopsy with immunofluorescence for mucous membrane pemphigoid.

Revisions to medications (43.1\%) frequently converted general statements into concrete regimens, for instance specifying fortified hourly vancomycin and ceftazidime in keratitis, detailing steroid pulses followed by taper in optic neuritis, or adding pyridostigmine when myasthenia was considered. Surgical and procedural interventions (38.2\%) were adjusted to align with revised diagnoses or to add ancillary procedures such as intravitreal anti-vascular endothelial growth factor (anti-VEGF) injections, lumbar puncture, or plaque brachytherapy. Referrals and multidisciplinary care (38.2\%) often added consultations, examples including neurologists, endocrinologists, ocular oncologists, and developmental pediatricians, which were often absent in the initial plans. Patient education and counselling (32.4\%) expanded to include items such as smoking cessation advice in thyroid eye disease and detailed discussions of surgical risks in congenital ptosis. Social and support services referrals (5.9\%) were least common, for example directing families to peer support groups or arranging psychological counselling after vision-threatening diagnoses.

Taken together, these patterns indicate that after viewing AMIE's structured response, raters most often increased the specificity and scope of investigations and treatments and, in a substantial minority of cases, augmented patient-centred elements including education, interdisciplinary collaboration and psychosocial support.

\subsection{Preference and quality comparison}

\begin{figure}[htbp!]
\centering
\includegraphics[width=0.6\textwidth]{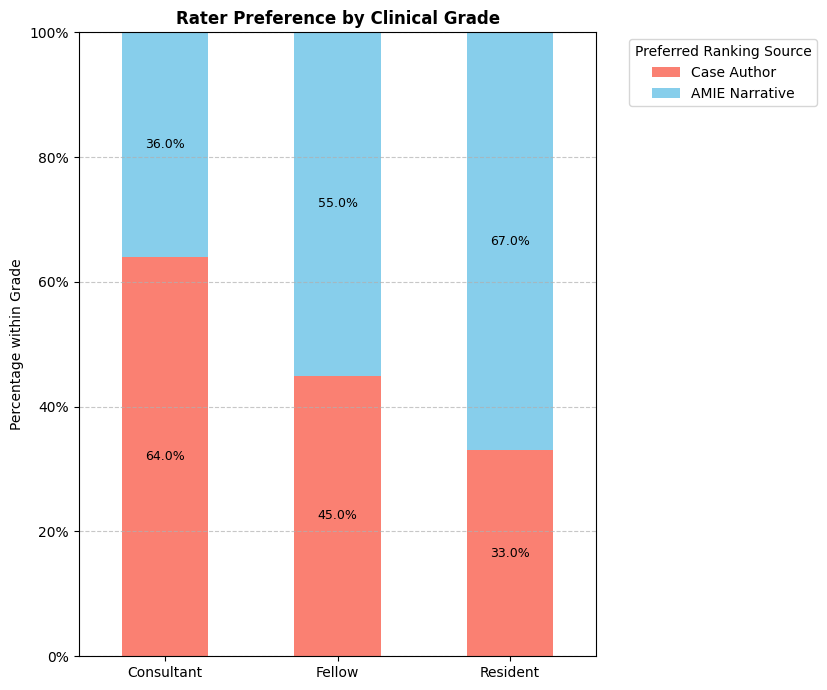}
\vspace{0.1cm}
\caption{\textbf{Rater preference by clinical grade (AMIE narrative vs. case author)} Stacked bar plot showing, within each grade (consultant, fellow, resident), the percentage of head-to-head comparisons preferring AMIE's narrative response versus the case author reference; bars sum to 100\% per grade.}
\label{fig:rater_preference_clinical_grade}
\end{figure}

In the masked, head-to-head comparison between AMIE's narrative responses and the case author reference responses, raters preferred AMIE's output in 52.7\% of pairings and the human reference in 47.3\%. Preference varied by clinical grade: Consultants selected AMIE in 36\% of comparisons, fellows 55\% and residents 67\%. The association between clinical grade and preference was significant ($p=0.0001$) (\cref{fig:rater_preference_clinical_grade}). 

\FloatBarrier

\begin{figure}[H]
  \centering

  \begin{subfigure}[t]{0.48\textwidth}
    \centering
    \includegraphics[width=\linewidth, height=0.25\textheight, keepaspectratio]{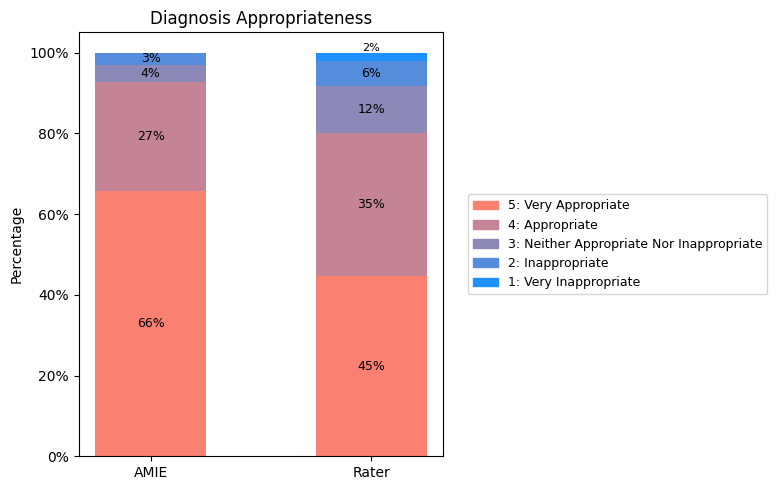}
    \caption{}\label{fig:1A}
  \end{subfigure}
  \hfill
  \begin{subfigure}[t]{0.48\textwidth}
    \centering
    \includegraphics[width=\linewidth, height=0.25\textheight, keepaspectratio]{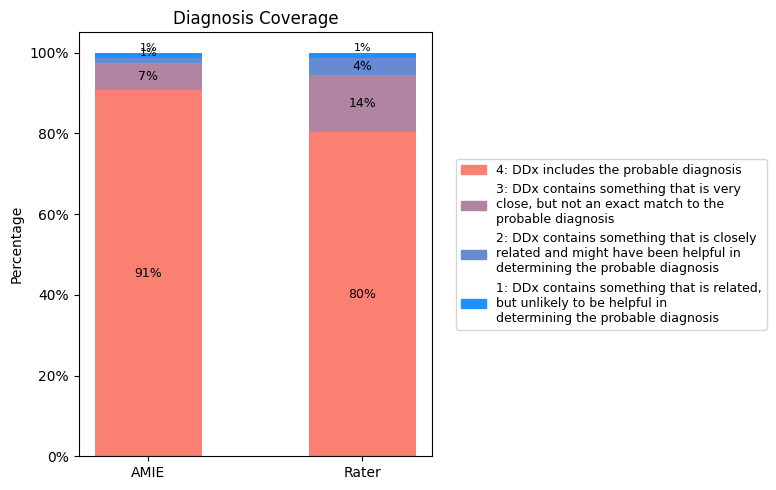}
    \caption{}\label{fig:1B}
  \end{subfigure}

  \vspace{1em}

  \begin{subfigure}[t]{0.48\textwidth}
    \centering
    \includegraphics[width=\linewidth, height=0.25\textheight, keepaspectratio]{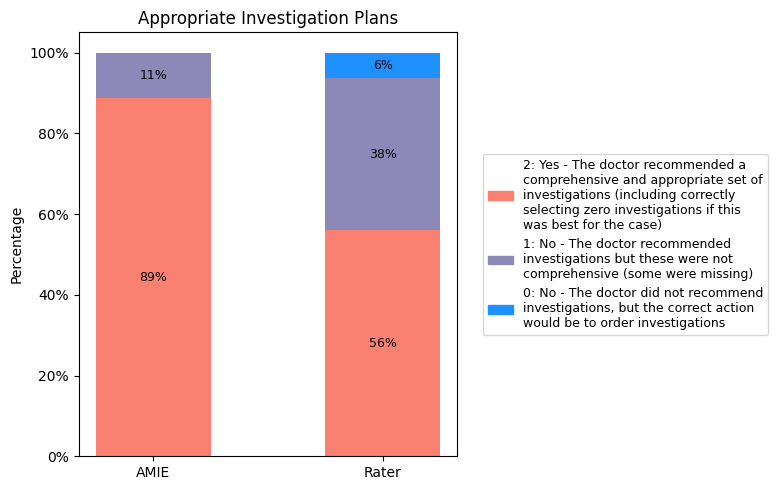}
    \caption{}\label{fig:1C}
  \end{subfigure}
  \hfill
  \begin{subfigure}[t]{0.48\textwidth}
    \centering
    \includegraphics[width=\linewidth, height=0.25\textheight, keepaspectratio]{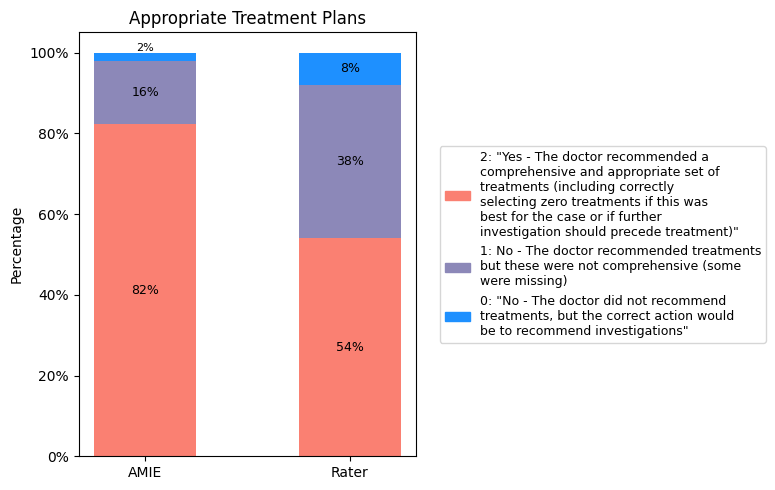}
    \caption{}\label{fig:1D}
  \end{subfigure}

  \vspace{1em}

  \begin{subfigure}[t]{0.6\textwidth}
    \centering
    \includegraphics[width=\linewidth, height=0.25\textheight, keepaspectratio]{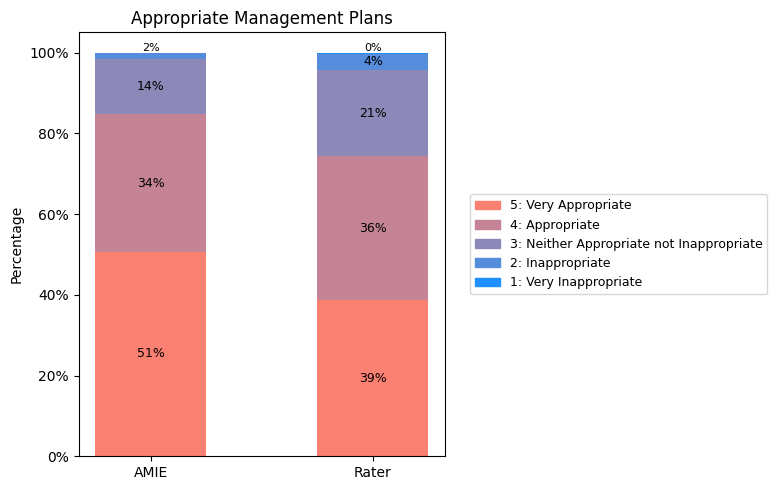}
    \caption{}\label{fig:1E}
  \end{subfigure}

  \caption{\textbf{Rubric-based evaluation of AMIE vs rater responses.} Five stacked bar plots show, for each item, the within-item distribution of ratings for AMIE and for raters. Panels: (A) Diagnosis appropriateness (5-point), (B) Diagnosis coverage (proximity of the DDx to the probable diagnosis) (5-point), (C) Investigation plans (3-point), (D) Treatment plans (3-point), and (E) Overall management plan appropriateness (5-point). Percentages in each panel sum to 100\%. Rating anchors follow the evaluation rubric in \cref{table:eval_rubrics}; DDx = differential diagnosis}
  \label{fig:rubric_scores}
\end{figure}

\vspace{-0.5cm}
Following the preference task, raters, still masked to authorship, scored both narratives on the five criteria rubric (\cref{table:eval_rubrics}). AMIE outperformed the human reference on all five graded domains: diagnostic appropriateness (Very Appropriate 65.7\% for AMIE vs 44.7\% for human), inclusion of the probable diagnosis (90.7\% vs 80.3\%), appropriate investigations (88.7\% vs 56.0\%), appropriate treatments (82.3\% vs 54.0\%), and overall management quality rated Appropriate or better (84.7\% vs 74.4\%) (\cref{fig:rubric_scores}). Paired Wilcoxon signed‑rank tests confirmed significant advantages for AMIE on the five graded domains (all $p <0.005$).

\renewcommand\thesubfigure{\Alph{subfigure}}
\captionsetup[subfigure]{labelformat=simple, labelsep=space}

\FloatBarrier

\clearpage
\subsection{Post-study feedback and recommendations for improvement.}

Upon completing all assigned cases, raters answered a four-item Likert questionnaire on their experience with AMIE (\cref{table:post_survey_feedback}). Most raters selected ‘consistently' for clarity (8/9), meaningful insights (7/9), and comfort considering recommendations generated by AMIE (8/9). Two-thirds (6/9) reported the responses consistently encouraged more critical reflection.

\begin{table}[htbp!]
\centering
\renewcommand{\arraystretch}{1.1}
\begin{tabular}{p{8cm}ccc}
\hline
\textbf{Statement} & \textbf{Consistently} & \textbf{Sometimes} & \textbf{Rarely} \\
\hline
AI responses were clear and easy to understand & 88.9\% & 11.1\% & 0\% \\
AI provided meaningful insights that complemented my clinical reasoning & 77.8\% & 22.2\% & 0\% \\
I would feel comfortable considering the AI's recommendations in a clinical discussion & 88.9\% & 11.1\% & 0\% \\
AI responses encouraged me to reflect more critically on my initial approach & 66.7\% & 33.3\% & 0\% \\
\hline
\end{tabular}
\vspace{0.2cm}
\caption{\textbf{Post-study feedback from raters on AMIE}. Counts and percentages (n=9) for a four-item, three-point Likert questionnaire completed after all cases.}
\label{table:post_survey_feedback}
\end{table}

Raters also answered two open-ended questions: (1) which aspects of the responses were most helpful, and (2) how they could be improved. An inductive thematic analysis of all comments identified eight themes:
\begin{itemize}
    \item Structured clinical reasoning with explicit positive/negative findings
    \item Comprehensive differentials and management plans
    \item Prioritisation and personalisation toward the most likely diagnosis
    \item Selective and staged testing (first-line vs second-line)
    \item Safety emphasis and clearer signposting of red-flag diagnoses
    \item Evidence, transparency and local context (citations, country-specific drugs)
    \item Concise, scannable formatting (subheadings/bullets)
    \item Human-AI interaction and risk of over-reliance
\end{itemize}

Raters consistently valued the structure and clinical logic of the responses, highlighting clear summaries of positives and negatives, likelihood-ordered differentials, and detailed investigation/management plans. Several described this as a practical cognitive scaffold: ``clear structure; weighed pros and cons'' that helped them not forget examinations and check medication doses outside their usual practice. The breadth and completeness of differentials and plans were frequently praised as reassuring and useful under time pressure (``consistently comprehensive differential \ldots{} welcome aid'').

Set against these strengths, raters asked for stronger prioritisation and personalisation. Lists were often perceived as over-inclusive, with a preference for a focused account of the leading diagnosis and a succinct explanation linking the pattern of signs and symptoms to that conclusion: ``too many diagnostics/treatments rather than missing them \ldots{} prefer more focused answers highlighting the most likely diagnosis and why.'' Relatedly, clinicians emphasised selectivity in testing and staged work-ups, requesting clear separation of essential first-line investigations from second-line options to avoid low-yield or duplicated testing (``discriminate between absolutely necessary investigations and those for a second step'').

Raters also called for clearer safety signposting. Even when unlikely, red-flag conditions should be flagged prominently and early; for example, in painful third-nerve palsy, microvascular aetiology may be plausible, but aneurysm and giant cell arteritis require explicit consideration. For evidence and transparency, clinicians asked for references or links to trusted guidelines and brief caveats about limitations, along with local adaptation (e.g., country-specific drug names and availability).

On usability, respondents preferred greater concision and consistent subheadings (Differentials, Investigation, Management), with bullets where appropriate, to support rapid scanning in clinical contexts. Finally, several reflected on human-AI interaction, noting occasional self-doubt or over-reliance when an AI suggestion conflicted with their judgement: ``when the AI feels inappropriate, I start to doubt my own judgement.''

Overall, clinicians viewed the responses as structured, comprehensive and practically helpful, while advocating for tighter focus on the probable diagnosis, staged and selective testing, clearer red-flag emphasis, transparent sourcing adapted to local context, and more concise presentation to maximise clinical utility.

\section{Discussion}

In this evaluation, we demonstrate that AMIE can serve as a support tool that complements ophthalmologists' performance. Across 100 clinical vignettes, AMIE's standalone accuracy was comparable to the raters' baseline, a notable result given the model was not fine-tuned on a dedicated ophthalmology dataset. Its primary utility was evident in its complementary effect: after consulting AMIE's structured analysis, ophthalmologists improved the ranking of correct diagnoses within their differentials, and inter-ophthalmologist agreement rose by approximately ten percentage points. Furthermore, this interaction prompted revisions to the differential diagnosis in 32.3\% of cases and the management plan in 38.2\% of cases, with changes most often increasing the specificity and scope of investigations and treatments. Precisely because AMIE had no ophthalmology specific fine‑tuning, the study also surfaced subspecialty blind spots (e.g., ocular oncology), providing a concrete blueprint for targeted domain adaptation in future versions.

Our work advances the evaluation of AMIE by moving beyond accuracy-only benchmarks that are common in ophthalmic literature. While previous studies have often focused on exam-style knowledge tests or model-versus-model comparisons, our study explores how AMIE interacts with expert reasoning. The outcomes we measured speak to the ranking of diagnoses, agreement among clinicians, and the content of the plan. Interpreting the model's impact across different levels of expertise reveals important nuances. A salient finding was the variation in preference by clinical grade: residents tended to favour AMIE's comprehensive narrative, whereas consultants preferred the more focused human reference response. The key implication is that a “one-size-fits-all” output may be suboptimal, highlighting an important area for future research into personalised prompting that can tailor the AI's response style to the user's level of experience and specific clinical needs. Consistent with \cref{table:diag_by_grade}, residents showed a numerically higher baseline and no significant gain after reviewing AMIE, whereas fellows and consultants improved. A plausible explanation is differences in skill set and scope of practice: residents are immersed in general ophthalmology and closer to exam-style learning, and may list wider differentials and recall guideline based steps that favour top-1 and top-3 metrics. In contrast, fellows and consultants, who often practise within narrower subspecialties, may benefit more when reasoning outside their routine domains.

Our thematic analysis of ophthalmologist feedback provides a clear blueprint for improvements. While raters valued the structured and comprehensive nature of AMIE's responses, they also provided concrete suggestions for enhancing clinical utility. Key themes included a desire for stronger prioritisation around the leading diagnosis, staged testing that separates first line from second line, earlier and clearer red flag signposting, and concise formatting that is easy to scan. Notably, ophthalmologists called for greater transparency and localisation, such as the inclusion of citations to clinical guidelines and the adaptation of recommendations to local formularies and country-specific drug names, which would be crucial for safe real-world deployment.

Our study used text vignettes that contained narrative descriptions of history, examination, and where relevant, imaging findings. This isolates text based reasoning. A next step could be to supply the actual images and other modalities alongside the clinical description, which better reflects real clinical data flow. Ophthalmology is well suited to this progression. Many diagnoses are strongly informed by the patient's story. Time course, which eye is affected, and distinctive symptom patterns often narrow the differential before examination. This makes the specialty a natural fit for LLMs that can elicit, structure, and reason over detailed narratives, and then pose targeted next questions. At the same time, ophthalmology is highly visual. Clinically useful imaging ranges from simple photographs of the external eye to colour fundus photography and optical coherence tomography. The research base for image centred AI is mature, and there are regulated AI as a medical device tools in retinal imaging \cite{ongScopingReviewArtificial2025}. There is also evidence that clinical deployment of these tools (diabetic retinopathy screening AI) can support care for underserved groups \cite{huangAutonomousArtificialIntelligence2024a}. Multimodal systems specific to eye care underline the opportunity. EyeFM is a multimodal vision language copilot trained across several ocular imaging modalities and paired clinical text \cite{wuEyecareFoundationModel2025}. In parallel, the Med-Gemini family extends Gemini to medical images and text, and in their evaluations the models performed strongly on ophthalmology image classification and approached the performance of specialist models trained for single tasks \cite{yangAdvancingMultimodalMedical2024b}. Med-Gemini has also been tested with multimodal inputs in other clinical settings such as photographs of the skin, electrocardiograms, and clinical documents \cite{saabAdvancingConversationalDiagnostic2025}. Taken together, these strands support the development and evaluation of a domain adapted multimodal AMIE for ophthalmology that pairs structured history with targeted ingestion of ocular images.

Translating such a system from vignettes to practice will require prospective studies that measure clinically meaningful endpoints. Recent work in African primary care provides a useful template. A large real world study in Nairobi primary care clinics reported fewer diagnostic and treatment errors when clinicians had access to LLM supported decision support \cite{koromAIbasedClinicalDecision2025}.  In addition, cluster based trials and other pragmatic studies are now being launched across the region to evaluate LLM supported care with frontline workers, including community health workers and designs that enable voice based interaction in local languages \cite{mateenTrialsLLMsupportedClinical2025}. These programmes focus on outcomes beyond accuracy and are intended to test implementation at scale. Ophthalmology is a natural next setting for similar studies. Prospective evaluations in eye services could include smartphone supported triage for red eye in primary care, community diabetic retinopathy screening with fundus photography, and macula and glaucoma pathways that combine examination with OCT. Outcomes should include change in diagnostic ranking, appropriateness and selectivity of investigations, time to referral, safety events such as missed red flags, patient adherence, and equity metrics across language, device, and access. Studies should also evaluate task sharing models in which trained non physician eye care providers operate within clear scopes of practice. Task sharing and task shifting are long standing public health approaches used to extend specialist capacity and improve access where workforce gaps are greatest \cite{worldhealthorganizationTaskShiftingRational2007}.

This study has several limitations. First, the high rate of plan “revisions” may partly reflect a documentation effect rather than a true change in clinical intent: busy clinicians often record only immediate next steps, and when shown a comprehensive plan they may elaborate items already intended; accordingly, revision rates are best interpreted as an interaction/usability signal rather than a direct measure of improved decision-making. Second, each vignette was authored to lead to a pre-specified principal diagnosis and to reinforce a discrete learning objective. While pedagogically useful, such “well-posed” cases may idealise presentations and under-represent the ambiguity, comorbidity, conflicting findings and missing data that characterise routine care; performance here may therefore not fully generalise to real-world encounters. Third, learning, order and fatigue effects are possible: raters answered many cases and could have adapted their responses over time independent of model input, potentially inflating or attenuating measured changes. Fourth, the rater and case author samples were modest (nine ophthalmologists; five case authors), and the evaluation relied on preference judgements and rubric-based ratings which, though structured, retain some subjectivity; together these factors may limit precision and generalisability across settings and subspecialties.

\section{Conclusion}

In summary, this study demonstrates that a domain-general clinical LLM can achieve diagnostic accuracy comparable to ophthalmologists using text-based vignettes and, more importantly, can alter expert behaviour by changing diagnostic ranking, increasing agreement between clinicians, and enriching the content of management plans. In masked, head-to-head comparisons, raters also preferred AMIE's narrative responses across multiple scored domains. These findings highlight the potential of conversational LLMs to complement clinical reasoning in ophthalmology. Future work should build on this foundation through targeted domain adaptation, multimodal extensions that incorporate ocular imaging, and prospective studies in real-world and resource-limited settings, including task-sharing models, to ensure that the technology advances both clinical quality and health equity.

\subsubsection*{Acknowledgments}

We thank Samuel Schmidgall and Sebastian Nowozin for their
comprehensive review and detailed feedback on the manuscript.

\subsubsection*{Data availability}
The clinical vignettes are available to view in the appendix or download from the arXiv page under ancillary files.

\subsubsection*{Code availability}
AMIE is an LLM-based research AI system for medical applications. We are not open-sourcing model code and weights due to the safety implications of unmonitored use of such a system in medical settings. In the interest of responsible innovation, we will be working with research partners, regulators, and providers to validate and explore safe onward uses of AMIE. For reproducibility, we have documented technical deep learning methods while keeping the paper accessible to a clinical and general scientific audience. Our work builds upon Gemini 1.5, for which technical details have been described extensively in the technical report~\cite{reid2024gemini}.

\subsubsection*{Competing interests}
This study was funded by Alphabet Inc and/or a subsidiary thereof (`Alphabet'). The following authors are employees of Alphabet (Google Research or Google DeepMind) and may own stock as part of their standard compensation package: Anil Palepu, Khaled Saab, Tao Tu, Ryutaro Tanno, Yong Cheng, Mike Schaekermann, S. Sara Mahdavi, Elahe Vedadi, David Stutz, Vivek Natarajan, Alan Karthikesalingam, and Wei-Hung Weng.

Pearse A. Keane is a cofounder of Cascader Ltd. and has acted as a consultant for insitro, Retina Consultants of America, Roche, Boehringer Ingelheim, and Bitfount, and is an equity owner in Big Picture Medical. He has received speaker fees from Zeiss, Thea, Apellis, and Roche, and grant funding from Roche. He has received travel support from Bayer and Roche. He has attended advisory boards for Topcon, Bayer, Boehringer Ingelheim, and Roche.

\clearpage
\newpage
\setlength\bibitemsep{3pt}
\printbibliography
\balance

\end{refsection}

\newpage
\begin{refsection}

\clearpage

\renewcommand{\thesection}{A.\arabic{section}}
\renewcommand{\thefigure}{A.\arabic{figure}}
\renewcommand{\thetable}{A.\arabic{table}} 
\renewcommand{\theequation}{A.\arabic{equation}} 
\renewcommand{\theHsection}{A\arabic{section}}

\setcounter{section}{0}
\setcounter{figure}{0}
\setcounter{table}{0}
\setcounter{equation}{0}


\noindent \textbf{\LARGE{Appendix}}\\

In the following sections, we report additional details. We provide:

\begin{itemize}[leftmargin=1.5em,rightmargin=0em]
  \item Example case vignettes (\cref{sec:vignette}).
  \item Prompting details for search, self-critique, and revision (\cref{sec:prompts}).
  \item Example of AMIE-generated responses (\cref{sec:ex_outputs}).
  \item Clinician post-survey feedback (\cref{sec:post_survey_feedback}).
  \item Diagnostic accuracy by subspecialty (\cref{sec:subspecialty_accuracy}).
  \item Clinical cases (\cref{sec:clinical_cases}).
\end{itemize}

\section{Case vignette}
\label{sec:vignette}

\begin{figure}[htbp!]
\begin{tcolorbox}[title=Case Vignette - Iris Melanoma]
A 51-year-old man presented for routine ophthalmologic exam. He denies any symptoms and has no vision loss, flashes, floaters, or eye pain. His best corrected visual acuity is 20/20 in each eye, intraocular pressures 15 and 16 by tonometry, and pupils equal, round, and reactive to light without afferent pupillary defect. On slit lamp exam, the lids and lashes are normal, the corneas are clear without opacity or irregularity, and the anterior chamber is deep and quiet in both eyes. The left iris shows a 4 mm irregular, elevated, darkly pigmented lesion at the 5:00 o'clock position, causing corectopia and ectropion uveae. Gonioscopy reveals extension of the lesion into the angle. The vitreous is normal in both eyes. The optic nerve has a normal appearance with a cup-to-disc ratio of 0.3 in each eye. The retina and the vessels are normal.
\end{tcolorbox}
\vspace{0.2cm}
\caption{\textbf{Example case vignette 1.}}
\label{fig:case_vignette_1}
\end{figure}

\begin{figure}[htbp!]
\begin{tcolorbox}[title=Case Vignette - Sickle cell retinopathy]
A 27-year-old male of African descent presented to the ophthalmology clinic complaining of a three day history of decreased vision and floaters in his right eye. He began experiencing these symptoms suddenly after coming home from work. He is an investment banker and finds his work very stressful. He is known for -2.50 myopia OU and wears glasses. He has no previous medical history. He has no history of intravenous drug use. His mother has a history of glaucoma and the family history of his father’s side of the family is unknown.
 
 On examination, there were no abnormalities noted on external exam. His corrected visual acuity was 20/100 OD and 20/20 OS. Extraocular movements were full and non-painful OU. Confrontational visual fields were full OU. Pupils were 3 mm OU and had normal pupil responses. On anterior segment examination, there were no abnormalities noted OU. On posterior segment examination, there was vitreous haemorrhage and superotemporal neovascularization noted OD. There was superotemporal neovascularization also noted OS but without vitreous haemorrhage. His intraocular pressures were 16 mmHg OU.
\end{tcolorbox}
\vspace{0.2cm}
\caption{\textbf{Example case vignette 2.}}
\label{fig:case_vignette_2}
\end{figure}

\clearpage
\section{Prompts}
\label{sec:prompts}

\begin{figure}[htbp!]
\begin{tcolorbox}[title=Draft Response Prompt]
\small
\textbf{Instructions}: You are an experienced ophthalmologist. You will be asked few questions about a clinical vignette. Read the case vignette, and think step by step carefully before answering. \\
\textbf{Answering rules}
\begin{itemize}
    \item First, read through the whole case and identify positive findings, negative findings, medications, social history, medical history, and family history. Try the best to identify at least one item in each category. If nothing identified, mark as "None".
    \item Then, list out at least five possible differential diagnoses (DDx), DDx reasonings, and the float confidence score ranged from 0.0 to 1.0 of the DDx being correct, and ranked them by likelihood. Do consider the "uncertain" as one option and give an explanation why you don't have enough confidence in other diagnoses. If no diagnosis is identified, assign the confidence score of 1.0 to the "uncertain" diagnosis.
    \item For each diagnosis, give reasons why the diagnosis is likely to be suspected (support), and why it is unlikely to be suspected (opposition), and also give the confidence score for this diagnosis.
    \item Confidence scores across all diagnoses and uncertain should be summed up to 1.0. To compute the confidence score of each diagnosis / uncertain: (1) score each diagnosis being correct from 0.0 to 1.0, (2) sum up the normalized score across all diagnoses, (3) divide each score from (1) by the sum from (2).
    \item Next, identify which diagnosis is the most likely predicted diagnosis.
    \item Finally, list out all necessary next step investigations to reach a final diagnosis, and along with the possible management plans if after the investigation. The format should be: "<investigation>: <if normal then do this>, <if abnormal then do that>." For example, "MRI: If normal, obtain lumbar puncture and start acetazolamide pills. If abnormal, consult neurosurgery.".
\end{itemize}
\textbf{Format}: Format your response as follows:
\footnotesize
\begin{verbatim}
=== Case Summary ===
positive findings: ...
negative findings: ...
medications:  ...
social history:  ...
medical history:  ...
family history:  ...
differential diagnosis:
  - diagnosis: 
    reasoning: ...
        support:
          - topic:
            explanation:
            importance:
        opposition:
          - topic:
            explanation:
            importance:
    confidence score:
  - ...
most likely diagnosis: 
next steps:  ...
\end{verbatim}
\small
\textbf{Case vignette}: \color{blue}[Case Content]\color{black} \\
\textbf{Questions}: What are the suspected differential diagnoses, and the reasons of including and excluding them? What is the most likely diagnosis and the reasons given all the information? What is the next step investigations and management plans for confirming the diagnosis?
\end{tcolorbox}
\vspace{0.2cm}
\caption{\textbf{AMIE draft initial response prompt.} AMIE used this prompt to draft a response to case questions.}
\label{fig:inference_prompt}
\normalsize
\end{figure}

\begin{figure}[htbp]
\begin{tcolorbox}[title=Search Retrieval Prompt]
\textbf{Instruction:} You are an expert ophthalmologist. Given the following patient information with eye conditions, please generate at least five critical search query terms that can be helpful to reach the final diagnosis.\\

\color{blue}[AMIE's first response] \color{black}\\

List at least 5 critical search query terms in a list like:
\begin{verbatim}
- SEARCH_QUERY_1
- SEARCH_QUERY_2
- SEARCH_QUERY_3
- SEARCH_QUERY_4
- SEARCH_QUERY_5
\end{verbatim}
\end{tcolorbox}
\vspace{0.2cm}
\caption{\textbf{Search retrieval prompt.} AMIE used this prompts to retrieve search results for each of clinical findings and management recommendations.}
\label{fig:search_prompts}
\end{figure}

\begin{figure}[htbp!]
\begin{tcolorbox}[title=Self-Critique Prompt]

\textbf{Instructions:} You are an expert ophthalmologist. Consider the following provided information:\\

\textbf{Patient Information: }\\
\color{blue}[Case Content]\color{black}\\

\textbf{Another ophthalmologist's differential diagnosis candidates and management plan: }\\
\color{blue}[AMIE's first response]\color{black}\\

\textbf{Google search results about the key terms in the patient information: }\\
\color{blue}[Web search results given five search queries]\color{black}\\

\textbf{Google search results about the differential diagnosis candidates: }\\ \color{blue}[Web search results given all diagnoses in AMIE's first response]\color{black}\\

Considering what you know about ophthalmology and the Google search results above, give a careful critique of the differential diagnosis candidates and management plan for the ophthalmologist's case summary. Specifically focusing on whether the patient info and the differential diagnosis candidates can be matched appropriately, and whether the management plans is suitable for the diagnosis. Explicitly explain why you agree or disagree with each diagnosis candidate provided by the ophthalmologist according to your knowledge and the web search results one by one. For those diagnoses you agree with, list them in the Good section, else list in the Bad section. List as much suggestions as you can. Also be sure to note any innaccuracies.

Format your response as follows:
\begin{verbatim}
Good: ...
Bad: ...
Summary: ...
\end{verbatim}
\end{tcolorbox}
\vspace{0.2cm}
\caption{\textbf{Self-critique prompt.} AMIE used this prompts to critique its drafted responses to the cases.}
\label{fig:critique_prompt}
\end{figure}

\begin{figure}[htbp]
\begin{tcolorbox}[title=Revise Response Prompt]
\textbf{Instructions}: You are an expert ophthalmologist. Consider the following provided information:

\textbf{Patient information: }\\
\color{blue}[Case Content]\color{black}\\

\textbf{First ophthalmologist's differential diagnosis candidates and management plan: }\\
\color{blue}[AMIE's first response]\color{black}\\

\textbf{Critique from the second opinion: }\\
\color{blue}[Critique output]\color{black}\\

Revise the answer to incorporate this feedback. Make sure to keep the original answer format, and just revise based on your judge. Only change the diagnoses and the ranking if the critique and the original answer do not agree with each other.
\end{tcolorbox}
\vspace{0.2cm}
\caption{\textbf{Revise Response Prompt.} AMIE used this prompts to revise its drafted responses based on its self-critique.}
\label{fig:revision_prompt}
\end{figure}

\begin{figure}[htbp]
\begin{tcolorbox}[title=Narrative Generation Prompt]
You are an English writer who knows how to rewrite the clinical diagnosis reasoning and management plans. Given the example of the expected rewrite output, and the original text of context of clinical diagnosis reasoning and management plans provided by the expert, please follow the structure of the examples of expected rewrite output to rewrite the original text without dropping any clinically important information, such as the rank of differential diagnoses, and the key topics and terms of the next step plans. You should start from the most likely diagnosis, followed by symptoms, signs, examinations, and remaining differential diagnoses, then the next investigations and plans. Your rewrite should keep all critical clinical information, but be narrative and descriptive, and should remove all itemized or bullet point structures in the original text, drop the model prediction confidence scores, and you must mention the rank of the differential diagnoses explicitly. However, you should also keep the length of rewrite output at most 20\% longer than the averaged length of examples provided and make the rewrite output descriptive and readable, but make sure to not omit necessary information, no significant errors, and still follow the case vignette well. If the length of rewrite output is more than 25\% longer than the averaged length of examples provided, shorten the explanations for the diagnoses that are not top-2 most likely diagnoses and remove the detailed descriptions of the management plans.\\

\textbf{5 examples of the expected rewrite output: }\\
\color{blue}[Rewrite examples]\color{black}\\

\textbf{Clinical diagnosis reasoning and management plans to rewrite: }\\
\color{blue}[Structured Response]\color{black}\\
\end{tcolorbox}
\vspace{0.2cm}
\caption{\textbf{Narrative generation prompt.} 
AMIE used this prompt to generate a narrative response from the structured response}
\label{fig:narrative_generation}
\end{figure}

\clearpage
\section{Example AMIE outputs}
\label{sec:ex_outputs}

\begin{figure}[hbtp!]
\begin{tcolorbox}[title=Example of AMIE Structured Response (Iris Melanoma)]
*** Case Summary ***\\
- Positive findings: 4 mm irregular, elevated, darkly pigmented lesion on the left iris at the 5:00 o'clock position, Corectopia in the left eye, Ectropion uveae in the left eye, Extension of the lesion into the angle on gonioscopy\\
- Negative findings: No vision loss, No flashes or floaters, No eye pain, Normal best corrected visual acuity (20/20 in each eye), Normal intraocular pressures (15 and 16), Pupils equal, round, and reactive to light without afferent pupillary defect, Normal lids and lashes, Clear corneas without opacity or irregularity, Deep and quiet anterior chamber in both eyes, Normal vitreous in both eyes, Normal optic nerve appearance with a cup-to-disc ratio of 0.3 in each eye, Normal retina and vessels\\
- Medications: None\\
- Medical history: None\\
- Social history: None\\
- Family history: None\\

- Differential diagnosis\\
  * Diagnosis: Iris Melanoma \\
    * Supports: [Clinical Presentation] The presence of a darkly pigmented, elevated, irregular lesion on the iris with corectopia, ectropion uveae, and extension into the angle is highly suggestive of an iris melanoma. These are classic features of this malignancy. The patient's age, while not the most common demographic, does not rule it out. The lack of symptoms is typical for early iris melanomas.\\
    * Oppositions: [Patient's Age] While iris melanomas can occur at this age, they are more common in older individuals. However, this is not a strong opposing factor.\\

  * Diagnosis: Iris Nevus \\
    * Supports: [Pigmented Iris Lesion] Iris nevi are common pigmented lesions of the iris and can sometimes be elevated and irregular. They can also cause corectopia and ectropion uveae if large enough.\\
    * Oppositions: [Extension into the Angle] Extension into the angle is less common for iris nevi and more suggestive of a malignant process like melanoma. The size and irregularity of the lesion also favor melanoma. Nevi are usually flat or minimally elevated, and less likely to cause corectopia, ectropion uveae, and angle involvement.\\

  * Diagnosis: Pigmented Epithelial Cyst \\
    * Supports: [Elevated Iris Lesion] Pigmented epithelial cysts can be elevated and cause corectopia and ectropion uveae. They can also be pigmented.\\
    * Oppositions: [Pigmentation and Irregularity] Iris cysts are typically smooth and round, and less likely to be darkly pigmented and irregular. They also do not typically extend into the angle.\\

  * Diagnosis: Iris Cyst \\
    * Supports: [Elevated Iris Lesion] Iris cysts can be elevated and cause corectopia and ectropion uveae. They can also be pigmented.\\
    * Oppositions: [Pigmentation and Irregularity] Iris cysts are typically smooth and round, and less likely to be darkly pigmented and irregular. They also do not typically extend into the angle.\\
\end{tcolorbox}
\vspace{0.2cm}
\caption{\textbf{Example structured response from AMIE.} A response generated by AMIE for a case vignette of iris melanoma, illustrating the structured format including summary of findings, differential diagnosis with reasoning and confidence scores, and proposed next steps/management plan.}
\label{fig:structured_response_1}
\end{figure}

\begin{figure}[hbtp!]
\begin{tcolorbox}[title=Example of AMIE Structured Response (Iris Melanoma)]

 * Diagnosis: Medulloepithelioma \\
    * Supports: [Pigmented Iris Lesion] Medulloepitheliomas can be pigmented and can occur in the iris. They can also cause corectopia and ectropion uveae.\\
    * Oppositions: [Age of Onset] Medulloepitheliomas are more common in children and are rare in adults. The patient's age makes this diagnosis less likely.\\
    
  * Diagnosis: Metastatic Lesion to the Iris \\
    * Supports: [Pigmented Iris Lesion] Metastatic lesions can present as pigmented lesions in the iris and can cause corectopia and ectropion uveae.\\
    * Oppositions: [Lack of Systemic Symptoms] The patient has no systemic symptoms or known history of cancer, making metastasis less likely. Metastatic lesions are also less likely to be the first presentation of cancer.\\

  * Diagnosis: Foreign Body \\
    * Supports: [Pigmented Iris Lesion] A retained intraocular foreign body can sometimes cause a pigmented lesion and associated changes.\\
    * Oppositions: [Lack of Trauma History] The patient has no history of trauma.\\

- Most likely diagnosis: Iris Melanoma\\
- Next steps\\
  * Urgent Referral to an Ocular Oncologist: Given the high suspicion for melanoma, this patient should be referred to an ocular oncologist immediately for further management.\\
  * Urgent Ultrasound biomicroscopy (UBM): To determine the extent of the lesion, its thickness, and involvement of the ciliary body and angle. This should be done as soon as possible.\\
  * Anterior segment photography: To document the lesion's appearance and size for future comparison.\\
  * Systemic evaluation: Including a thorough history and possibly a basic physical exam to rule out any underlying systemic malignancy.\\
  * Excisional biopsy or iridectomy: If the lesion is small and localized, perform excisional biopsy or iridectomy for histopathological diagnosis and treatment. Given the angle involvement, a careful surgical approach is needed to ensure complete removal and avoid seeding.\\
  * Plaque Radiotherapy or Enucleation: If the lesion is large or involves the ciliary body, consider plaque radiotherapy or enucleation.\\
  * Genetic testing: If melanoma is confirmed, consider genetic testing for prognostic information and potential targeted therapies.\\
  * Close Follow-up: Regardless of the initial treatment, close follow-up is essential to monitor for recurrence or metastasis.\\
  * Patient Education: Educate the patient about the high suspicion for melanoma, the need for further testing, and the potential treatment options.\\
\end{tcolorbox}
\vspace{0.2cm}
\caption{\textbf{Example final output from AMIE (continued)}}
\label{fig:structured_response_2}
\end{figure}

\begin{figure}[hbtp!]
\begin{tcolorbox}[title=Example of AMIE Narrative Response (Iris Melanoma)]
The primary concern is an iris melanoma, given the presence of a 4mm, irregular, elevated, darkly pigmented lesion on the left iris at the 5 o'clock position, accompanied by corectopia and ectropion uveae in the same eye, and extension of the lesion into the angle. These findings are highly suggestive of iris melanoma, a malignancy that often presents with such characteristics. Although iris melanomas are more common in older individuals, the patient's age does not rule out this diagnosis. The absence of symptoms is also typical of early iris melanomas.\\

The second most likely diagnosis is an iris nevus. While iris nevi can be pigmented, elevated, and irregular, and can sometimes cause corectopia and ectropion uveae if large enough, their extension into the angle is less common, making melanoma more likely. Additionally, nevi are usually flat or minimally elevated, and less likely to cause corectopia, ectropion uveae, and angle involvement. Other less likely possibilities include a pigmented epithelial cyst, which can be elevated and cause corectopia and ectropion uveae, but these are typically smooth and round, and less likely to be darkly pigmented and irregular, and do not typically extend into the angle. An iris cyst is also a possibility, but similar to pigmented epithelial cysts, they are typically smooth and round, and less likely to be darkly pigmented and irregular, and do not typically extend into the angle. Medulloepithelioma is another differential, but it is more common in children and rare in adults, making it less likely in this case. Metastatic lesions to the iris are also a possibility, but the patient has no systemic symptoms or known history of cancer, making metastasis less likely. Finally, a foreign body could cause a pigmented lesion, but the patient has no history of trauma.\\

Given the high suspicion for iris melanoma, the patient should be urgently referred to an ocular oncologist. An urgent ultrasound biomicroscopy (UBM) should be performed to determine the extent of the lesion, its thickness, and involvement of the ciliary body and angle. Anterior segment photography should be done to document the lesion's appearance and size for future comparison. A systemic evaluation, including a thorough history and possibly a basic physical exam, should be done to rule out any underlying systemic malignancy. If the lesion is small and localized, an excisional biopsy or iridectomy should be performed for histopathological diagnosis and treatment. If the lesion is large or involves the ciliary body, plaque radiotherapy or enucleation should be considered. If melanoma is confirmed, genetic testing should be considered for prognostic information and potential targeted therapies. Close follow-up is essential to monitor for recurrence or metastasis. The patient should be educated about the high suspicion for melanoma, the need for further testing, and the potential treatment options.
\end{tcolorbox}
\vspace{0.2cm}
\caption{\textbf{Example of AMIE narrative response.} A narrative  response generated by AMIE for iris melanoma, aligning the case author writing style.}
\label{fig:narrative_response}
\end{figure}

\begin{figure}[hbtp!]
\begin{tcolorbox}[title=Example of case author reference response (Iris Melanoma)]
The suspected diagnosis is iris melanoma. The presence of a raised iris lesion, particularly in the inferior portions of the iris, with corectopia and ectropion uvea and angle seeding should raise suspicion for the diagnosis. The differential diagnosis for pigmented lesions on the iris includes freckle, iris nevus, melanocytoma, Lisch nodule, and melanoma. A freckle is often small (<1-2 mm) and sits on the surface of the iris. A nevus can penetrate deeper into the iris stroma. A melanocytoma often appears dome-shaped. Iris melanomas, importantly, are more often found in the inferior quadrant, often in middle-aged adults, with a basal mean diameter of 6.2 mm in one retrospective study. If the patient’s iris lesion was non-pigmented, the differential is broad and would include, most commonly, non-neoplastic mimics (iridocorneal endothelial syndrome, iris atrophy, foreign body, coloboma, congenital heterochromia, to name the most common examples), as well as true neoplastic lesions such as metastasis from another site (most commonly breast, lung, kidney, and skin), choristomatous tumors, vascular tumors, fibrous tumors, neural tumors, myogenic tumors, epithelial tumors, xanthomatous tumors, lymphoid tumor, or leukemic tumors. Cystic lesions of the iris may also be included on the differential diagnosis, and these include primary iris cysts such as stromal cysts and pigment epithelial cysts. 
\end{tcolorbox}
\vspace{0.2cm}
\caption{\textbf{Example of case author reference response.} A reference response generated by the case author for iris melanoma.}
\label{fig:ex2}
\end{figure}

\clearpage

\section{Clinician Post-Survey Feedback}
\label{sec:post_survey_feedback}

\begin{figure}[htbp]
\begin{tcolorbox}[
  enhanced, breakable, title filled,
  title=Post-Survey Feedback,
  colback=gray!05, colframe=black!65,
  colbacktitle=black!75, coltitle=white, fonttitle=\bfseries,
  width=\linewidth-6pt, left=2pt, right=2pt, top=2pt, bottom=2pt
]

\setlength{\tabcolsep}{5pt}
\renewcommand{\arraystretch}{1.15}
\small
\newcolumntype{Y}{>{\RaggedRight\arraybackslash}X}
\newcommand{\choice}{$\bigcirc$}

\begin{tabularx}{\linewidth}{@{} Y c c c @{}}
\toprule
\textbf{Statements} & \textbf{Consistently} & \textbf{Sometimes} & \textbf{Rarely} \\
\midrule
The AI-generated responses were clear and easy to understand. & \choice & \choice & \choice \\
The AI responses provided meaningful insights that complemented my clinical reasoning. & \choice & \choice & \choice \\
I would feel comfortable considering the AI’s recommendations as part of a clinical discussion. & \choice & \choice & \choice \\
The AI responses encouraged me to reflect more critically on my initial approach. & \choice & \choice & \choice \\
\bottomrule
\end{tabularx}

\medskip

{\bfseries What aspects of the AI responses did you find most helpful?}\par
\textit{Feel free to share your thoughts openly. No specific format required.}
\begin{tcolorbox}[colback=white,colframe=black!60,boxsep=3pt,
                  left=4pt,right=4pt,top=4pt,bottom=4pt,
                  width=\linewidth,height=1cm]
\end{tcolorbox}

{\bfseries How could the AI responses be improved to better support clinical decision-making?}\par
\textit{Feel free to share your thoughts openly. No specific format required.}
\begin{tcolorbox}[colback=white,colframe=black!60,boxsep=3pt,
                  left=4pt,right=4pt,top=4pt,bottom=4pt,
                  width=\linewidth,height=1cm]
\end{tcolorbox}

\end{tcolorbox}
\vspace{0.2cm}
\caption{\textbf{Post-survey feedback.}
Clinicians evaluated the AMIE structured response using a three-option Likert scale and provided open-ended comments on helpful aspects and improvements.}
\label{fig:post_survey_feedback}
\end{figure}

\clearpage
\section{Diagnostic Accuracy by Subspecialty}
\label{sec:subspecialty_accuracy}

\begin{figure}[htbp!]
\centering
\includegraphics[width=0.7\textwidth]{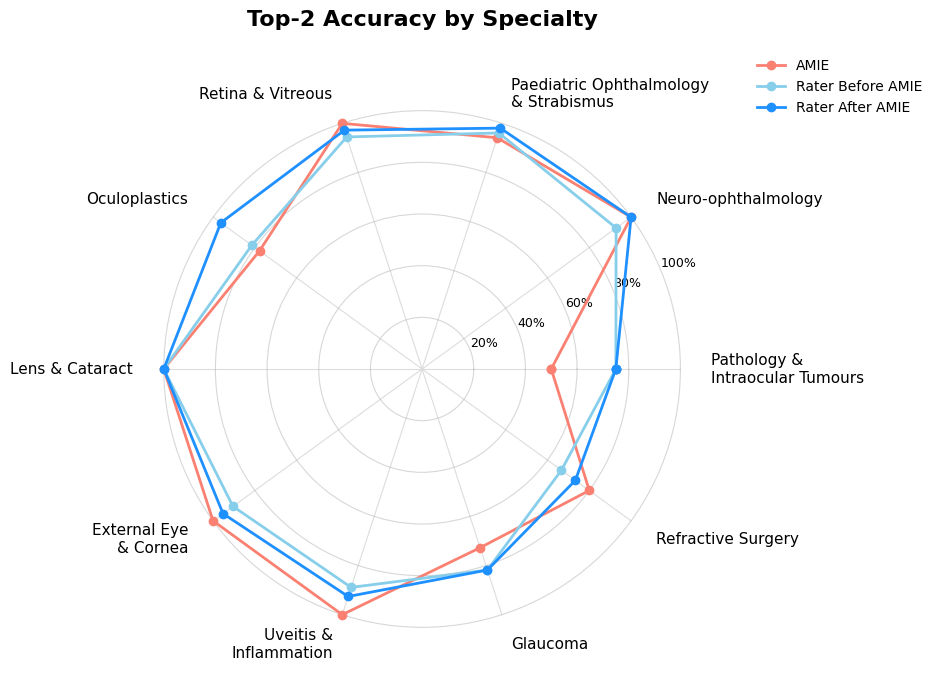}
\vspace{0.1cm}
\caption{\textbf{Top-2 accuracy by subspecialty (Rater Before AMIE, Rater After AMIE, AMIE)} Radar plot showing Top-2 diagnostic accuracy within each subspecialty for AMIE, raters before seeing AMIE, and the same raters after review. Radial axis: 0-100\%.}
\label{fig:radar_plot_top_2}
\end{figure}

\begin{figure}[htbp!]
\centering
\includegraphics[width=0.7\textwidth]{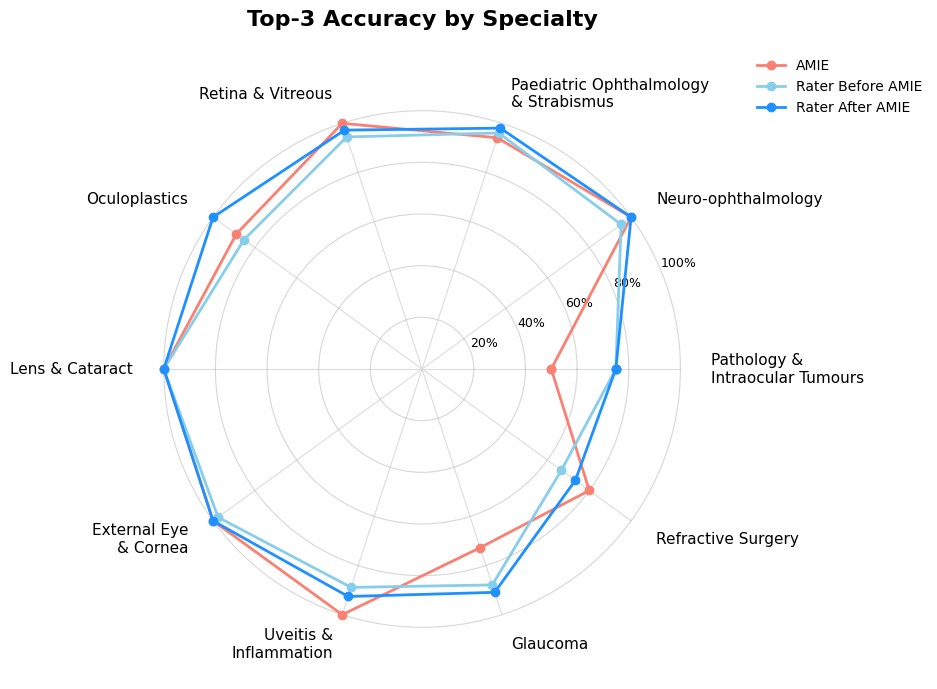}
\vspace{0.1cm}
\caption{\textbf{Top-3 accuracy by subspecialty (Rater Before AMIE, Rater After AMIE, AMIE)} Radar plot showing Top-3 diagnostic accuracy within each subspecialty for AMIE, raters before seeing AMIE, and the same raters after review. Radial axis: 0-100\%.}
\label{fig:radar_plot_top_3}
\end{figure}

\begin{table}[htbp]
\centering
\renewcommand{\arraystretch}{1.1}
\resizebox{\textwidth}{!}{%
\begin{tabular}{llccc}
\toprule
\textbf{Subspecialty} & \textbf{Metric} & \textbf{Rater before AMIE} & \textbf{Rater after AMIE} & \textbf{AMIE} \\
\midrule
\multirow{3}{*}{Ophthalmic pathology \& intraocular tumours} & Top-1 & 66.7\% (41.7--91.7\%) & \textbf{75.0\% (50.0--100.0\%)} & 50.0\% (25.0--75.0\%) \\
 & Top-2 & \textbf{75.0\% (50.0--100.0\%)} & \textbf{75.0\% (50.0--100.0\%)} & 50.0\% (25.0--83.3\%) \\
 & Top-3 & \textbf{75.0\% (50.0--100.0\%)} & \textbf{75.0\% (50.0--100.0\%)} & 50.0\% (16.7--75.0\%) \\
\midrule
\multirow{3}{*}{Neuro-ophthalmology} & Top-1 & 92.9\% (83.3--100.0\%) & 97.6\% (92.9--100.0\%) & \textbf{100.0\% (100.0--100.0\%)} \\
 & Top-2 & 92.9\% (83.3--100.0\%) & \textbf{100.0\% (100.0--100.0\%)} & \textbf{100.0\% (100.0--100.0\%)} \\
 & Top-3 & 95.2\% (88.1--100.0\%) & \textbf{100.0\% (100.0--100.0\%)} & \textbf{100.0\% (100.0--100.0\%)} \\
\midrule
\multirow{3}{*}{Paediatric ophthalmology \& strabismus} & Top-1 & 92.2\% (84.3--98.0\%) & \textbf{96.1\% (90.2--100.0\%)} & 94.1\% (86.3--100.0\%) \\
 & Top-2 & 96.1\% (90.2--100.0\%) & \textbf{98.0\% (94.1--100.0\%)} & 94.1\% (86.3--100.0\%) \\
 & Top-3 & 96.1\% (90.2--100.0\%) & \textbf{98.0\% (94.1--100.0\%)} & 94.1\% (86.3--100.0\%) \\
\midrule
\multirow{3}{*}{Retina \& vitreous} & Top-1 & 91.7\% (80.6--100.0\%) & \textbf{97.2\% (91.7--100.0\%)} & 83.3\% (69.4--94.4\%) \\
 & Top-2 & 94.4\% (86.1--100.0\%) & 97.2\% (91.7--100.0\%) & \textbf{100.0\% (100.0--100.0\%)} \\
 & Top-3 & 94.4\% (86.1--100.0\%) & 97.2\% (91.7--100.0\%) & \textbf{100.0\% (100.0--100.0\%)} \\
\midrule
\multirow{3}{*}{Oculoplastics} & Top-1 & \textbf{70.4\% (55.5--85.2\%)} & \textbf{70.4\% (51.9--88.9\%)} & 66.7\% (48.1--81.5\%) \\
 & Top-2 & 81.5\% (66.7--96.3\%) & \textbf{96.3\% (88.9--100.0\%)} & 77.8\% (59.3--92.6\%) \\
 & Top-3 & 85.2\% (70.4--96.3\%) & \textbf{100.0\% (100.0--100.0\%)} & 88.9\% (77.8--100.0\%) \\
\midrule
\multirow{3}{*}{Lens \& cataract} & Top-1 & 93.3\% (80.0--100.0\%) & \textbf{100.0\% (100.0--100.0\%)} & 80.0\% (60.0--100.0\%) \\
 & Top-2 & \textbf{100.0\% (100.0--100.0\%)} & \textbf{100.0\% (100.0--100.0\%)} & \textbf{100.0\% (100.0--100.0\%)} \\
 & Top-3 & \textbf{100.0\% (100.0--100.0\%)} & \textbf{100.0\% (100.0--100.0\%)} & \textbf{100.0\% (100.0--100.0\%)} \\
\midrule
\multirow{3}{*}{External eye disease \& cornea} & Top-1 & 85.7\% (73.8--95.2\%) & 92.9\% (83.3--100.0\%) & \textbf{100.0\% (100.0--100.0\%)} \\
 & Top-2 & 90.5\% (81.0--97.6\%) & 95.2\% (88.1--100.0\%) & \textbf{100.0\% (100.0--100.0\%)} \\
 & Top-3 & 97.6\% (92.9--100.0\%) & \textbf{100.0\% (100.0--100.0\%)} & \textbf{100.0\% (100.0--100.0\%)} \\
\midrule
\multirow{3}{*}{Uveitis \& ocular inflammation} & Top-1 & \textbf{85.2\% (70.4--96.3\%)} & \textbf{85.2\% (70.4--96.3\%)} & 66.7\% (48.1--85.2\%) \\
 & Top-2 & 88.9\% (77.7--100.0\%) & 92.6\% (81.5--100.0\%) & \textbf{100.0\% (100.0--100.0\%)} \\
 & Top-3 & 88.9\% (77.8--100.0\%) & 92.6\% (81.5--100.0\%) & \textbf{100.0\% (100.0--100.0\%)} \\
\midrule
\multirow{3}{*}{Glaucoma} & Top-1 & \textbf{72.7\% (57.6--87.9\%)} & \textbf{72.7\% (54.5--87.9\%)} & \textbf{72.7\% (57.6--87.9\%)} \\
 & Top-2 & \textbf{81.8\% (69.7--93.9\%)} & \textbf{81.8\% (66.7--93.9\%)} & 72.7\% (57.6--87.9\%) \\
 & Top-3 & 87.9\% (75.8--97.0\%) & \textbf{90.9\% (78.8--100.0\%)} & 72.7\% (57.6--87.9\%) \\
\midrule
\multirow{3}{*}{Refractive surgery} & Top-1 & 53.3\% (26.7--80.0\%) & 53.3\% (33.2--73.5\%) & \textbf{60.0\% (40.0--86.7\%)} \\
 & Top-2 & 66.7\% (40.0--86.7\%) & 73.3\% (53.3--93.3\%) & \textbf{80.0\% (60.0--100.0\%)} \\
 & Top-3 & 66.7\% (40.0--86.7\%) & 73.3\% (53.3--93.3\%) & \textbf{80.0\% (60.0--100.0\%)} \\
\bottomrule
\end{tabular}}
\vspace{0.2cm}
\caption{\textbf{Diagnostic accuracy by subspecialty and Top-N (Rater Before AMIE, Rater After AMIE, AMIE).}
Accuracy (\%) with 95\% confidence intervals (CIs) for raters before seeing AMIE, the same raters after reviewing AMIE, and AMIE itself. 
Bold marks the highest accuracy within each row (ties bolded for all). 
Top-N indicates that the correct diagnosis appears within the top N choices.}
\label{tab:diagnostic_accuracy}
\end{table}

\section{Clinical cases}
\label{sec:clinical_cases}


\begingroup
\small
\setlength{\parskip}{0.25em}

\newcounter{rowctr}\setcounter{rowctr}{0}
\def\prevSubspecialty{} 

\begin{filecontents*}{supplementary_cases.tsv}
Subspecialty	Diagnosis	Vignette	Reasoning
Ophthalmic pathology and intraocular tumours	Iris melanoma	"A 51-year-old man presented for routine ophthalmologic exam. He denies any symptoms and has no vision loss, flashes, floaters, or eye pain. His best corrected visual acuity is 20/20 in each eye, intraocular pressures 15 and 16 by tonometry, and pupils equal, round, and reactive to light without afferent pupillary defect. On slit lamp exam, the lids and lashes are normal, the corneas are clear without opacity or irregularity, and the anterior chamber is deep and quiet in both eyes. The anterior chamber is deep and quiet in both eyes. The left iris shows a 4 mm irregular, elevated, darkly pigmented lesion at the 5:00 o’clock position, causing correctopia and ectropion uveae. Gonioscopy reveals extension of the lesion into the angle. The vitreous is normal in both eyes. The optic nerve has a normal appearance with a cup-to-disc ratio of 0.3 in each eye. The retina and the vessels are normal."	"The suspected diagnosis is iris melanoma. The presence of a raised iris lesion, particularly in the inferior portions of the iris, with correctopia and ectropion uvea and angle seeding should raise suspiscion for the diagnosis. The differential diagnosis for pigmented lesions on the iris includes freckle, iris nevus, melanocytoma, Lisch nodule, and melanoma. A freckle is often small (<1-2 mm) and sits on the surface of the iris. A nevus can penetrate deeper into the iris stroma. A melanocytoma often appears dome-shaped. Iris melanomas, importantly, are more often found in the inferior quadrant, often in middle-aged adults, with a basal mean diameter of 6.2 mm in one retrospective study. If the patient’s iris lesion was non-pigmented, the differential is broad and would include, most commonly, non-neoplastic mimics (iridocorneal endothelial syndrome, iris atrophy, foreign body, coloboma, congenital heterochromia, to name the most common examples), as well as true neoplastic lesions such as metastasis from another site (most commonly breast, lung, kidney, and skin), choristomatous tumors, vascular tumors, fibrous tumors, neural tumors, myogenic tumors, epithelial tumors, xanthomatous tumors, lymphoid tumor, or leukemic tumors. Cystic lesions of the iris may also be included on the differential diagnosis, and these include primary iris cysts such as stromal cysts and pigment epithelial cysts."
Ophthalmic pathology and intraocular tumours	Choroidal metastases	"A 63-year-old man presented to the ophthalmology clinic complaining of blurry vision and a painful eye OS. The patient noticed his blurry vision a week ago and his eye pain has progressively been getting more severe. He had a history of cardiovascular disease but no history of malignancy. He had a 40-pack-year history of cigarette smoking.   On examination, there were no obvious signs of facial trauma and the patient’s globe was tender to palpation. His corrected vision was 20/20 OD and 20/80 OS. His extraocular movements were full but painful on extreme gaze. Hertel exophthalmometer measurement was 19mm OU. Pupils were 3mm in the dark and 2 mm in the light OU. The pupil response was normal bilaterally. The anterior segment examination was normal OU. In the dilated fundus examination, there was a superotemporal exudative retinal detachment and a superotemporal amelanotic subretinal mass OS. Intraocular pressures were 14 mmHg OD and 15 mmHg OS."	"The suspected diagnosis is choroidal metastases. Metastases are the most common cause of intraocular malignancies. Although the patient does not have a history of malignancy, metastases can be the initial presenting sign of cancer. Given the patient's significant smoking history and examination finding of exudative retinal detachment and an amelanotic subretinal mass, choroidal metastases secondary to a common cancer such as lung cancer is the suspected diagnosis.  The differential for a choroidal mass includes benign intraocular tumours such as a choroidal naevus or haemangioma, malignant tumours such as a choroidal melanoma, and choroidal lesions secondary to an autoimmune or vascular cause such as posterior scleritis, choroid granuloma or a haematoma/neovascular membrane. The most important differential is a primary malignancy such as a choroidal melanoma. These are usually pigmented subretinal masses but can be amelanotic on fundus exam.   Next, the patient should have an ocular ultrasound performed to differentiate choroidal metastases from a choroidal melanoma. Choroidal metastases demonstrate greater reflectivity than a choroidal melanoma on A-scan ultrasound. Choroidal metastases also demonstrate a significantly lower height-to-base ratio than choroidal melanoma on B-scan ultrasound. If choroidal metastases are suggested from the findings on ultrasound, the patient should have a systemic assessment performed to define the primary lesion, which will then guide further investigations and management. "
Ophthalmic pathology and intraocular tumours	Bilateral diffuse uveal melanocytic proliferation (BDUMP)	"A 60-year-old man was referred to the ophthalmology clinic for bilateral decreased vision. The patient had no known past ocular history. He had type 2 diabetes and hypertension treated with metformin, and lisinopril, respectively. He was a heavy smoker and was diagnosed with metastatic lung carcinoma three months prior to presentation. He was being treated with chemotherapy, but did not recall the details of this treatment. He started having a decrease in vision two weeks prior to presentation. He described blurry central vision and door frames appearing distorted. He denied any ocular pain or discharge.   On examination, his visual acuity was 20/60 in the right eye and 20/70 in the left eye. His intraocular pressures were 12 mmHg in the right eye and 13 mmHg in the left. His anterior chambers were deep and quiet. There were mild anterior cortical changes in both eyes, as well as 1+ posterior subcapsular cataracts. On dilated fundus examination, he had bilateral multiple nevi-like melanocytic lesions in both eyes with diffuse retinal pigment epithelial (RPE) stippling. On optical coherence tomography, he had multiple pockets of subretinal fluid in both eyes as well as choroidal and RPE thickening. Fundus autofluorescence showed a nummular pattern of hypo- and increased autofluorescence."	"The suspected diagnosis is bilateral diffuse uveal melanocytic proliferation (BDUMP), a rare paraneoplastic syndrome. The disease is typically characterised by multiple elevated uveal melanocytic tumours, diffuse choroidal thickening, serous retinal detachments, and rapidly progressing cataracts. Fundus autofluorescence reveals a characteristic 'leopard' or 'giraffe' pattern. The patient has all these features.  The differential diagnosis of BDUMP includes choroidal nevi and choroidal melanoma. However, the rapid development of multiple tumours strongly suggests BDUMP rather than a primary ocular malignancy. A differential diagnosis for subretinal fluid must also be considered. It includes central serous chorioretinopathy and choroidal neovascularization, as well as vitelliform lesions, inflammatory diseases, primary ocular tumours, hematologic malignancies (e.g., choroidal lymphoma), genetic diseases, ocular developmental anomalies (e.g., optic pit), and medication-induced retinopathies (e.g., MEK inhibitor-associated retinopathy).  Immune-related adverse events from systemic immunotherapy must also be considered, because it has been associated with conditions that resemble BDUMP, like uveal effusion and Vogt-Koyanagi-Harada-like syndromes.  BDUMP is thought to be due to the secretion by primary non-ocular tumours of a serum factor that selectively causes choroidal melanocytic proliferation. BDUMP is associated primarily to genitourinary tumours in women and lung cancer in men. This rare disease demands swift recognition for two critical reasons: to provide urgent vision-saving treatments and to investigate underlying cancer, which may be unknown in up to 50% of BDUMP cases. There is no conclusive treatment for BDUMP but plasmapheresis has been reported to have a good effect on the resolution of retinal fluid."
Ophthalmic pathology and intraocular tumours	Enophthalmos and ptosis secondary to metastatic breast cancer.	"A 58-year-old female presented to the ophthalmology clinic with a droopy left eyelid. She first noticed her eyelid drooping one month ago and believes that it is gradually progressing. She has hypertension that is controlled with medication. For her family history, her mother was diagnosed with breast cancer at 62 and her father was diagnosed with colon cancer at 66. She has a 50-pack-year history of cigarette smoking but does not smoke anymore. She has no history of ocular trauma or surgery.   On ocular examination, there was ptosis of the left eye noted. Her corrected visual acuities were 20/20 OU. Pupils were 2mm OU and there was no RAPD noted OU. Exophthalmometer measurements were 11mm OS and 15mm OD. Levator function was 5mm OS and 10mm OD. Extraocular movements were full and non-painful OU. Confrontational visual fields were full OU. On anterior segment exam, there were no other findings noted OU. On posterior segment exam, there were no findings noted OU. Her intraocular pressures were 18 mmHg OU."	"The suspected diagnosis in this patient is enophthalmos and ptosis secondary to metastatic breast cancer. Breast cancer is the most prevalent primary tumour that has metastases to the orbit. Orbital breast metastases typically localise within orbital fat or extraocular muscles. The classic triad of enophthalmos, ptosis and restricted ocular motility can be present in up to 1/3rd of patients with orbital metastatic spread of breast cancer. Breast cancer is suspected due to the patient’s 50-pack-year history of cigarette smoking and strong family history.   Enophthalmos is an uncommon and subtle exam finding. It can be appreciated subjectively and objectively by a measured difference of more than 2 mm on the exophthalmometer. It can be caused by two mechanisms: increases in the size of the orbit or decrease in the volume of the orbits contents. The size of the orbit is most commonly increased by trauma (a large orbital floor fracture) and congenital abnormalities. The volume of the orbital contents may be decreased by microphthalmos, orbital fat atrophy, scirrhous tumours (breast cancer), and post-inflammatory or post-surgical changes.   Next the patient should have a CT/MRI scan of the orbit and a biopsy to confirm the diagnosis. More detailed imaging studies such as a PET scan should also be done, but CT/MRI scan and a biopsy are the initial diagnostic tests that will confirm the suspected diagnosis. For treatment orbital external beam radiotherapy is the mainstay therapy. Patients require a systemic workup to evaluate for metastases to other organ systems, and a multidisciplinary team including oncology, radiation oncology, and ophthalmology."
Neuro-ophthalmology	Optic neuropathy	"A 72-year-old male is seen at the clinic one month following cataract surgery of his right eye. He had a white cataract that prevented visualisation of the optic nerve or fundus preoperatively. He noted a good improvement following surgery, but still finds his vision blurry. He is known for dyslipidemia, hypertension and diabetes mellitus for which he is treated. He is an active smoker with a smoking history of 30 pack years and has sleep apnea, although he rarely uses his BiPap machine.    On examination, visual acuity was 20/100 in the right eye and 20/20 in the left. There was a relative afferent pupillary defect in the right eye. Intraocular pressure was 18 mmHg in both eyes. Color vision was decreased in the right eye (5/11) compared to the left eye (10/11) on the Hardy Rand and Rittler (HRR) test. Anterior segment examination revealed a well positioned posterior chamber intraocular lens in the right eye and a mild cataract in the left eye. On dilated fundus examination, the right optic nerve had diffuse pallor with a cup-to-disc ratio of 0.2. The left optic nerve had a normal appearance with a similar cup-to-disc ratio of 0.2. The retinal vessels and macula appeared normal in both eyes."	"The suspected diagnosis is optic neuropathy, also referred to as optic atrophy. The presence of unilateral optic disc pallor, decreased visual acuity, impaired color vision, and a relative afferent pupillary defect strongly suggest this diagnosis. The primary cause is ischemic optic neuropathy, which can be subdivided into non-arteritic anterior ischemic optic neuropathy (NAION) and arteritic ischemic optic neuropathy (AION) associated with giant cell arteritis (AION). These initially present with optic disc edema in the acute phase, but the optic nerve undergoes atrophy after a few weeks, resulting in pallor.  NAION is associated with cardiovascular risk factors, sleep apnea, smoking, and sleep apnea while AION is secondary to giant cell arteritis. Inflammatory conditions, such as multiple sclerosis, sarcoidosis, and systemic lupus erythematosus, are part of the differential. Optic nerve atrophy can also be secondary to compressive etiologies, including intracranial tumors like meningiomas and pituitary adenomas. Toxic and nutritional deficiencies, such as chronic alcohol consumption, tobacco use, and vitamin deficiency are potential causes. Other etiologies include infectious (e.g. tuberculosis, syphilis), congenital (e.g. Leber’s hereditary optic atrophy, autosomal dominant optic atrophy) and traumatic. While the diagnosis of optic neuropathy is made clinically, investigations are required to identify the underlying etiology. Although the patient has multiple risk factors of NAION (sleep apnea, smoking, etc.) and possibly had an event in the past that was not noted because of the white cataract, it is crucial to obtain a thorough medical history and perform a basic work-up to rule out other possible pathologies. The best next step would therefore be to obtain an MRI of the head and orbit and basic blood work for inflammatory and infectious causes such as tuberculosis, syphilis and sarcoidosis. Additional testing can also be done if the history or examination is suggestive of a certain condition. Optic neuropathy can be unilateral or bilateral and the pattern of involvement can help narrow down the underlying etiology. The main goal of management is to identify the cause in order to prevent additional damage. The treatment will vary according to the etiology. "
Neuro-ophthalmology	Papilledema due to increased intracranial pressure	"A 30-year-old woman was referred to the ophthalmic clinic by her family doctor as part of an investigation for headaches. She has no known past medical or surgical history. She had -2.00 myopia but had refractive surgery two years prior. She takes no medications. She has two children, aged 2 and 4. She complains of daily headaches not associated with light or sound sensitivity. The pain typically lasts 2 to 4 hours every day and is only partly alleviated by over-the-counter ibuprofen. Occasionally, when she picks up her children from the ground, her vision darkens for a few seconds. She also sometimes experiences an abnormal whooshing sound in her ears, particularly before falling asleep. She does not report nausea, vomiting, vertigo, or double vision.  On examination, the patient appears overweight. Her pupil examination was normal without relative afferent pupillary defect or anisocoria. Her extraocular movements were full and non-painful. Visual acuity was 20/25 in both eyes and colour vision was normal (11/11 in both eyes) on the Hardy Rand and Rittler (HRR) test. Humphrey visual fields showed bilateral increased blind spots but not arciform deficits. Anterior segment examination was normal bilaterally and intraocular pressures were 20 mmHg in both eyes. Dilated fundus examination showed a circumferential halo of disc edema in both eyes leading to blurring of the disc margins. The macula appeared normal."	"The suspected diagnosis is papilledema due to increased intracranial pressure. The appearance of bilateral disc edema with normal visual function (good visual acuity and colour vision, absence of an afferent pupillary) suggests papilledema more than papillitis or optic neuropathy.   The differential diagnosis for disc edema is broad, including pseudopapilledema (where there appears to be edema without actual swelling, due to optic disc drusen or congenitally anomalous discs), papillitis (typically presenting unilaterally and affecting vision), optic neuropathy (which may be ischemic, hypertensive, diabetic, infiltrative, or genetic), and edema associated with uveitis.   Once papilledema is diagnosed, determining the aetiology of increased intracranial pressure is crucial. The aetiologies can include: primary or metastatic brain tumours, hydrocephalus, haemorrhage and hematomas, brain abscesses and meningitis. Intracranial hypertension can also be idiopathic (idiopathic intracranial hypertension or IIH) which is typically seen in young overweight females, like in this case. To make the diagnosis of IIH, negative MRI and MRV of the brain are needed to rule out other causes. The MRV is useful to rule out cerebral venous sinus thrombosis. After imaging is performed, a lumbar puncture that shows increased opening pressure (25 cm) with normal CSF composition is usually diagnostic.   Treatment is generally indicated to preserve visual acuity and prevent visual field loss. Weight loss alone can be effective, and can be augmented with acetazolamide to improve the degree of edema and improve quality of life in patients with mild vision loss due to IIH. CSF shunting procedures are reserved for severe cases with unsuccessful treatments."
Neuro-ophthalmology	Typical optic neuritis	"A 28-year-old woman presented to the ophthalmology clinic for sudden decrease in vision in the right eye. The patient describes three days of periocular pain and pain on extraocular movements preceding the vision loss. The patient is healthy with no known past medical or surgical history. Her ocular history is also unremarkable. On review of systems, the patient denies any arm, face or leg weakness or paresthesias.   On examination, visual acuity was 20/80 in the right eye and 20/20 in the left. There was an evident relative afferent pupillary defect in the right eye. Extraocular movements were full but painful in the right eye. Colour vision was asymmetrical, with 4/11 in the right eye and 11/11 in the left on the Hardy Rand and Rittler (HRR) test. Anterior segment examination was normal in both eyes. On dilated fundus examination, the optic nerve, retinal vessels and macula appeared normal. Intraocular pressure was 14 mmHg in both eyes."	"The suspected diagnosis is typical optic neuritis in the right eye. This diagnosis is clinical and is characterised by acute unilateral vision loss associated with optic nerve dysfunction (visual acuity and dyschromatopsia). Normal fundus examination in a young patient increases the likelihood of this diagnosis. A normal fundus examination suggests retrobulbar neuritis, which is seen in 65% of cases.   Typical optic neuritis must be distinguished from atypical optic neuritis. Atypical features include: lack of pain, prolonged vision loss beyond 2 months, significant optic nerve swelling, uveitis, bilateral vision loss and involvement of other cranial nerves. The most important differential diagnosis for optic neuritis with atypical features are neuromyelitis optica (NMO) and myelin oligodendrocyte glycoprotein (MOG) antibody disease. Those conditions have unique imaging features and are managed differently.  Most cases of typical optic neuritis are isolated or associated with demyelinating disorders like multiple sclerosis, and do not require urgent workup. This patient is expected to show some improvement within 1 month, with or without treatment, as is seen in the majority of cases (>90% recover 20/40 or better). Since this patient is not known for multiple sclerosis, non-urgent MRI of the brain is recommended to evaluate for periventricular white matter lesions. The absence or presence of those lesions determines the risk of future multiple sclerosis and guides subsequent decisions on the use of immunomodulatory treatments.                                                                                                      Treatment with corticosteroids can be considered, but it has no long-term benefit on vision. Intravenous methylprednisolone, 250 mg every 6 hours for 3 days, followed by oral prednisone, 1 mg/kg/day for 11 days, can speed recovery by 1-2 weeks. Oral corticosteroids alone are not recommended due to the high risk of disease recurrence with this route."
Neuro-ophthalmology	Right Horner syndrome	"A 56-year-old woman presented to the ophthalmology clinic for headache and a change in the appearance of her left eye. She has no known past medical or surgical history. She is known for -3.00 myopia and wears glasses. The patient mentions that her headache started five days ago after she was involved in a motor vehicle accident. She mentions that she left the scene before being evaluated by the paramedics. While applying eyeliner this morning, she noticed that her left pupil was larger than the right one and decided to consult.   On examination, she had no obvious signs of facial trauma. Her corrected visual acuity was 20/20 in both eyes. There was a 1-to-2 mm subtle ptosis of her right upper eyelid. Her extraocular movements were full and non-painful. There was anisocoria: her right pupil appeared smaller than her left one. The difference was more significant in the dark. Her anterior segment and dilated fundus examinations were normal. Intraocular pressures were 15 mmHg and 17 mm Hg."	"The suspected diagnosis is right Horner syndrome. There is anisocoria that is worse in dark conditions and a mild ptosis on the same side, suggesting Müller muscle dysfunction. Together, those signs are suggestive of Horner syndrome. The differential diagnosis of anisocoria in general is broad, and the most important differential diagnoses are third nerve palsy and pharmacologic dilation. Both of those diagnoses are associated with anisocoria that is worse in bright conditions. The differential diagnosis for ptosis is also broad, but a mild ptosis with anisocoria worse in the dark is Horner syndrome until proven otherwise.   To confirm the Horner syndrome, pharmacologic testing can be carried out using apraclonidine 1% or cocaine 10%. Drops are instilled in both eyes and after 30 to 45 minutes, a test is considered positive if: for cocaine, the Horner pupil fails to dilate while the normal pupil does; for apraclonidine, the Horner pupil dilates, while the normal pupil does not (reverse anisocoria).  Once confirmed, the differential diagnosis of Horner syndrome is broad and can be categorised by the anatomic level of pathology across the sympathetic pathway: first order neurons (stroke, multiple sclerosis, spinal cord tumours, etc.), second order neurons (pancoast tumour, iatrogenic from neck dissection) and third order neurons (internal carotid dissection, carotido-cavernous fistula).  An acute right Horner syndrome associated with headache, in the context of a recent neck trauma should raise the suspicion for a cervical artery dissection. As such, urgent noninvasive imaging of the brain and neck should be obtained: magnetic resonance imaging (MRI) and head and neck magnetic resonance angiography (MRA), or head computed tomography (CT) and computed tomography angiography (CTA) of the head and neck. This will help confirm an initial diagnosis of cervicocephalic dissection and to guide serial treatment decisions."
Neuro-ophthalmology	Bilateral aponeurotic ptosis	"A 64-year-old woman is referred by her primary care physician for progressive bilateral ptosis. For the past 2 years, she’s noted that both her upper eyelids are droopy, slightly more on the right side than the left. The patient has no past medical or ocular history. Her eyelids are not droopier in the evening or when she is tired. She denies any double vision, muscle weakness, dysphagia or fatigue. There is no history of trauma or previous ocular surgery. On examination, visual acuity is 20/25 in both eyes. Pupils are symmetrical and reactive to light without a relative afferent pupillary defect. Extraocular movements are full in both eyes without any pain or diplopia. Slit-lamp examination of both the anterior and posterior segments is within normal limits. Ptosis of both upper eyelids is noted, with a palpebral fissure distance of 6 mm on the right eye and 7 mm on the left eye. The marginal reflex distance-1 (MRD1) of 1 mm in the right eye and 2 mm in the left eye. Levator function in both eyes is measured to be 12 mm. There was no worsening of the ptosis after having the patient look in extreme upgaze for one minute."	"The patient has bilateral aponeurotic ptosis. The presence of ptosis in both eyes without any pupil or extraocular abnormalities, as well as the absence of systemic symptoms, is suggestive of this diagnosis. It is important to distinguish between aponeurotic blepharoptosis and other types of acquired ptosis. Thus, a complete ophthalmological examination with particular attention to pupils and extraocular movements is required. Myogenic ptosis can be found in myasthenia gravis, chronic progressive external ophthalmoplegia, and other muscular dystrophies. Patients with myasthenia gravis will have  a variable muscle weakness and will report worsening of the ptosis in the evening. On examination, they will have a positive fatigability test or improvement of the ptosis with rest or the ice-pack test. Neurogenic ptosis is a result of a third nerve palsy or Horner’s syndrome. In addition to the ptosis, patients will have extraocular movement deficits in third nerve palsy and/or mydriasis. In Horner’s syndrome, the affected side will have a minimal ptosis secondary to the involvement of Müller muscle and a miosis. Mechanical ptosis is secondary to an eyelid mass such as a chalazion or a tumor. Traumatic blepharoptosis can be seen following an eyelid laceration that damages the levator palpebrae. Pseudoptosis is often seen in elderly patients with dermatochalasis. The redundant upper eyelid skin gives the impression of a ptosis. The mainstay management of aponeurotic ptosis is surgical repair and various procedures are possible based on the severity and the function of the levator palpebrae.  These include levator advancement, Müller muscle resection and frontalis muscle suspension."
Neuro-ophthalmology	Thyroid eye disease	"A 37-year-old woman is sent by her optometrist because of asymmetry in eyelid positions. The patient has noted that over the past few months, her right upper eyelid seems higher than the left one. She feels like her right eye is “bigger”. She also has been having foreign body sensation in that eye with frequent tearing. There is no pain or photophobia. She notes mild redness occasionally in that eye. There is no diplopia and symptoms do not seem to be worst at night.  Her past medical history is non-contributory. Upon review of systems, she mentions having lost weight recently and has been having a hard time sleeping at night. On examination, there is right upper eyelid retraction. The marginal reflex distance-1 (MRD1) is 6 mm in the right eye and 4 mm in the left eye, with sclera visible above the limbus in the right eye. MRD2 is 5 mm in both eyes. When asked to close her eyes, there is a 1 mm lagophthalmos in the right eye without corneal show. Visual acuity is 20/30 in the right eye and 20/20 in the left eye. Her extraocular movements are full and pupils are reactive and symmetric. There is no proptosis on Hertel examination. On slit lamp examination, there is mild conjunctival injection in the right eye with superficial punctate keratitis inferiorly. The posterior segment exam is normal in both eyes."	"The presence of unilateral eyelid retraction with systemic symptoms suggestive of thyroid function abnormalities points toward a diagnosis of thyroid eye disease (also known as Graves’ disease). Thyroid eye disease is the most common cause of both unilateral and bilateral eyelid retraction. This malposition of the eyelid is initially due to increased sympathetic activity, but over time, the levator palpebrae becomes fibrotic and adherent to the orbital tissues. Other causes of eyelid retraction include unilateral ptosis that can cause contralateral retraction because of Hering’s law of equal innervation. The need to strain to elevate the ptotic eyelid, this results in increased firing to the contralateral levator palpebrae leading to eyelid retraction. Cutaneous scarring from burns, neoplasms, or dermatitis can also result in retraction. Surgeries for strabismus and blepharoptosis can potentially lead to abnormal eyelid positions postoperatively. Bilateral upper eyelid retraction is a classic sign of Parinaud syndrome (dorsal midbrain syndrome), which is mainly caused by pineal gland tumors or stroke. Eyelid retraction can cause lagophthalmos, which can then cause dry eyes and exposure keratopathy. Treatment of upper eyelid retraction is aimed at correcting the underlying condition. In this patient, given that thyroid eye disease is suspected, the best next step would be to order a blood workup to assess thyroid hormone levels and thyrotropin receptor antibodies. Orbital imaging is also necessary to evaluate the severity of the disease and involvement of extraocular muscles. Artificial tears, ointments or punctal plugs can be used to relieve irritation in mild cases of corneal exposure. Multiple surgical techniques can be used to correct eyelid retraction if the condition persists or is severe. "
Neuro-ophthalmology	Transient monocular vision loss	"An 80-year-old woman presented with transient vision loss in her left eye. She mentions losing vision completely in that eye twice over the last week, with recovery thereafter. Each episode lasted between 10 to 30 minutes and was not accompanied by any headache or other symptoms. She describes the loss of vision as a dark curtain coming down on her vision. She has a history of diabetes, for which she takes metformin, and was a heavy smoker for 40 years. She quit smoking last year after her granddaughter's birth. Upon review of systems, she mentions having chronic back and hip pain. She denies any recent weight loss, fever, jaw pain or headache.   On examination, her extraocular movements were full, her pupils were normal and her visual fields by confrontation were normal. Her corrected visual acuity was 20/40 in the right eye and 20/50 in the left eye. Her anterior segment examination was normal apart from bilateral moderate nuclear sclerosis and cortical changes. The fundus examination of both eyes appeared normal with no evidence of retinal whitening, haemorrhages or macular edema. There were no visible cholesterol emboli. The retinal periphery was normal."	"This patient has transient monocular vision loss in the left eye, also known as amaurosis fugax. To make this diagnosis, it is important to distinguish, by careful patient history, monocular versus binocular vision loss. The differential diagnosis for bilateral causes includes vertebrobasilar insufficiency (with neurological symptoms) and migraine with aura. For monocular causes, consider papilledema, giant cell arteritis (GCA), pigmentary dispersion, intraocular haemorrhage and intermittent angle closure.  In an elderly patient describing a monocular “dark curtain coming down”, acute retinal ischemia should be suspected. Possible etiologies would include embolus from the ipsilateral carotid artery and GCA. In patients over 50 years of age, particularly if no retinal embolus is seen, ruling out GCA with immediate C-reactive protein, erythrocyte sedimentation rate and platelet count is mandatory. This should be done even in patients without typical symptoms of GCA such as jaw claudication or headache because up to 20% of cases can be “occult”.   If GCA is not considered, patients must be referred to the nearest stroke centre because amaurosis fugax is considered a type of transient ischemic attack. The rationale for referral is to identify high risk patients needing hospitalisation and to identify the source of embolism. Urgent MRI (with DWI) is needed for suspected asymptomatic cerebral infarctions (CT head is acceptable). Subsequent workup should include carotid ultrasound, echocardiography, and prolonged cardiac monitoring to evaluate for sources of occlusion or thromboembolism.  Treatment depends on the aetiology and may include steroids for GCA, lifestyle changes, aspirin, cardiovascular risk factor control (for carotid or cardiac disease), and carotid revascularisation (per vascular surgery assessment)."
Neuro-ophthalmology	Parinaud syndrome	"A 42-year-old man presents to the clinic with a 2-week history of vertical diplopia. He also mentions having difficulty looking up since the onset of double vision. He has the feeling that his vision is more blurry than usual, despite having adjusted his glasses recently. On questioning, he reports headaches and frequent vomiting in the past months that he believed were due to stress. He has no past medical or ocular history. His family history is non-contributory. On examination, corrected visual acuity is 20/30 in both eyes. Pupils have poor constriction to light but preserved constriction with convergence. There is no relative afferent pupillary defect. Intraocular pressure is 15 mmHg in both eyes. Extraocular motility testing reveals a bilateral limitation of upward gaze. Horizontal movements and downward gaze are preserved. There are oscillatory movements of the eyes on attempted upward gaze accompanied by globe retraction.  Anterior and posterior segments exam was unremarkable. There is no ptosis or eyelid abnormalities."	"The suspected diagnosis is Parinaud syndrome, also known as dorsal midbrain syndrome. In this pathology, damage to the various structures in the dorsal midbrain will lead to the classic findings observed. Involvement of the vertical gaze centers such as the riMLF and the interstitial nucleus of Cajal will cause vertical gaze palsy. Convergence-retraction nystagmus, which presents with irregular, oscillatory movements of the eyes particularly on upward gaze, is secondary to damage to the supranuclear fibers. There is also involuntary globe retraction and convergence. Loss of the supranuclear input to the third nerve nucleus will lead to bilateral eyelid retraction. Finally, involvement of the Edinger-Westphal and pretectal nuclei will lead to light-near dissociation, where pupils react poorly to light but maintain the near reflex. The classic triad of this syndrome consists of upgaze palsy, convergence-retraction nystagmus as well as light-near dissociation, and is seen in 65% of cases. The most common causes of Parinaud syndrome include pineal gland tumor, brainstem hemorrhage and stroke. Neoplastic pathologies such as pineocytoma and germinoma are seen in younger patients whereas vascular and ischemic causes are found in the elderly population. The diagnosis is primarily clinical, but neuroimaging is suggested in all patients presenting with symptoms suggestive of Parinaud syndrome to identify any underlying structural abnormalities. The best next step would therefore be to obtain an MRI of the brain. Treatment is directed toward the underlying condition, which can vary significantly based on the etiology. In the meantime, symptoms can be improved with occlusion and visual training exercises. Persistent upgaze palsy or symptoms can be corrected surgically."
Neuro-ophthalmology	Third cranial nerve palsy	"A 55-year-old man presented with severe right eye eyelid ptosis starting two days prior to presentation. He reported a brow ache of 7/10 on the same side but no systemic neurological symptoms, arm or leg weakness, or paresthesia. He was a 5 pack-year smoker with a history of diabetes treated with metformin, gliclazide, and semaglutide. He also had hypertension managed with amlodipine and valsartan. His ocular history included bilateral cataract surgery three years prior.   On examination, with lifting of the right upper lid, his vision was 20/40 in the right eye and 20/20 in the left. He had significant right upper eyelid ptosis. Orthoptic exam revealed large-angle exotropia with significant hypotropia. Extraocular movements were completely limited in all directions except in abduction in the right eye. The left eye had normal extraocular movement.. There was no evident anisocoria and both pupils reacted normally to light. There was no afferent pupillary defect and the rest of the cranial nerves were normal. Anterior segment examination mild bilateral superficial punctate keratitis, bilateral  pseudophakia and fundus examination showed normal retina and optic nerves."	"The patient has right third cranial nerve palsy. The patient has external ophthalmoplegia, characterised by complete palsy limiting extraocular movements in all fields of gaze except temporally. The patient has no internal ophthalmoplegia as the pupil was spared. This constellation of findings must be considered together and is highly suggestive of third nerve palsy. In this case, the palsy is isolated, not associated with other cranial nerve palsies (fourth or sixth) and not associated with systemic syndromes.  Determining the aetiology of isolated third nerve palsy is crucial. Pupil-involving palsies are typically associated with aneurysms of the posterior communicating artery. Less common conditions also include tumour, trauma, uncal herniation, cavernous sinus lesions, orbital disease and infiltrative conditions. Pupil-sparing palsies are typically associated with ischemic microvascular disease, as suspected in this vignette.   Immediate CNS imaging to rule out mass and aneurysm is indicated in all third nerve palsies regardless of pupil examination. This patient must be urgently referred for contrast-enhanced CT and CTA. If a mass or aneurysm is to be found, urgent consultation with neurosurgery or interventional radiology must be made. In this case, if the imaging is normal, the palsy can be attributed to ischemic microvascular disease given the multiple cardiovascular risk factors. Those palsies typically improve within 3 months. If no improvement is seen with time, alternate diagnoses must be considered and additional investigations must be carried out. No treatments are indicated, except for controlling the underlying cardiovascular risk factors. Patching may be necessary, as the ptosis improves, to alleviate the diplopia."
Neuro-ophthalmology	Bilateral cranial nerve IV palsy secondary to head trauma	"A 52-year-old man presented to the ophthalmology clinic complaining of painless diplopia following head trauma the week before. He was standing on a stool and changing a light bulb in the kitchen when he lost his balance, fell down and struck his head on the tiled kitchen floor. He lost consciousness for less than a minute. He immediately noticed diplopia upon awakening but had no other neurological symptoms. He presented to the local emergency department and was evaluated. MRI of the brain was unrevealing and the CT did not show any orbital fractures. He has no previous medical history. He has 20/20 vision normally and does not wear glasses.  On examination, there was no deviation in primary gaze. His uncorrected visual acuity was 20/20 OU. Confrontational visual fields were full OU. Ocular motility examination showed an esotropia in downgaze OU. An infraduction deficit and a supraduction excess in adduction were also seen bilaterally. Right hypertropia was noted on right head tilt and left hypertropia was noted on left head tilt. His anterior segment examination was normal OU. On dilated fundus examination, fundus excyclotorsion was noted bilaterally. Pupil’s were 3 mm in the dark and 2 mm in light OU. Pupil responses were normal. His intraocular pressures were 13 mmHg OD and 12 mmHg OS."	"The suspected diagnosis is a bilateral cranial nerve IV (CNIV) palsy secondary to head trauma. CNIV palsy causes include traumatic injury, vascular infarcts, demyelinating disease and congenital CNIV palsy.   Isolated, acute onset, painless diplopia in the absence of positive neuroimaging and vascular risk factors (such as diabetes and hypertension) in the setting of recent closed head trauma is likely to be an isolated CNIV palsy secondary to trauma. The Parks-Bielschowsky 3-step test is used to isolate CNIV and diagnose CNIV palsy. A positive 3-step test involves (1) hypertropia in primary gaze, (2) worsened hypertropia in adduction, (3) worsened hypertropia in ipsilateral head tilt.   Absent hypertropia in primary gaze and reversible hypertropia in left and right gaze and head tilt is atypical; presence of these signs suggests a bilateral CNIV palsy. This patient’s V-pattern esotropia and excyclotorsion of both fundus are also suggestive of bilateral CNIV palsy.  For confirmation, strabismus measurements should be taken in all positions of gaze and with head tilt. V-pattern esotropia is typically greater than 25 PD in bilateral CNIV palsy.  Double Maddox rod testing should also be performed to measure total excyclotorsion, which is typically greater than 10 degrees in bilateral CNIV palsy."
Neuro-ophthalmology	Sixth cranial nerve palsy	"A 67-year-old man was referred from the emergency department for painless double vision since yesterday morning. His past medical history includes hypertension and diabetes, for which he takes medications. However, he is not very compliant with his treatment. His family history is non-contributory. He complains of constant horizontal double vision, that is greater at distance than at near. He finds it very incapacitating. He denies headache, weakness or any other neurological symptom. There is no history of head trauma. On examination, his corrected visual acuity was 20/25 in both eyes. His pupillary exam was normal without an afferent pupillary defect. His intraocular pressures were 14 mmHg OD and 15 mmHg OS. Orthoptic examination revealed a 20-prism diopter esotropia of the right eye in primary position. The esotropia would improve on left gaze, but worsen on right gaze. Ocular motility examination showed a severe limitation of abduction of the right eye. Other extraocular movements were normal. There was no ptosis or proptosis. Color vision and visual field by confrontation were normal in both eyes."	"The suspected diagnosis is a right sixth nerve palsy. The patient has severe limitation of abduction of the right eye with an esotropia in primary gaze, which is consistent with the diagnosis. The most common aetiology of sixth nerve palsy in the elderly population is microvascular ischemia. This patient has microvascular risk factors such as diabetes and high blood pressure, which were not well controlled. Other possible aetiologies include stroke, tumour, giant cell arteritis, multiple sclerosis or increased intracranial pressure. It is also crucial to perform a complete neuro-ophthalmological examination to determine if other neurological abnormalities are present, which could suggest a brainstem or cavernous sinus pathology. The best next step would be reassurance. In fact, in a case of an ischemic-isolated mononeuropathy in a patient over 50 years-old, the disease is self-limited and usually resolves within 3 months. The patient can temporarily cover one of his eyes to prevent diplopia. He should be monitored for other neurological symptoms. If there is no complete resolution within 3 months, then neuroimaging is indicated to rule-out another aetiology. In patients under 50 years of age or those with non-isolated sixth nerve palsy, it is recommended to obtain a brain and orbital MRI.  Treatment of sixth nerve palsy depends on the underlying aetiology.  Strabismus surgery can be performed for persistent abducens nerve palsies that demonstrate stable measurements over a 6-month period"
Neuro-ophthalmology	Internuclear ophthalmoplegia (INO)	"A 30-year-old woman presented with double vision and difficulty reading. She has no past medical or ocular history. Her family history is non-contributory. She reports intermittent horizontal double vision beginning four months ago, primarily during reading. She denies headache, weakness, or paresthesia. She does not feel that the symptoms are worse at the end of day.  On examination, her vision was 20/20 in both eyes. Colour vision and visual fields by confrontation were normal. Her pupillary examination was also normal, without anisocoria or afferent pupillary defect. Extraocular movements were abnormal. On saccade to the left, the right eye adduction was slow and left eye horizontal jerk nystagmus appeared. This unique examination feature was reproducible. There were no significant abnormalities of vertical eye and eyelid movements."	"The patient has right internuclear ophthalmoplegia (INO). This unique ophthalmoplegia is due to a lesion in the medial longitudinal fasciculus. It is characterised by weakness of adduction with horizontal jerk nystagmus of the abducting eye. The INO is located to the side of weak adduction, the right side in this case. A skew deviation (a comitant vertical deviation) can also be seen.  The most common causes of INO are demyelination and stroke. In a young apparently healthy female, like in this case, INO is typically due to demyelination from multiple sclerosis. The most important differential diagnosis for INO is myasthenia gravis. This condition can lead to attenuated adduction but is typically associated with ptosis and orbicularis oculi weakness. Myasthenia gravis patients also typically describe fluctuating symptoms, worsening at the end of the day. INO can also be associated with one-and-half syndrome from a pontine lesion. This syndrome is characterised by a horizontal gaze palsy and ipsilateral INO. The only horizontal eye movement remaining is abduction of the eye contralateral to the lesion.   This patient must be referred for neuroimaging with MRI with stroke protocol and contrast, in order to discover where the damage is located and the underlying aetiology. In general, the prognosis of INO is good and most cases improve over time. If multiple sclerosis is suspected, referral to neurology is warranted."
Neuro-ophthalmology	Migraine with aura	"A 27-year-old woman is sent by the emergency department for positive visual symptoms. Yesterday, while watching television, she started seeing a small scotoma in her peripheral vision with multiple colorful flashes of light and “zig zags” slowly expanding over 5 minutes. She says it looked like a kaleidoscope. The lights were still present when covering one eye at a time. The symptoms resolved after half an hour but were followed by a headache that lasted a few hours. She had to turn off the lights and television because they were aggravating her pain. It was the first time that she experienced such symptoms. Her personal and family history are non-contributory. Besides wearing contact lenses for mild myopia, she doesn’t have any ophthalmological condition. She takes the oral contraceptive pill. She hasn’t had a recurrence of symptoms. Corrected visual acuity in both eyes was 20/20. Pupils were symmetric and reactive to light without a relative afferent pupillary defect. Intraocular pressure was 14 mm Hg bilaterally. Visual fields by confrontation were complete and extraocular movements were normal. Anterior segment exam was unremarkable besides some mild superficial keratopathy. Dilated fund examination did not reveal any retinal tear or detachment."	"The suspected diagnosis is migraine with aura. The presence of positive visual phenomena (colors, flashes, zigzags) that are fully reversible, the presence of photophobia and the time duration (less than 30 minutes) are highly suggestive of this diagnosis. As per the International Classification of Headache Disorders (ICHD-3), the aura symptoms usually spread gradually over ≥5 minutes and last between 5-60 minutes. It can be subdivided into typical aura with headache or typical aura without a headache. In typical aura with headache, the aura is accompanied, or followed within 60 minutes, by a headache, whereas it is absent in typical aura without a headache. The differential diagnosis for migraine with aura includes transient ischemic attack and seizure. It is crucial to ask about the presence of other neurological symptoms. The distinguishing features are the gradual onset of symptoms, duration and presence of other symptoms such as photophobia and phonophobia. When no aura is present, it is important to differentiate migraines from other primary causes of headaches such as cluster headache and tension headache. The initial step in the management of migraines is the application of prophylactic measures such as lifestyle modification (e.g. diet, physical activity) and avoidance of triggers (e.g. alcohol, caffeine, chocolate). Medications can be used both in the acute setting to break the attacks as well as prophylactically to prevent episodes. Management of migraines is usually handled by a primary care physician or neurologist. It is important to ask about medication history as estrogen-based combined oral contraceptives are contraindicated in patients with migraine with aura over the age of 35. They are also contraindicated in patients under 35 if they smoke, as there is increased risk of cardiovascular disease. Neuroimaging is not necessary for stable patients who meet the diagnostic criteria for migraine. Indications for imaging include: headache duration >72 hours, new onset in patients older than 50, and change in the frequency/pattern/severity of headaches."
Neuro-ophthalmology	Optic disc drusen	"A 10-year-old male was referred to the ophthalmology clinic for strange appearing optic discs bilaterally. The patient went to the optometrist to receive a new glasses prescription. He had recently also been experiencing visual disturbances so the optometrist did a fundus exam on both eyes. The patient claims that he has spontaneous and transient losses of vision that last for a few seconds. He received a new prescription of -1.75 OU. He has no previous medical history. He has no family history of ocular disease.   On ocular examination, there were no findings noted on external exam. His corrected visual acuities were 20/20 OU. Pupils were 3mm OU and there was no RAPD noted OU. Extraocular movements are full and non-painful OU. Confrontational visual fields were also full OU. On anterior segment examination, there were no findings noted OU. On posterior segment exam, the optic discs had indistinct margins, no disc hyperaemia and the overlying vessels were clearly visible OU. His intraocular pressures were 16 mmHg OU."	"The suspected diagnosis in this patient is pseudopapilloedema due to optic disc drusen (ODD). There are two main types of ODD: visible drusen and “buried” drusen. This patient most likely has “buried” optic disc drusen that are causing pseudopapilloedema. “Buried” drusen are most common in the pre-teenage years and become more visible as patients get older and have a characteristic “lumpy bumpy” appearance. ODD are commonly asymptomatic and discovered incidentally, however they can also cause transient visual obscurations or visual field defects.   The most important differential for pseudopapilloedema from any cause is true papilloedema. Pseudopapilloedema is suspected due to the visibility of the overlying blood vessels and lack of disc hyperaemia. Nerve fibre layer oedema causes blurring of the optic disc margins in true papilloedema. Other conditions that can cause optic disc oedema such as papillitis, ischaemic optic neuropathy and Leber’s hereditary optic neuropathy usually cause visual loss. Conditions that cause infiltration of the optic disc such as sarcoidosis and tuberculous granuloma etc. would present with other ocular or systemic abnormalities so are also unlikely.   A number of investigations can be done to distinguish between pseudopapilloedema and true papilloedema. Although this patient most likely has pseudopapilloedema due to ODD, given the obscure nature of his presentation of “buried” ODD, confirmatory diagnostic testing should be done next. B-scan ultrasonography or fluorescein angiography/fundus autofluorescence is the recommended investigation for confirmation. CT imaging can also be done but the previously mentioned investigations are generally less costly and have a higher sensitivity.  "
Paediatric ophthalmology and strabismus 	Spamus nutans	"A 14-month-old boy is brought to the ophthalmology clinic by his mother due to concerns about new “twitching” of his eyes and frequent head shaking. She reports that over the past two weeks, she has observed movements of his eyes and abnormal head shaking. These eye movements are variable and sometimes differ between the two eyes. The mother denies any changes in his vision or overall health and notes that he is meeting all developmental milestones appropriate for his age. On examination, the child appears well. Intermittent, small-amplitude, high-frequency nystagmus is observed, varying in direction and occasionally dissociated between the two eyes. Head shaking is also noted during the exam. Visual acuity was ""fix and follow"" in both eyes. Intraocular pressures are measured at 15 mmHg in the right eye and 16 mmHg in the left eye. Pupils are equal, round, and reactive to light without an afferent pupillary defect. Examination of the anterior segment reveals normal lids, lashes, conjunctiva, sclera, cornea, and anterior chamber in both eyes. Dilated fundus examination shows normal optic discs and retina, with no lesions, edema, or hemorrhage."	"This child likely has spasmus nutans syndrome (SNS) which is an acquired nystagmus in infancy with intermittent small-amplitude high-frequency variable or dissociated nystagmus, variable torticollis, head shaking/bobbing, and usually a normal-appearing fundus. With time, the nystagmus resolves. The pattern of nystagmus can resemble the Heimann-Bielschowsky phenomenon which is a monocular vertical nystagmus associated to profound vision loss. Generally, the SNS diagnosis is made after normal MRI imaging.    In general, the differential diagnosis of childhood nystagmus is broad, and workup requires a thorough history and review of symptoms. Nystagmus should be considered as a clinical sign rather than diagnosis and should be worked up appropriately. It is defined by the preseence of uncontrolled, repetitive movements of the eyes, and is described based on the amplitude, frequency and direction of oscillations. History, clinical examination and review of systems will guide next investigations than can include optical coherecne tomography and electroretinography, as well as brain imaging and genetic testing."
Paediatric ophthalmology and strabismus 	Congenital esotropia	"A 5-month-old girl is brought to the pediatric ophthalmologist by her mother due to concerns about her daughter’s “crossed eyes.” The mother reports that since the age of 2 months, she has noticed that her daughter’s eyes turn inward most of the time. The crossing appears constant, and the child seems to fixate on objects using either eye while the other eye remains deviated inward. There are no complaints of discomfort, and the infant appears unbothered by the condition. The mother notes that her daughter is otherwise healthy, feeding well, and meeting all developmental milestones. On examination, the infant is alert and interacts appropriately for her age. There is a large-angle constant esotropia of approximately 40 prism diopters in primary gaze. Alternate cover testing reveals equal fixation preference; the child readily fixes and follows with either eye when the other is occluded. Ocular motility is full in all directions without any restrictions. No nystagmus is observed. Pupils are equal, round, and reactive to light with no afferent pupillary defect. Intraocular pressures are measured at 12 mmHg in the right eye and 13 mmHg in the left eye. Anterior segment examination shows normal lids, lashes, conjunctiva, sclera, cornea, and anterior chamber in both eyes. Cycloplegic refraction reveals minimal hyperopia (+2.00 diopter in both eyes). Dilated fundus examination reveals healthy optic nerves and retinas with no masses, lesions, or edema in either eye."	"This child has an examination consistent with congenital esotropia. This disease manifests by age 6 months and is characterized by a large angle esotropia and mostly equal at near and distance. Refractive error is normal for age, generally slightly hyperopic. Amplyopia is uncommon because children will cross-foxate. However, surgery for esotropia is generally recommended to allow the development of binocular vision. Bilateral medial rectus recessions are most commonly performed. The differential diagnosis includes: acquired nonacommodative esotropia, accomodative esotropia, sensory-deprivation esotropia and sixth cranial nerve palsy. Acquired nonaccomodative esotropia usually develops after 6 months of age, has smaller deviation than seen in congenital esotropia. Accomodative esotropia is a common condition, typically presenting after 6 months of age with an average age of 2.5 years. Children will typically show high hyperopia in the range of +3.00 to +10.00 and a moderate deviation (20 to 30-prism diopters)."
Paediatric ophthalmology and strabismus 	Priimary congenital glaucoma	"A 7-month-old boy presents with increasing tearing and abnormal appearance of both eyes. The patient’s mother has noticed the eyes appear darker, with increasing tearing and blinking of both eyes. There is no known family history of ophthalmic disease.   On exam, the patient has blepharospasm of both eyelids. The eyes are buphthalmic-appearing. The patient fixes and follows. Pupils are equal, round, and reactive to light in both eyes without afferent pupillary defect. Intraocular pressures are 32 and 30 mmmHg in the right and left eye, respectively. Portable slit lamp examination reveals corneal edema and striae in the endothelium of both corneas. On dilated exam, the patient has cup/disc ratio of 0.6 in the right and 0.5 in the left."	"The patient has primary congenital glaucoma, as evidenced by epiphora, blepharospasm, buphthalmos, and increased intraocular pressure. Additionally, the patient’s corneal exam and optic nerve exam are consistent with this diagnosis. In this disease, which may present at any point during infancy or childhood but often presents between ages 3-9 months, the primary management is surgical with angle surgery (goniotomy or trabeculotomy), with consideration of ciliary body destruction or tube-shunt surgery if initial surgical management is not effective. In other causes of childhood glaucoma, such as juvenile open-angle glaucoma and secondary glaucoma with open angles, medication is often used, with timolol (beta-blocker) or a topical carbonic anhydrase inhibitor (dorzolamide or brinzolamide) used first. Prostaglandin analogues such as latanoprost are not FDA-approved in the United States for childhood glaucoma, although they are approved for use in Europe. Additionally, in uveitic childhood glaucoma, surgical management with goniotomy has also been shown to be effective."
Paediatric ophthalmology and strabismus 	Shaken baby syndrome	"A 10-week-old male infant is brought for ophthalmic examination. The father reported that the patient suddenly lost consciousness while they were playing together. The infant was brought to the emergency department by ambulance. Paramedics found the patient apnoeic and pulseless. CPR was continued and the patient was intubated. Arterial blood gas analysis was performed upon patient arrival to the hospital and showed that the patient was severely acidotic with a 7.07 blood pH. Non-contrast head CT showed bilateral subdural haematomas. The patient was admitted to the paediatric intensive care unit. Medical records showed that the patient had a spontaneous, uneventful birth and that he had no medical history. The father reported that the patient has not had any suspected or known head trauma.   On examination, the patient was intubated and unresponsive. Scattered bruises were noted on the patients trunk and upper extremities. He had no reaction to light OU. Pupils were 6 mm OU and unreactive to light. Anterior segment examination was normal. On fundus examination, diffuse retinal haemorrhages that extended to the periphery were noted OU. Intraocular pressures were normal."	"The suspected diagnosis is shaken baby syndrome or non-accidental traumatic brain injury. The differential includes accidental head trauma, normal birth trauma and a rare cause such as a coagulopathy or leukaemia. This can be a difficult diagnosis to make in a clinical setting because of having to differentiate between accidental/normal birth trauma and non-accidental head trauma.   The father reports that the patient has not had any suspected or known head trauma. This raises suspicion as in even the most severe accidental head trauma, retinal haemorrhages are an uncommon exam finding. For the infant to present with retinal haemorrhages would mean that he suffered a severe and traumatic accidental head injury. Due to the head injury the infant would have most likely been in an highly obtunded state which should have raised concern in the father, instead the father claims that they were “playing” together before the infant suddenly lost consciousness. In addition, the retinal haemorrhages seen in this patient are severe/diffuse, which makes accidental head trauma even more unlikely. In normal birth trauma, extensive retinal haemorrhages can be seen but typically disappear within 4-6 weeks. The infant also had an uneventful birth, making birth trauma less likely to be the cause of the retinal haemorrhages.  The next investigation should be blood tests. A Full Blood Count (FBC), Prothrombin Test (PT)/International Normalized Ratio (INR), and Partial Thromboplastin Time (PTT) should be ordered to rule out a coagulopathy or leukaemia. A skeletal survey can also be ordered to examine for fractures. For suspected shaken baby syndrome, care should be coordinated between paediatrics, neurosurgery, ophthalmology and social services."
Paediatric ophthalmology and strabismus 	Juvenile idiopathic arthritis (JIA)	"An 11-year-old girl presents for unilateral left eye pain and decreased vision in the left eye for the past 1 week. She has had 2 similar episodes in the past. On further review of systems, she has had intermittent pain and swelling in her knees and elbows, worse upon waking in the morning, for the past 2 months.   On exam, the patient’s vision is 20/20 in the right eye and 20/40 in the left eye. Both pupils are equal, round, and reactive to light without afferent pupillary defect. Intraocular pressures are 12 mmHg in the right eye and 17 mmHg in the left eye. External exam is normal, and slit lamp and dilated exam of the right eye is normal. Slit lamp examination of the left eye reveals 2+ cell and 1+ flare in the anterior chamber of the left eye. The left iris shows 180 degrees of posterior synechiae. There is no vitreous cell, and dilated exam shows no lesions or edema in the retina."	"This patient most likely has juvenile idiopathic arthritis (JIA), as evidenced by her anterior uveitis and joint symptoms lasting at least 6 weeks in multiple joints. She should undergo rheumatologic evaluation for further workup, testing, and systemic treatment. JIA is the most common systemic disease associated with pediatric uveitis. Other systemic associations to uveitis can include tubulointerstitial nephritis and uveitis (TINU), sarcoidosis, infectious anterior uveitis, Fuchs heterochromic iridocyclitis, postinfectious autoimmune uveitis, traumatic uveitis, and idiopathic uveitis. Additionally, must-not-miss diagnoses include neoplastic uveitic masquerades such as leukemia and retinoblastoma.   In the management of juvenile idiopathic arthritis, systemic immunomodulatory therapy (IMT) is usually necessary to control inflammation. IMT often used includes methotrexate +/- an anti-tumor necrosis factor biologic, such as adalimumab. These patients should be followed closely, especially when changing or tapering therapy, and should receive ophthalmologic exams 1 month after tapering or discontinuing topical corticosteroid therapy and 2 months after tapering or discontinuing systemic therapy. If cataract surgery is needed, surgery should be planned after inflammation has been controlled for at least 3 months, and strict inflammatory control should be pursued in the postoperative period."
Paediatric ophthalmology and strabismus 	Intermittent exotropia	"A 4-year old girl presented with her parents to the ophthalmology clinic. The parents mention that the child's eyes have been ""wandering"" particularly at the end of the day or when she is distracted. She is overall healthy, has reached all her developmental milestones. They feel that the eye is deviating more frequently than before and decided to consult. On examination, the child appears well. Her pupils are normal and reactive to light. The visual acuity is 20/20 in both eyes on Allen cards. On the cover-uncover test, the uncovered eye moves from the out-turned position to the midline to fixate when the previously fixating eye is covered. Her anterior segment examination is normal. On dilated fundus examination, her optic nerve and retina appear normal. Her cycloplegic refraction shows +1.00 of hyperopia in both eyes."	"The child has intermittent exotropia, the most common type of exodeviation in children. The onset is usually before age 5 with frequency increasing over time. The disease is categorized by phases. Phase 1 is when one eye truns out spontaneously or when it is covered, usually occuring when the patient is not attentive, as in this case. The eyes become straight again with one or two blinks. Phase 2 is when frequcny increases, and phase 3 is when there is constant exotropia at distance and near. The disease can also be subclassified based on the amount of deviation at near and at distance: basic, true divergence excess, simulated divergence excess and convergence insufficiency. The differential diagnosis include constant exotropia (sensory-deprivation, congenital exotropia, consecutive exotropia following previous strabismus surgery). Other differentials include neuromuscular anomalies like third nerve palsy and myasthenia gravis. This patient should be followed closely, but there is no indication for surgery at this time. The child also has no amblyopia or significant refractive errors to treat. Phase 2 and 3 patients may be considered for exotropia surgery, typically consisting of bilateral lateral rectus recession."
Paediatric ophthalmology and strabismus 	Retinoblastoma	"A 2-year-old girl presented to the paediatric eye clinic with her parents. The chief complaint of the parents was the abnormal appearance of the child’s right eye and its deviation inwards. Pregnancy and birth were normal. The girl had reached all her developmental milestones and her vaccinations were up to date. She has no siblings.  On examination, there were 20 prism diopters of right eye esotropia. The extraocular movements appeared full. Visual acuity could not be assessed due to the child's fussiness, especially when covering the left eye. There was a white pupillary reflex in the right eye and a normal reflex in the left eye. Anterior segments were normal. Dilated fundus examination of the right eye revealed a single, white, nodular mass, approximately one disc diameter in width, extending into the vitreous along the superior vascular arcade. The optic nerve and macula appeared normal. Left eye examination was unremarkable."	"This child presents with leukocoria, esotropia, and a retinal tumour, which is highly suggestive of retinoblastoma. Retinoblastoma is the most common malignant intraocular tumour of childhood, occurring in 1:20,000 live births. It is equally common in both sexes and has no racial predilection. The most common initial sign of this condition is leukocoria (white pupil), which is evident in this case and often noticed by parents. Strabismus, as also seen here, presents in up to 25% of cases.  The differential diagnosis of leukocoria is broad, but retinoblastoma remains the single most important disease to consider. The differential diagnosis includes toxocariasis (a nematode infection associated with a white retinal granuloma and diffuse endophthalmitis), persistent foetal vasculature (a developmental abnormality associated with microphthalmia, an abnormal white membrane behind the lens), paediatric cataract, retinopathy of prematurity, and other rare genetic conditions with distinct features. Coats disease is the most common retinal condition that can be mistaken for retinoblastoma. It is a retinal vascular abnormality leading to the presence of crystalline material, extensive subretinal fluid, and peripheral vascular abnormalities. Coats disease typically affects boys and, unlike retinoblastoma, does not cause calcifications.  This child likely has retinoblastoma. The most important thing to do is to urgently refer this child to a centre with expertise in managing retinoblastoma. Examination under anaesthesia will be required to perform a thorough dilated fundus examination looking for multiple tumours and tumours in the left eye. This will allow disease classification according to the International Classification of Retinoblastoma which guides treatment. Imaging of the head and orbits should be performed to confirm the diagnosis and to evaluate for extraocular extension and intracranial disease. MRI and ultrasonography, which avoid the use of radiation, are recommended. Biopsy is avoided to avoid malignant cell dissemination. Referral to genetic counselling is crucial to aid in determining the risk of subsequent cancers in the affected child and the risk of retinoblastoma in other family members. Treatments will vary based on disease classification and laterality and can include focal laser and cryotherapy treatment, intra-arterial chemotherapy, enucleation and systemic chemotherapy."
Paediatric ophthalmology and strabismus 	Retinopathy of prematurity	"You are consulted to examine a preterm boy, born at 28 weeks of gestation and weighing 900 grams, in the neonatal intensive care unit. The neonatal history was significant for prolonged oxygen supplementation but no other complications reported during the NICU stay. On examination, the anterior segments of both eyes appeared normal. The pupils were round and reactive to light, with no relative afferent pupillary defect. However, dilated fundus examination revealed abnormal retinal vascularization in both eyes, with a demarcation line and fibrovascular proliferation. This pathology was noted in the posterior retina within 60° from the optic disc. There was also arterial tortuosity and venous dilation in the posterior pole."	"This child demonstrates retinopathy of prematurity, stage 3 in Zone I as well as Plus disease which could be consistent with ""threshold"" ROP or at least Prethreshold type 1 ROP. This disease stage will require treatment. Treatment can include laser photocoagulation to ablate the avascular retina, which helps to reduce abnormal vessel growth and prevent further progression. In some cases, intravitreal anti-VEGF injections may be considered as an adjunct or alternative, particularly in posterior disease or when there is a risk of systemic complications. The differential diagnosis for ROP includes familial exudative vitreoretinopathy which appears similar to ROP but there is usually no history of prematurity or oxygen supplementation. Incontinentia pigmenti is another possible differential diagnosis but is lethal in males."
Paediatric ophthalmology and strabismus 	Bilateral optic neuritis	"A 7-year-old boy is brought in by his parents due to sudden vision loss in both eyes. The child has been complaining of progressively worsening blurry vision over the past few days, with the right eye being more severely affected. He reports pain with eye movements and headaches. He has also had a cough, fever, and myalgias over the past week. There is no associated nausea, vomiting, or hiccups. He has not experienced any limb weakness or tingling, and he has not received any recent vaccinations. The family has a dog but no cats, as the patient's younger brother is allergic. There is no family history of autoimmune or demyelinating diseases. The boy is of healthy weight and has not taken any new medications. On examination, his visual acuity was 20/200 in the right eye and 20/100 in the left eye. Pupils were reactive and symmetric, with a very mild relative afferent pupillary defect in the right eye. Extraocular movements were complete but painful, particularly on lateral gaze bilaterally. Color vision was decreased in both eyes, with scores of 5/11 in the right eye and 7/11 in the left eye on the Hardy Rand and Rittler (HRR) test. Fundus examination revealed optic nerve disc edema, slightly more pronounced in the right eye. There were no associated hemorrhages or exudates. The macula and periphery were normal, but rare vitreous cells were noted in both eyes. An urgent cerebral CT scan was performed and did not show any mass or hemorrhage. A lumbar puncture revealed normal cerebrospinal fluid composition and an opening pressure of 13 mm Hg."	"The suspected diagnosis is bilateral optic neuritis. The presence of optic nerve swelling, in the context of decreased visual acuity, dyschromatopsia, pain on extraocular movements, and a history of flu-like symptoms, strongly suggests this diagnosis. The normal CT head scan and lumbar puncture findings further support this condition. Optic disc edema in the pediatric population has different etiologies and prognoses compared to adults. Potential causes include optic neuritis, papilledema, pseudopapilledema, neuroretinitis, optic nerve tumor, infiltrative processes, and others. In this patient, given the acute onset of simultaneous bilateral disc edema with vision loss, the differential diagnosis is narrowed to bilateral optic neuritis, papilledema, and an infiltrative process. Optic neuritis remains the most likely diagnosis given the normal CT scan, lumbar puncture results, and the onset of vision loss following a flu-like illness. The best next step would be to perform a cerebral and orbit MRI to evaluate the pattern and extent of the optic neuritis. Unlike in adults, optic neuritis in children often presents bilaterally and is associated with optic disc swelling, whereas it is more often retrobulbar in adults. It is frequently considered a post-infectious condition that does not necessarily indicate multiple sclerosis. The most common cause is idiopathic, but other conditions such as acute disseminated encephalomyelitis, neuromyelitis optica, myelin-oligodendrocyte glycoprotein antibody disease, and multiple sclerosis can present similarly. In addition to the MRI, a chest X-ray and blood workup for inflammatory and infectious etiologies are often performed. Despite the severity of visual loss, the prognosis for recovery is generally excellent, as most children will regain good visual acuity. Treatment will vary based on the underlying condition, but most patients are treated with high-dose corticosteroids, and plasma exchange and/or intravenous immunoglobulin can be used in cases that do not respond well to corticosteroids."
Paediatric ophthalmology and strabismus 	Dermoid cyst	"A 4-month-old male infant was brought into the ophthalmology clinic for evaluation of a left upper eyelid mass. The parents report that they first noticed the mass two months ago and they believe it has been slowly increasing in size but are unsure. The mass is subcutaneous and has not changed colour since appearing. The infant is not bothered when the mass is touched by the parents. The infant’s birth was spontaneous and uneventful. He has no previous medical history. There is no family history of ocular disease.   On ocular examination, there was a 2.5cm x 2.0cm, uniformly smooth, firm and fixed subcutaneous mass at the superolateral orbital rim. There was no oedema, erythema, tenderness to palpation or globe displacement noted. The globes were also soft to palpation OU. His fixation was central, steady and maintained OU. Pupils were 6mm OU in dark and 4mm OU in light, there was no RAPD noted OU. Extraocular movements were full OU. On anterior segment examination, there were no findings noted. On posterior segment exam, there were no abnormalities noted OU. His intraocular pressures were 9 mmHg OU."	"The suspected diagnosis in this patient is a dermoid cyst. Dermoid cysts are benign tumours and one of the most common tumours in children. Dermoid cysts are usually smooth, firm and mobile masses, however they can also be immobile due to affixation with the periosteum.   The differential diagnosis for dermoid cysts depends on the location of the lesion. For lateral lesions such as this one, lacrimal gland tumours are the most important differential diagnoses. Given that the patient is an infant, a lacrimal gland tumour is unlikely. The mass is fixed, which can suggest malignancy, but it is also uniformly smooth, which is suggestive of a cyst. Infantile haemangiomas can also present similarly. However, haemangiomas usually cause a blue discolouration with/without an associated strawberry naevus on the skin and are found most commonly in the superonasal quadrant of the orbit.   The next step in the management of this patient should be surgical excision of the dermoid cyst. Small and asymptomatic cysts may not require treatment but this patient has a large cyst that has the potential for traumatic rupture. Surgical excision is the mainstay of treatment and if possible the surgeon should aim to remove the cyst in total without rupture."
Paediatric ophthalmology and strabismus 	Brown syndrome	"An 8-year old girl presents with her parents for an ophthalmologic exam. The patient’s parents think they have occasionally noticed a long-standing “lazy eye.” The child is asymptomatic, is healthy and her review of systems is negative.  On exam, visual acuity is 20/20 in each eye. Intraocular pressures are 13 and 15. Pupils are equal, round, and reactive to light. On primary gaze, the eyes are aligned. On right gaze, there is full adduction but -4 elevation deficit of the left eye. In all other gazes, there is no restriction of extraocular movement. Slit lamp examination shows normal exam of the anterior and posterior segments of both eyes."	"The patient likely has Brown syndrome, a form of vertical strabismus characterized by restriction of upgaze during adduction. The main differential diagnosis includes an inferior oblique palsy, which can also present with limited elevation in adduction. Distinguishing between these two conditions is crucial for appropriate management. Brown syndrome will show positive forced ductions and typically a V-pattern of strabismus. Isolated inferior oblique palsy is generally rare. Brown syndrome is generally congenital but can be idiopathic or acquired secondary to trauma, surgery or inflammation in the trochlear region. Acquired causes of Brown syndrome should undergo workup including imaging for mass or other mechanical cause of strabismus, as well as detailed history taking and laboratory workup for thyroid eye disease, myasthenia gravis, and inflammatory conditions such as juvenile idiopathic arthritis. As a first step, it would be reasonable to observe this patient given the negative review of systems.    Brown syndrome can be managed with observation if mild. If a medical cause is found, the underlying disease should be treated, and the strabismus is expected to resolve. However, if no clear etiology is found, surgical management is indicated if the patient is reporting significant unresolving diplopia, has hypotropia in primary gaze, or has significant orbital pain, compromised binocular vision, or significant head tilt."
Paediatric ophthalmology and strabismus 	Type 1 Duane retraction syndrome	"A 5-year-old boy presented to the paediatric eye clinic with his parents. The chief complaint of the parents was that the child’s eye movements were abnormal. He was not known for any past medical or surgical history. Pregnancy and birth were normal. He had reached all his developmental milestones and his vaccinations were up to date. He has two sisters, aged 10 and 7, who do not have any ocular problems. The parents mention that the child’s eyes seem to wander inwards.  On examination, the child had a mild left head turn. He also had 18 prism diopters of esotropia and his extraocular movements were abnormal. His left eye showed limited abduction but normal adduction. Interestingly, on adduction, his left eye appears smaller. His eyelids and pupil examination were otherwise normal. Visual acuity was 20/25 in both eyes. On cycloplegic refraction, he had bilateral mild hyperopia (+2.00) without astigmatism. Anterior segment examination and dilated fundus examination were normal without signs of optic nerve hypoplasia or atrophy."	"This patient has Type 1 Duane retraction syndrome (DRS) involving the left eye. Type 1 DRS is characterised by esotropia and limited abduction. Co-contraction of the medial and lateral rectus muscles on attempted adduction causes the globe to retract (thus making the palpebral fissure appear smaller). Type 1 is the most common form of DRS. Type 2 is characterised by limited adduction and mild exotropia, while Type 3 is associated with limited abduction and adduction.   The most important differential diagnosis is sixth nerve palsy. Differentiating those conditions is crucial because their management differs significantly, with sixth nerve palsy requiring neurologic investigation. Typically, observation of globe retraction confirms the diagnosis of Type 1 DRS. Another sign that the condition is not sixth nerve palsy is the modest primary position esotropia (less than what is seen with sixth nerve palsy for the same degree of abduction deficit). Upshoots or downshoots can also occur in the affected eye and are highly suggestive of the condition.  This child may require surgery, but no surgical approach can normalise the eye movements. Surgery is typically reserved for children with primary position deviations and a head turn (such as in this case). Surgery is also indicated with marked globe retraction and large upshoots and downshoots. Amblyopia is not an indication for surgery. While surgeon preferences may vary, for this patient, recession of the medial rectus muscle on the left eye side may correct the primary position deviation and eliminate the head turn. Augmenting the procedure with right medial rectus recession may be needed if the deviation is larger than 20 prism diopters. Lateral rectus resections are usually avoided because they may worsen the globe retraction."
Paediatric ophthalmology and strabismus 	Vernal keratoconjunctivitis	"A 9-year-old male presented to the ophthalmology clinic following 4 weeks of decreased visual acuity and discharge in both eyes. Since coming back to school from spring break, he has been having difficulty seeing the teacher’s handwriting on the blackboard. He is known for -1.50 myopia OU and infrequently wears his glasses. He has a previous history of asthma, atopic dermatitis and seasonal allergies. He is currently being treated with oral antihistamines for allergic rhinitis. He has no personal or family history of ocular disease.    On ocular examination, there was bilateral mucous discharge noted. Corrected visual acuity was 20/40 OD and 20/25 OS. Extraocular movements were full and non-painful OU. Pupils were 2mm OU and had normal pupil responses. On anterior segment exam, there was bulbar conjunctival hyperaemia and a superior limbal lesions noted OU. Palpebral papillae were also noted OD. There were no corneal abnormalities noted OU. Dilated fundus exam was normal OU. Intraocular pressures were 16 mmHg OU."	"The suspected diagnosis is vernal keratoconjunctivitis (VKC). VKC is an allergy associated with inflammatory disease that has a predilection for pre-pubescent males. It is associated with seasonal allergies, as seen in this patient as he began experiencing symptoms after returning from spring break. On examination, superior limbal lesions could be noted which given the other clinical signs, are most likely Horner-Trantas dots, which are white dots of eosinophils and epithelial debris and characteristic of VKC.   The main differential for VKC is atopic keratoconjunctivitis (AKC). Although AKC is associated with atopy, VKC is also associated with atopy and especially with seasonal allergies. AKC typically affects older individuals between the 2nd-5th decades of life. AKC also more commonly affects the lower tarsus. Other differentials include seasonal/perennial conjunctivitis and giant papillary conjunctivitis. Seasonal/perennial conjunctivitis is not associated with limbal lesions that are most likely Horner-Trantas dots and giant papillary conjunctivitis occurs most commonly in contact lens wearers.   The diagnosis is made clinically so next the patient should be treated with topical antihistamines. Antihistamine drops should be continued for the duration of the season and can be re-initiated a few weeks before next spring to prevent recurrence of VKC. Topical steroid drops are used for severe inflammation or when patients have associated corneal “shield” ulcers. VKC is a benign and self-limiting disease that commonly resolves with age or spontaneously with puberty, so long-term management is not required but should be managed in the short-term with medical treatment."
Paediatric ophthalmology and strabismus 	Optic nerve glioma	"A 15-year-old male is brought by his parents for progressive unilateral proptosis of his right eye. Over the past few months, they noticed that his right eye is bulging. This is accompanied by gradual loss of vision in the last 2 weeks. He denies any pain, redness or photophobia. The patient is known for neurofibromatosis type 1 with multiple café-au-lait spots and freckling in the axillary region. He recently immigrated and didn’t have regular exams and follow-ups in his home country. On examination, visual acuity was 20/50 in the right eye and 20/20 in the left eye. Pupils were reactive and symmetric, with a right relative afferent pupillary defect. Extraocular movements were complete. Visual fields by confrontation were difficult to do but seemed complete. Color vision was decreased in the right eye (7/11) compared to the left eye (11/11) on the Hardy Rand and Rittler (HRR) test. A proptosis of 4 mm was measured using the Herel exophthalmometer. On slit-lamp examination, around 5 Lisch nodules were present in each eye. There was right optic nerve swelling without hemorrhages or exudates. The left optic nerve was normal. The macula and posterior pole exam were normal in both eyes."	"The suspected diagnosis is right optic nerve glioma. The presence of gradual proptosis, vision loss and optic nerve edema in a patient with neurofibromatosis type 1 is highly suggestive of this diagnosis. Optic nerve glioma is a slow-growing tumour that typically affects children and has a prevalence of 15% in patients with NF1. It is one of the criteria for the diagnosis of this phacomatosis. They comprise around 1% of all intracranial tumours and they can be associated with intracranial hypertension.  The best next step would be to obtain neuroimaging to confirm the diagnosis and differentiate optic nerve glioma from other tumours, such as optic nerve sheath meningioma (ONSM). CT scan will show calcifications in the optic nerve sheath and adjacent bony hyperostosis in meningioma, whereas those signs are absent in gliomas. On MRI, there will be thickening of both the nerve and sheath by the tumour in gliomas. On the other hand, only the optic nerve sheath will be thickened with a relative sparing of the optic nerve substance, creating the “tram track/railroad” sign. While OSNM is mainly seen in adult female patients, it can also be seen in the paediatric population, specifically in patients with neurofibromatosis type 2. NF2 is associated with vestibular schwannomas. Management of these two tumours varies depending on the severity and progression of the tumour. Based on individual patient factors, options can include observation, chemotherapy, and, in rare cases, surgery or radiation therapy. Regular follow-up with a multidisciplinary team, including ophthalmologists, neurologists, and dermatologists, is essential for monitoring tumour presence or progression in patients with neurofibromatosis."
Paediatric ophthalmology and strabismus 	Rhabdomyosarcoma	"An 8-year-old male presented to the ophthalmology clinic for left eye swelling. He has had eyelid oedema for the last two months and it has been progressively increasing. His parents believed that it would decrease with time but they recently also noticed slight displacement of the patient’s eye. He is experiencing no pain or decreased vision. He has no previous medical history. He has no family history of ocular disease. He also has no history of ocular trauma or surgery.   On ocular examination, there is eyelid oedema and inferior displacement of the globe noted OS. The patient’s eyelids were soft, oedematous and non-tender to palpation, there was no palpable mass OS. Exophthalmometer measurements were 17mm OD and 24mm OS. Uncorrected visual acuities were 20/20 OU. Pupils were 3mm OU and there was no RAPD noted OU. Extraocular movements are full and non-painful OU. On anterior segment examination, there were no findings noted OU. On posterior segment exam, there were no abnormalities noted OU. His intraocular pressures were 13 mmHg OU."	"The suspected diagnosis in this patient is rhabdomyosarcoma. Rhabdomyosarcoma is the most common paediatric orbital malignancy. It presents with rapidly developing exophthalmos but can also have a chronic presentation of eyelid oedema. It occurs most commonly in the first decade of life and is most common in the superior/superonasal quadrant of the orbit, causing inferior/inferotemporal displacement of the globe. This patient has inferior displacement, proptosis and eyelid oedema which are characteristic of rhabdomyosarcoma. Limited extraocular movements and ptosis can be seen if the lesion is large enough but is not seen in this case.   Rhabdomyosarcoma can present similarly to many inflammatory conditions and conditions that can cause rapidly developing masses. Even though there is eyelid oedema noted, an inflammatory condition is unlikely due to the lack of inflammatory symptoms, this patient has symptoms of a rapidly developing mass. Conditions such as neuroblastoma, lymphangiomas, dermoid cysts and infantile haemangiomas are reasonable differentials that could be suspected, however rhabdomyosarcoma has a high index of suspicion as prompt diagnosis can be crucial for saving the life of the affected patients and thus is the suspected diagnosis.   The next step should be urgent imaging and biopsy. Based on the findings from clinical exam and imaging, the decision to perform and excisional/incisional biopsy should be made. Biopsy is necessary for establishing a diagnosis and managing the patient further. Following histopathological confirmation, referral to a paediatric oncologist should also be done for appropriate management with radiotherapy and chemotherapy. "
Paediatric ophthalmology and strabismus 	Isolated aniridia	"An 10-year-old male presented to the ophthalmology clinic complaining of poor vision and he is hoping to get a glasses prescription today. He has trouble seeing the board at school and his friend recently got a glasses prescription. He was adopted as an infant and his family history is unknown. He has never seen an ophthalmologist before. The adoptive parents report that he has had normal development and has no previous medical history. There is also no previous history of ocular disease or trauma.   On examination, there were no findings on external exam OU. His uncorrected visual acuities were 20/100 OU. Extraocular movements were full and non-painful OU, there was no nystagmus noted. Pupils were 8mm OD and 7mm OS in both light and dark, there were also minimal pupil responses noted OU. On anterior segment examination, iris stumps with iris processes were noted OU. There were also bilateral anterior subcapsular opacities noted. There were no corneal abnormalities noted OU. On posterior segment exam, there were no abnormalities noted OU. His intraocular pressures were 15 mmHg OD and 17mmHg OS."	"The suspected diagnosis is isolated aniridia. Aniridia is a rare familial or sporadic condition that can result in abnormalities of the iris, cornea, angle structures, lens, optic nerve and fovea. Isolated aniridia is inherited in autosomal dominant inheritance in 2/3rds of cases and sporadic in 1/3rd of cases. Isolated aniridia is suspected in this patient due to hypoplasia of the iris, reduced visual acuity, lens opacities and normal development. Inheritance could be autosomal dominant or sporadic in this patient.   The main differential is aniridia due to WAGR syndrome, Gillespie syndrome and other anterior segment developmental abnormalities. WAGR syndrome involves deletion of the PAX6 gene and causes Wilm’s tumour, Aniridia, Genital abnormalities, and mental Retardation (WAGR). Since this patient has had normal development, WAGR syndrome is less likely. Gillespie syndrome is associated with cerebellar ataxia and intellectual disability, which is also not seen in this patient. Many of the anterior segment syndromes such as Rieger syndrome and Peter’s anomaly result in many systemic findings on exam, which is not seen in this patient and is unlikely to be the diagnosis.   The next step in this patient should be to have an OCT scan to detect foveal hypoplasia. Foveal hypoplasia is almost always present in aniridia and should be confirmed with OCT. Following OCT, cataract extraction should also be considered if the lens opacities are severe and significantly reducing the patient’s vision. Children with sporadic aniridia also have a 30% chance of Wilm’s tumour development and require screening, typically with renal ultrasound. From now on this patient should also have regular eye examinations for monitoring the many eye conditions that can develop from this condition."
Paediatric ophthalmology and strabismus 	Orbital cellulitis	"An 8-year-old boy presented with his parents to the emergency department for a swollen right eyelid and inability to open his eye. He was not known for any past medical or surgical history. Pregnancy and birth were normal. He had reached all his developmental milestones and his vaccinations were up to date. The parents recall that the boy recently had a cold, and on the 4th day, he developed redness and swelling around his right eye. Over the last 24 hours, it has gotten worse and he can no longer open his right eye. According to the mother, the boy is eating well but appears lethargic.  On examination, he had no obvious signs of facial trauma. The right upper eyelid was shut closed, there was redness and swelling of the upper and lower eyelids. Visual acuity could not be measured without manual opening of the eyelid. Visual acuity was measured at 20/40 in the right eye and 20/20 in the left. The right conjunctiva was slightly red. Adduction of the right eye was painful and appeared mildly limited. There was possible mild proptosis, but it could not be measured with an exophthalmometer. His anterior segment and dilated fundus examinations were normal. Intraocular pressures were 21 mmHg and 17 mm Hg."	"The suspected diagnosis is right orbital cellulitis. A painful, red, inflamed orbit in a child should raise suspicion for this infectious condition. The differential diagnosis includes other less common causes including autoimmune (thyroid eye disease), granulomatosis (sarcoidosis) and tumours (rhabdomyosarcoma, neuroblastoma).   The most important step when evaluating this child is to determine if the infection is preseptal or post-septal (orbital), given the differing management implications of both conditions and the life-threatening nature of orbital cellulitis. Both conditions present with eyelid swelling and erythema. Pain on extraocular movements, proptosis, limitation of extraocular movements and decreased vision are suggestive of orbital cellulitis. Orbital cellulitis is most commonly due to infection extending from the paranasal sinuses (sinusitis), as in this case, but can also be associated to dental infections, dacryocystitis and occasionally exogenous causes like trauma and foreign bodies.   This child will require broad-spectrum antibiotics, typically intravenous and in an inpatient setting. The most common complication is a subperiosteal or orbital abscess which is seen in 10% of cases. It would be reasonable to obtain a CT scan of the orbits, sinuses and brain to evaluate the degree of sinusitis and to rule out this complication. If no imaging is obtained, with daily follow-up, any systemic compromise, failure of improvement of the orbital signs or vision loss, should warrant urgent imaging and surgical consultation. The Harris-Garcia criteria recommend surgical management for children aged 9 and older, those with frontal sinusitis, non medial or large abscesses, suspected anaerobic infections (evidenced by gas on CT), optic nerve or retinal compromise, chronic sinusitis, recurrent abscesses, and infections originating from dental sources."
Retina and vitreous	Geographic atrophy secondary to age-related macular degeneration	"An 85-year-old lady is referred to the eye clinic with a three year history of a gradual deterioration in vision in both eyes. She describes her vision as ""blurry and dim,"" and reports increasing difficulty reading and recognizing faces, particularly in low light. She has a history of hypertension and diabetes and is an ex-smoker, having quit 10 years before (40 pack-year history). Past ocular history is significant for bilateral cataract surgery 5 years before.  Visual acuities were 20/120 OD and 20/80 OS. Intraocular pressures were 15mmHg OD and 18mmHg OS. Anterior segment examination was unremarkable, with clear corneas and quiet pseudophakia present. Dilated fundus examination showed scattered macular drusen and well-demarcated hypopigmented areas through which the underlying choroidal vessels can be seen; these areas involve the fovea on the right and correspond with hypoautofluorescent areas on autofluorescence imaging. Optical coherence tomography scans confirm complete retinal pigment epithelium and outer retinal atrophy (cRORA) in the right eye and incomplete outer retinal atrophy (iRORA) in the left eye. There were no signs of neovascular activity."	"The suspected diagnosis is geographic atrophy secondary to age-related macular degeneration (GA-AMD). It is important to consider and exclude other conditions that can masquerade as geographic atrophy, for example inherited retinal diseases (Stargardt’s disease, pattern dystrophy, cone dystrophy, mitochondrial retinopathy) or other degenerative causes (chronic central serous chorioretinopathy).   GA-AMD tends to affect older individuals in their fifth decade of life, be less asymmetric than inherited retinal diseases, and be associated with drusen. Multimodal imaging including fundus autofluorescence is helpful in identifying masquerades related to inherited dystrophies, and electrophysiological testing may be indicated if the clinical suspicion arises. Making this distinction is important due to developments in the field of GA-AMD treatment.  At the present time, treatments to reduce the progression of GA-AMD are limited. Pegcetacoplan (Syfovre) intravitreal injections are currently the only treatment licensed to reduce the progression of GA-AMD based on the results of OAKS and DERBY trials, which demonstrated that monthly or two-monthly intravitreal injections resulted in a 20% reduction in progression of GA-AMD over 5 years. However, this has only been approved by the Food and Drug Administration (FDA) in the United States, and not in other jurisdictions in Europe or Asia because of concerns about the lack of clear, meaningful functional benefit. Further phase II and phrase III trials exploring other intravitreal and systemic therapies for GA-AMD are ongoing. In the meantime, a recent post-hoc analysis of the AREDS and AREDS2 trials demonstrated that the antioxidant supplement could also slow the expansion of atrophic regions towards the foveal region in patients with GA-AMD.   Beyond medical treatments, it is also important to consider social support in terms of referral to the low vision services, where low vision aids and environmental changes to improve the patient’s vision-related quality of life can be discussed. Support from eye charities or support groups may also be beneficial."
Retina and vitreous	Myopic choroidal neovascularization	"A 40-year-old woman presents to the emergency eye clinic with a 2-day history of blurred central vision and ‘wavy lines’ in her right eye. Her past medical history is unremarkable. Past ocular history is significant for high myopia (-11DS OD and -14DS OS).   Best corrected visual acuities were 20/60 OD and 20/20 OS. Anterior segment examination was unremarkable, with clear corneas and minimal lens opacities present. Dilated fundus examination showed bilateral tilted discs and peripapillary atrophy associated with a tesselated fundus appearance. The right macula featured a small greyish subretinal lesion centrally associated with a sliver of haemorrhage, while the left featured irregular yellowish bands in the posterior pole suggestive of lacquer cracks."	"The suspected diagnosis is myopic choroidal neovascularization (CNV) secondary to pathologic myopia. CNV is most commonly associated with degenerative conditions (e.g. age-related macular degeneration) in the elderly population, but may also result from multiple other causes such as myopia in younger patient groups, or inflammatory conditions (e.g. punctate inner choroidopathy, presumed ocular histoplasmosis syndrome), trauma (choroidal rupture), choroidal tumours (e.g. choroidal osteoma), inherited retinal dystrophies (e.g. Best disease), or be idiopathic in nature.  Multimodal imaging can be performed to confirm the diagnosis of a CNV. Optical coherence tomography (OCT) tends to show a hyperreflective area under the retinal pigment epithelium which may be associated with intraretinal fluid, subretinal fluid, or haemorrhage. An active myopic CNV may feature as a well-demarcated area of early hyperfluorescence on fluorescein angiography (‘classic’ pattern of CNV). If non-invasive imaging is preferred, OCT angiography can be performed, and myopic CNV will demonstrate as a ‘lacy’ vascular network.  Intravitreal anti-vascular endothelial growth factor (anti-VEGF) injections are the mainstay of treatment, and are often administered on a pro-re-nata (as required) basis. The REPAIR and RADIANCE trials demonstrated the efficacy and safety of intravitreal ranibizumab injections for patients with myopic CNV, while the MYRROR trial confirmed this in Asian populations. Intravitreal bevacizumab is not approved for the treatment of myopic CNV, but is often used on an off-license basis. Close monitoring is required."
Retina and vitreous	Centre-involving diabetic macular edema 	"A 40-year-old woman was referred to the eye clinic following a routine sight test at her optometrists. She had attended for screening on the advice of her primary care physician after a recent diagnosis of type 2 diabetes, with a HbA1c of 12.1%. Her past medical history also includes hypertension and hyperlipidaemia. Past ocular history was unremarkable, and she denies any ocular symptoms.   Visual acuities were 20/25 OD and 20/20 OS. Anterior segment examination was unremarkable. Dilated fundus examination featured circinate exudates and retinal thickening in the macula of the right eye, which corresponded to centre-involving macular oedema on optical coherence tomography, with a central retinal thickness of 454 microns. There was no macular oedema in the left eye. There were a few scattered microaneurysms in both eyes in keeping with mild non-proliferative diabetic retinopathy."	"The suspected diagnosis is centre-involving diabetic macular edema (CI-DME). Differential diagnoses may include retinal vein occlusion, pseudophakic macular oedema, or radiation retinopathy; none of these are likely from the history and examination.  DME can develop with any level of diabetic retinopathy. Risk factors include poor glycaemic control, hypertension, hyperlipidaemia, and the duration of diabetes. Patients may be asymptomatic or may report blurred vision, metamorphopsia, and difficulty reading.   Both systemic and ocular management should be considered. In terms of systemic management, working with the patient’s primary care physician to optimise their glycaemic control and cardiovascular risk factors is crucial in reducing the risk of progression as well as systemic complications of diabetes.   In terms of ocular management, the DRCR.net Protocol V study is particularly relevant to this patient. This study was designed to guide the management of people with centre-involving DME and good vision (20/25 or better). At the two-year study endpoint, patients treated with observation, intravitreal aflibercept, or macular laser had comparable vision, suggesting that observation is entirely appropriate with the option of adding in anti-VEGF if there is clinical deterioration. However, if the patient had been symptomatic with a visual acuity worse than 20/25, consideration of intravitreal anti-vascular endothelial growth factor (anti-VEGF) injections as a first-line treatment would have been clinically appropriate as well. Intravitreal corticosteroids such as dexamethasone implants are another option for patients who prefer lower frequency of treatment or demonstrate a poor response to anti-VEGF agents; however, this needs to be balanced against the risks of cataract formation and steroid-induced intraocular pressure response."
Retina and vitreous	Sickle cell retinopathy	"A 27-year-old male of African descent presented to the ophthalmology clinic complaining of a three day history of decreased vision and floaters in his right eye. He began experiencing these symptoms suddenly after coming home from work. He is an investment banker and finds his work very stressful. He is known for -2.50 myopia OU and wears glasses. He has no previous medical history. He has no history of intravenous drug use. His mother has a history of glaucoma and the family history of his father’s side of the family is unknown.   On examination, there were no abnormalities noted on external exam. His corrected visual acuity was 20/100 OD and 20/20 OS. Extraocular movements were full and non-painful OU. Confrontational visual fields were full OU. Pupils were 3mm OU and had normal pupil responses. On anterior segment examination, there were no abnormalities noted OU. On posterior segment examination, there was vitreous haemorrhage and superotemporal neovascularisation noted OD. There was superotemporal neovascularisation also noted OS but without vitreous haemorrhage. His intraocular pressures were 16 mmHg OU."	"The suspected diagnosis is sickle cell retinopathy. Sickle cell retinopathy is the ocular manifestation of the spectrum of sickle cell disease. Given this patient's African descent and examination finding of a neovascularisation and vitreous haemorrhage, sickle cell retinopathy is the most likely diagnosis. Retinopathy is also most common and severe in patients with sickle cell trait rather than sickle cell disease. Patients with sickle cell trait are commonly undiagnosed due to the lack of systemic health complications.   In this case, sickle cell retinopathy is the main suspected diagnosis. However other conditions can also present similarly. Sarcoidosis is common among people with African descent and can cause peripheral neovascularisation with vitreous haemorrhage. However, systemic symptoms and uveitis are prominent in sarcoidosis and not seen in this patient. Eales disease is an idiopathic occlusive vasculitis that most commonly affects young males, which causes recurrent vitreous haemorrhages and can cause neovascularisation. However, it is more common in people of Indian and Middle Eastern descent and is a diagnosis of exclusion.   The next step in this patient is to perform fluorescein angiography to examine the blood flow to the retina and choroid. This is needed to gain more insight into the severity of the patient's retinopathy. Neovascularisation with associated vitreous haemorrhage should be treated with pan-retinal photocoagulation to the avascular area, which should be done in the patients right eye. Isolated retinal neovascularisation, as seen in the patients left eye, does not require immediate treatment. The patient should also be referred to a haematologist for evaluation and establishment of care."
Retina and vitreous	Macula-on rhegmatogenous retinal detachment 	"A 50-year-old gentleman presented to the eye emergency clinic with a 1-day history of increased floaters and describes a “curtain-like shadow” in the bottom part of his visual field in his right eye. He had noticed increased flashes in this eye for 1 week prior to this. Past medical history was unremarkable. Past ocular history was significant for high myopia (-12DS OD and -10.5DS OS). He denies any antecedent trauma.   Visual acuity was 20/20 OD and 20/25 OS. Intraocular pressures were 15mmHg OD and 18mmHg OS. Anterior segment examination was unremarkable and he had early lens opacities present only. Dilated fundus examination revealed tobacco dusting in the right eye, a Weiss ring, and a superior rhegmatogenous retinal detachment with two horseshoe tears at 11 and 12 o’clock. The macula was unaffected. The left eye featured an incomplete posterior vitreous detachment and no peripheral retinal lesions. Scleral indentation did not identify any further breaks in either eye."	"This patient has a superior macula-on rhegmatogenous retinal detachment in his right eye which requires urgent intervention within 24 to 48 hours.   Given the presence of a posterior vitreous detachment, a pars plana vitrectomy with cryotherapy/ retinopexy and a gas tamponade could be considered. However, he would also qualify for pneumatic retinopexy as he meets the PIVOT trial criteria (a single retinal break or a group of breaks in the detached retina within 1 clock hour above the 8- and 4-o'clock meridia, for both macula-off and macula-on cases).   The PIVOT trial was a randomised controlled trial comparing the outcomes of pneumatic retinopexy versus vitrectomy in patients meeting these recruitment criteria. Key findings included better visual acuity outcomes, less vertical metamophopsia, and fewer intraoperative and postoperative complications for patients who underwent pneumatic retinopexy as compared to vitrectomy. This suggests that pneumatic retinopexy could be considered as first-line treatment in this group of patients, subject to the surgeon’s experience with this procedure and the patient’s ability to comply with strict postoperative posturing."
Retina and vitreous	Elevated intraocular pressure related to postoperative anterior chamber inflammation	"A 74-year-old gentleman attends the eye clinic for a postoperative check 2 weeks after pars plana vitrectomy, cryopexy, and SF6 gas tamponade for a macula-off rhegmatogenous retinal detachment in his left eye. He reports that he has been experiencing a mild headache in the past few days and that his vision remains slightly blurred. He is on a postoperative regimen of topical antibiotics and a tapering course of topical steroids as well. Past medical history was significant for hypertension and type 2 diabetes. Past ocular history was unremarkable apart from uncomplicated cataract surgeries in both eyes.    Visual acuities were 20/20 OD and 20/200 OS. Intraocular pressures were 15mmHg OD and 38mmHg OS. On examination, the left eye was mildly injected with a resolving subconjunctival haemorrhage present as well. The cornea was clear with a few Descemet membrane folds, and 2+ cells were present in the anterior chamber. There was quiet pseudophakia in the right eye. The anterior chambers were deep bilaterally. Dilated fundus examination of both eyes featured flat retinas with faint cryotherapy scarring superiorly in the left eye and a 50% gas fill."	"This patient has elevated intraocular pressure (IOP) after retinal detachment repair, which is most likely to be related to anterior chamber inflammation. In the early postoperative phase, elevated IOP may result from gas overfill or steroid response - the former is unlikely given that the gas tamponade has already started to resolve, and while the latter is possible, steroid-induced IOP rise typically occurs three to six weeks after initiation of topical steroids although it can sometimes occur earlier. The deep anterior chambers rule out the (rare) possibility of aqueous misdirection. Patients with pre-existing glaucoma are at a higher risk of postoperative IOP elevation due to the compromised trabecular meshwork; however this is not the case here.   Prolonged surgery and a history of diabetes may have contributed to the significant anterior chamber inflammation. In these cases, the inflammatory cells can clog up the trabecular meshwork, contributing to elevated IOP. In terms of management, it is important to first perform careful examination to exclude significant differentials such as endophthalmitis and other causes of IOP elevation as outlined above. The patient requires an intensive regimen of topical steroids to treat the inflammation and IOP-lowering medications to control the IOP while the inflammation settles, which will reduce the risk of peripheral anterior synechiae formation and thereby secondary angle closure glaucoma. Close monitoring is important to ensure that the IOP normalises.  Other common or serious complications of pars plana vitrectomy include cataract formation, suprachoroidal haemorrhage, secondary glaucoma, cystoid macular oedema, retinal tear or detachment, uveitis, and endophthalmitis. Proliferative vitreoretinopathy may also occur after a retinal detachment. "
Retina and vitreous	Retinitis pigmentosa	"A 15-year-old male patient presented to the ophthalmology clinic because of having trouble seeing at night. He has noticed a gradual decrease in his night vision for the last few months and his parents are concerned that something might be wrong. He has not noticed a decrease in vision during the day. He is known for -2.50 myopia OU and wears glasses. He has no previous medical history and does not take any regular medications. He has no family history of ocular disease.   On examination, there were no abnormalities noted on external exam. His corrected visual acuities were 20/20 OU. Extraocular movements were full and non-painful OU. Pupils were 2mm OU and had normal pupil responses. On anterior segment examination, there were no abnormalities noted OU. On posterior segment examination, there was arteriolar attenuation and hypopigmentation noted in the peripheral retina OU. His intraocular pressures were 15 mmHg OU."	"The suspected diagnosis is retinitis pigmentosa (RP). This patient has been experiencing progressive night blindness (nyctalopia) and has associated arteriolar attenuation and retinal pigmentary changes on fundus exam, which are highly suggestive of retinitis pigmentosa. Although he has no family history of ocular disease, autosomal recessive inheritance is the most common form of inheritance.   There are several other genetic disorders that cause pigmentary retinopathy and nyctalopia which include gyrate atrophy and choroideremia. Both diseases also commonly present in childhood or adolescence. However, in gyrate atrophy there is chorioretinal atrophy with hyperpigmentation of healthy RPE, and in choroideremia there are normal retinal vessels and an X-linked inheritance pattern in affected families is prominently seen. Infectious diseases such as syphilis and rubella can also cause pigmentary retinopathy that resembles RP. However in syphilis infection there would be many other ocular findings noted and in rubella infection the patient would have normal retinal vessels and normal visual function, including night vision.   The next step in this patient should be to have a full-field electroretinogram (ERG) and formal Goldmann visual field testing done. ERG is essential for the diagnosis of RP as it measures the electrical potential generated by rods and cones. Classically, ERG findings in RP would show an “extinguished” signal in RP. A formal Goldmann visual field test is also needed as RP commonly affects the mid-peripheral visual field. It is also useful for documenting progression of disease on follow-up appointments. Following confirmatory diagnostic testing, genetic counselling would also be beneficial to educate the patient and his family on the inheritance and health implications of RP."
Retina and vitreous	Macular telangiectasia type 2 (MacTel)	A 53-year-old woman is referred to the eye clinic with several years of increasing difficulty with reading. Past medical history includes type 2 diabetes mellitus. She denies any previous ocular history.   Visual acuities were 20/20 OU. Retinoscopy was -0.5DS OD and -0.25 DS OS. Both anterior segments were within normal limits with minimal lens opacities present. Dilated fundus examination featured blunting of the foveal reflex and subtle greyish discolouration of the temporal perifoveal retina; no signs of diabetic retinopathy were present.	"The suspected diagnosis is macular telangiectasia type 2 (or MacTel). Fundoscopic changes are subtle in early stages of the condition (as described above) and may be easily missed, resulting in delayed diagnosis. More advanced stages may feature perifoveal telangiectactic capillaries, right-angled venules, and pigmentary changes relating to intraretinal retinal pigment epithelial (RPE) cell migration.   Multimodal retinal imaging is therefore essential in helping to confirm the diagnosis,  Optical coherence tomography demonstrates asymmetry of the foveal pit early in the disease course, and may subsequently feature loss of the outer nuclear layer and ellipsoid zone temporal to the fovea which develop into outer retinal cavitations. This is often associated with the ‘internal limiting membrane (ILM) drape’ sign characteristic of MacTel, where the ILM is preserved and overlies the areas of cavitation. Blue-light reflectance imaging shows increased reflectance (typically temporal) in the areas of macular pigment loss, and can be helpful early in the disease course. Fluorescein angiography can be used to highlight the vascular changes - typically telangiectatic parafoveal vessels temporal to the fovea in the early stages and diffuse hyperfluorescence in more advanced stages.  Although several treatment trials are ongoing, no proven treatments for MacTel exist at this time. Instead, clinicians may opt to monitor patients for complications - these are uncommon but may include secondary macular neovascularisation (which can be treated with anti-vascular endothelial growth factor injections) or macular hole (for which surgery confers a much poorer prognosis than idiopathic cases)."
Retina and vitreous	Primary vitreoretinal lymphoma (PVRL)	"A 67 year old man presents with decreased vision in the left eye for the past 1 month. He has also seen an increase of floaters. He denies flashes, eye pain, and eye redness. There are no symptoms in the right eye.   On exam, visual acuity is 20/20 in the right eye and 20/60 in the left eye. Pupils are equal, round, and reactive to light without afferent pupillary defect. Intraocular pressure is 12 in the right eye and 17 in the left eye. Extraocular motility is full in each eye, and visual fields are full to confrontation in each eye. On slit lamp exam, the anterior exam is normal in both eyes. On dilated exam, the left eye demonstrates 2+ mixed vitreous cells. There are few small white vitreous opacities inferiorly. The optic nerve is sharp and perfused, with cup/disc ratio 0.3. The macula has two 1DD-sized creamy pale infiltrates with preserved overlying retinal vessels. The periphery has many scattered similar infiltrates of varying size, from 1mm in diameter to 2DD large. The posterior examination of the right eye is normal. Vitreous tap is performed, and flow cytometry of the vitreous demonstrates IL-10:IL-6 ratio of 2:1 and increased monoclonal lymphocytes."	"As evidenced by his flow cytometry results, this patient has primary vitreoretinal lymphoma (PVRL), a type of CNS lymphoma, which is itself a subtype of diffuse large B-cell lymphoma. The major risk factors for PVRL are age (>50) and immunocompromised status.   Intraocular lymphoma often presents as anterior or posterior uveitis; PVRL should be suspected in patients, especially older or immunocompromised patients, who have a posterior uveitis with an otherwise negative workup for infectious or other inflammatory causes of posterior uveitis. PVRL classically presents with subRPE infiltrates, but these were only present in 20% of patients in one series. Thus, suspicion should remain high for any posterior uveitis without a clear cause. The most common method of diagnosis is via cytology of vitreous or aqueous cells, which would demonstrate a mix of normal lymphocytes (indicating an inflammatory response), necrotic cells, and large abnormal lymphocytes with abnormal nuclear characteristics, including lymphoma. An increased IL-10:IL-6 ratio (>1), a population of monoclonal lymphocytes, and a positive L265P mutation of the MYD88 gene are positive diagnostic criteria for PVRL."
Retina and vitreous	Full-thickness macular hole	"A 64-year-old female presented to the ophthalmology clinic complaining of a decrease in vision and image distortion in her left eye that has been gradually increasing over the last few months. She believed this was common in people when they got older, but is now finding her symptoms debilitating. She has a history of hypertension and diabetes that is controlled with medication. She has no family history of ocular disease. She has no history of ocular trauma and had cataract surgery on both eyes two years ago.   On examination, there were no abnormalities noted on external exam. Her corrected visual acuity was 20/20 OD and 20/200 OS. Extraocular movements were full and non-painful OU. Confrontational visual fields were full OU. Pupils were 2mm OU and had normal pupil responses. On anterior segment examination, there were no abnormalities noted OU. On posterior segment examination, there was a round, red spot in the centre of the macula. On slight lamp examination, when shining a thin beam of light on the macula, a break in the line can be noted OS. There was no posterior vitreous detachment noted OU. His intraocular pressures were 14 mmHg OD and 13 mmHg OS."	"The suspected diagnosis is a full-thickness macular hole. Macular holes are commonly idiopathic and seen in women between the 6th and 8th decades of life. It can present with variably decreased vision, metamorphopsia or a central scotoma. The patient has a round, red spot in the centre of the macula and a break in the beam of light on slit lamp exam which is known as a positive Watzke-Allen sign and suggests a full-thickness macular hole.   The differential includes pseudoholes and lamellar macular holes, which can be difficult to distinguish. Pseudoholes do not involve loss of foveal tissue and would be seen as an abnormal, hollow foveal contour on dilated fundus exam due to an epiretinal membrane. Pseudoholes would also not present with a positive Watzke-Allen sign. Lamellar macular holes present similarly to full-thickness macular holes but the loss of visual acuity is typically less severe and would not present with a positive Watzke-Allen sign.   The next step in this patient should be to have an OCT, which is the gold standard for diagnosis and management of this condition. The International Vitreomacular Traction Study (IVTS) Group has formed a classification scheme of vitreomacular traction and macular holes based on OCT findings. OCT is critical for evaluating the vitreoretinal interface and determining the degree of traction from vitreous or epiretinal membranes. OCT will also be used to stage the macular hole, which will guide further medical management and surgical management with a pars plana vitrectomy with tamponade +/- peeling of the internal limiting membrane."
Retina and vitreous	Central serous chorioretinopathy	"A 53-year-old man presented to the ophthalmology clinic complaining of a sudden decrease in vision in his right eye. Two days ago he noticed that images from his right eye were blurry and  distorted. He has never experienced these symptoms before. He has no refractive error or previous history of ocular disease. He has a history of arthritis and regularly uses oral steroids, most recently a week ago. He is a former smoker with a 10-pack-year history of cigarette smoking.  On ocular examination, there were no findings on external exam. His uncorrected visual acuities were 20/20 OS and 20/40 OD. Ocular movements were full and non-painful OU. Confrontational visual fields were full OS and there was a small central scotoma and central distortion on Amsler grid testing noted OD. Pupils were 3mm OU and there was no RAPD noted. On anterior segment exam, there were no findings noted OU. On dilated fundus exam, there was subretinal fluid noted beneath the fovea with adjacent hypopigmentary changes inferiorly OD. Intraocular pressures were 13mmHg OU."	"The suspected diagnosis is central serous chorioretinopathy (CSCR). CSCR is a relatively common and idiopathic retinopathy. It causes central visual loss, metamorphopsia, hyperopic/myopic shift and reduced sensitivity to contrast sensitivity and colour saturation. The pathophysiological mechanisms are not fully understood but it is known to cause a serous detachment of the neurosensory retina. This patient's symptoms and exam findings are consistent with the CSCR. In addition, this patient recently used steroids for his arthritis and steroid use has the strongest association with CSCR.  Age-related macular degeneration (AMD) can present similarly to CSCR. AMD is usually seen in patients older than 50 and can also blurry vision with metamorphopsia. It can also cause a macular neurosensory detachment with pigmentary changes. However, drusen is a prominent feature in AMD and not seen on examination of this patient so is less likely, especially considering this patient’s recent steroid use which is highly associated with CSCR.   Next, diagnostic testing such as fluorescein angiography and OCT should be performed to rule out other differential diagnosis such as AMD. Acute CSCR is usually self-limiting and has a good prognosis, so the patient can be observed. If possible, steroid treatments should also be withheld until there is resolution of the CSCR."
Retina and vitreous	Epiretinal membrane (ERM)	"A 74-year-old man presents with a 3-month history of worsening metamorphopsia and blurred vision in his left eye. He denies any previous ocular history or surgery. There is no past medical history of note.   On examination, best corrected visual acuities were 20/25 OD and 20/60 OS. Intraocular pressures were 15 OD and 17 OS. Both anterior segments were within normal limits with minimal lens opacities present. Dilated fundus examination was unremarkable in the right eye and featured a whitish sheen overlying the macula and dragging of the perifoveal vessels in the left eye."	"The most likely diagnosis is an epiretinal membrane (ERM). This can be primary (idiopathic) or secondary to previous retinal treatment or pathology, such retinal tears, retinal laser, or previous vascular occlusions. It is therefore important to perform a dilated fundus examination (with consideration of scleral indentation)  to exclude secondary causes of epiretinal membranes requiring additional treatment.  Optical coherence tomography (OCT) scans are often helpful in documenting the preoperative appearance, assessing the vitreoretinal interface (e.g. for vitreomacular traction), or preoperative counselling regarding visual prognosis (e.g. in chronic cases with severe disruption of the outer retinal layers, where the visual prognosis is guarded). Fluorescein angiography may be considered in cases where a previous retinal vascular occlusion is suspected.    Management options include conservative treatment (watchful waiting) for asymptomatic patients or those with mild ERM, or surgical intervention (pars plana vitrectomy with epiretinal membrane peel) for symptomatic patients who have been adequately counselled regarding the risks and benefits of surgery. There is no clear evidence that concurrent internal limiting membrane peel improves surgical outcomes or reduces the risk of recurrence, which has been hypothesised previously. Patients should be counselled regarding their expectations of surgery - most patients experience some degree of improvement in distortion by 12 months postoperatively, while improvement in visual acuity may not always occur depending on the preoperative anatomical appearances."
Oculoplastics	Involutional ectropion	"A 74-year-old man is referred to the ophthalmology clinic with a 6-month history of right eye irritation and watering. He reports that his vision is intermittently blurry as well. He does not have any past medical history of note. He wears contact lenses for myopia and his past ocular history is otherwise unremarkable.  Visual acuities were 20/20 OD and 20/25 OS. Intraocular pressures were 15mmHg OD and 16mmHg OS. On examination, there were signs of periocular dermatitis, and his right lower lid appeared to be turned outwards. Punctal eversion was present as well. There was significant horizontal and vertical lower lid laxity on the right and very mild changes on the left. There were no ocular masses or periocular scarring, and facial nerve examination was unremarkable. Anterior segment examination featured mild punctate epitheliopathy in the right cornea. Dilated fundus examination was unremarkable."	"This patient has a right lower lid ectropion which is most likely involutional in nature. This is the most common cause of an ectropion, but it is important to exclude paralytic, mechanical, and cicatricial causes in the history and examination as well. Risk factors in this patient comprise age, periocular dermatitis, and this history of contact lens wear (which can involve lid pulling). Other risk factors to consider may include eyelid rubbing, a history of trauma, and the use of certain classes of intraocular pressure lowering eyedrops.   In terms of management, lubricating eye drops can provide some symptomatic relief and protect the ocular surface. The periocular skin disease should be adequately treated in the first instance, as this may reduce or reverse the degree of ectropion prior to considering definitive treatment with surgery. This is particularly indicated for patients who are very symptomatic or have corneal changes. Surgical treatment should address the underlying anatomical problem and may depend on the severity and the individual surgeon’s preferences. For example, a lateral tarsal strip procedure may be undertaken to address the horizontal lower lid laxity, combined with a medial spindle to address the punctal eversion, and/or the Jones procedure for lower lid retractor disinsertion."
Oculoplastics	Orbital compartment syndrome secondary to a retrobulbar haemorrhage	"A 48-year-old male presents to the emergency department with acute reduction in vision and periocular pain in his left eye 2 hours following blunt trauma to his face from a soccer ball. Past medical and ocular history was unremarkable. He denies the use of any medications such as anticoagulants.  Visual acuity was 20/20 OD and light perception OS. Intraocular pressures were 12mmHg OD and 47mmHg OS. The left eye was proptosed (3mm on Hertel exophthalmometry) with oedema, ecchymosis, and tightness of the left upper lid and complete ophthalmoplegia. There was a relative afferent pupillary defect in the left eye. The right eye had a normal pupillary reflex and full ocular motility."	"The suspected diagnosis is orbital compartment syndrome secondary to a retrobulbar haemorrhage from blunt trauma. This is a sight-threatening ocular emergency which occurs when the intraorbital pressure (for example from the haemorrhage) exceeds the perfusion pressure of the ophthalmic artery due to rigid walls of the bony orbit and tight orbital septum. Lateral canthotomy and cantholysis should be undertaken urgently to restore vision before there is irreversible structural damage to the retina and/or optic nerve. The inferior crus of the lateral canthal tendon is cut first, and if there is no improvement in the clinical signs or symptoms, the superior crus is then addressed. Early and prompt treatment is essential. This should not be delayed by attempts to obtain neuroimaging to confirm the diagnosis. The patient should be closely monitored postoperatively for signs of recurrence from a rebleed.   It is important to note that the presence of retrobulbar haemorrhage does not necessarily require treatment with lateral canthotomy and cantholysis. This is a common misconception. In the absence of orbital compartment syndrome, the risk of the retrobulbar haemorrhage causing compressive optic neuropathy and/or elevated intraocular pressure compromising the central retinal artery and therefore permanent visual loss is very low. Retrobulbar haemorrhage without orbital compartment syndrome may occur in the context of concurrent orbital fractures.  Other causes of retrobulbar haemorrhage include orbital surgery, or spontaneous bleeding from coagulopathies , or bleeding from intraorbital lesions such as an arteriovenous malformation. "
Oculoplastics	Orbital inflammatory disease	"A 60-year-old woman presented with a 1 week history of progressive severe right-sided retro-orbital ache which was subsequently followed by right upper eyelid swelling associated with ptosis. She has no past medical or ocular history and denied any antecedent trauma. She was systemically well and afebrile.   On examination, her best corrected visual acuity was 20/20 in both eyes. Intraocular pressures were 15mmHg on the right and 11mmHg on the left. There was no relative afferent pupillary defect or anisocoria. Colour vision and confrontation visual fields by confrontation were full. There was a complete ptosis on the right without associated fatiguability or periocular erythema. A subtle right-sided proptosis was present. Extraocular movements revealed incomplete ophthalmoplegia on the right; cranial nerve examination was otherwise unremarkable. The anterior segments were quiet bilaterally, and fundoscopic examination was within normal limits."	"An acute painful, unilateral ophthalmoplegia raises the suspicion of orbital inflammatory disease. This can be attributed to systemic inflammatory diseases such as granulomatosis with polyangiitis, sarcoidosis, thyroid eye disease, or other connective tissue disorders; local inflammatory processes related to neoplasms or infections (orbital cellulitis), or idiopathic orbital inflammatory disease (Tolosa Hunt syndrome), which is a diagnosis of exclusion.  The clinical presentation of orbital inflammatory disease can vary depending on the orbital structure(s) affected. Radiologic imaging is essential in determining the extent of inflammation and to exclude any underlying local processes. The findings will depend on the structure(s) affected and the underlying pathology. CT and MRI scans are helpful in determining the extent of the inflammation and whether important structures such as the optic nerve are threatened, for example in cases with orbital apex or cavernous sinus involvement. B-scan ultrasonography may be considered to examine for a characteristic ‘T-sign’ suggestive of posterior scleritis.   Further testing should be guided by clinical presentation following a detailed history and examination. For example, laboratory tests may include full blood count, inflammatory markers (erythrocyte sedimentation rate and C-reactive protein), serum protein electrophoresis, a vasculitic screen, serum ACE, and thyroid function tests and autoantibodies.  Treatment depends on the underlying cause. "
Oculoplastics	Rhino-orbital-cerebral-mucormycosis	"You are consulted regarding a 54 year old man in the emergency department with a past medical history of type 2 diabetes mellitus on insulin presenting with headache, blurred vision in the right eye, and dark-appearing rash over the bridge of the nose. There are scattered necrotic-appearing lesions in the nasal mucosa. The patient is lethargic and febrile. Laboratory workup is significant for a white blood cell count of 16,000. CT of the face and orbits with contrast shows fat stranding, edema, and loss of tissue architecture in the nasal cavity, maxillary and ethmoid sinuses, as well as the right orbit.   Examination is limited by the patient’s mental status, but the patient displays -1 abduction and -1 supraduction of the right eye. Pupils are equal, round, and reactive to light in both eyes. Intraocular pressures are 18 and 15 in the right and left eyes, respectively."	"This patient likely has rhino-orbital-cerebral-mucormycosis, a devastating invasive fungal infection most commonly occurring in immunocompromised hosts and most commonly caused by the aseptate, filamentous fungus Rhizopus oryzae. This fungus is spread locally via intravascular dissemination and angiogenesis and commonly invades the nasal mucosa and adjacent structures, as its spores are inhaled through the nose or mouth. The patient’s toxic appearance, examination with black eschar lesions of the mucosa, apparent invasion of the maxillary and ethmoid sinuses and the orbit all suggest mucormycosis. For definitive diagnosis, biopsy and culture of affected tissue is required.   Treatment is surgical, with wide debridement as the goal with complete removal of necrotic tissue. Antifungal therapy such as systemic liposomal amphotericin B is also suggested and will be started in conjunction with an infectious disease or medical team. The surgical approach commonly involves otolaryngology/head and neck surgery, neurosurgery (if intracranial spread is suspected), and plastic surgery. Ophthalmology may be involved if there is orbital involvement on exam or imaging. Long term survival outcomes are poor, and surgical excision of necrotic tissue may even necessitate exenteration if the infection has involved much of the orbit and globe."
Oculoplastics	Thyroid eye disease	"A 43-year-old female presented to the ophthalmology clinic for dry eyes. She has had a gritty sensation in her eyes for the last two months and has had no pain or changes in vision. She goes for runs every weekend but has recently begun experiencing more fatigue and weight loss than usual, even though there has been no change in her diet or exercise. She has no previous medical history. She has a family history of pancreatic cancer in her mother and colorectal cancer in her father that was diagnosed in their late 60’s. She also has no history of ocular trauma or surgery.    On ocular examination, there was mild eyelid oedema noted bilaterally. Her uncorrected visual acuity was 20/20 OU. Pupils were 2mm OU and there was no RAPD noted OU. Extraocular movements were limited on upgaze in both eyes. Confrontational visual fields were full OU. Exophthalmometer measurements were 22.5mm OS and 23mm OD. Margin reflex distances 1 were 4mm OU. On anterior segment exam, there were no findings noted OU. On posterior segment exam, there were no findings noted OU. Her intraocular pressures were 13 mmHg OU."	"The suspected diagnosis in this patient is thyroid eye disease (TED). TED commonly causes non-specific eye complaints in patients such as a gritty sensation. Clinical examination is the most important factor for making the diagnosis. The European Group of Graves' Orbitopathy (EUGOGO) classification assesses TED activity and severity. TED is the most common cause of unilateral and bilateral axial proptosis, which is seen in this patient evidenced by the exophthalmometer measurements. Proptosis and mild eyelid oedema are the only findings on physical exam and according to the EUGOGO classification gives the patient an activity level of inactive TED and severity level of mild TED. Although lid retraction is the most common exam finding in TED and not seen in this patient, the patient has signs such as fatigue and weight loss that are suggestive of systemic hyperthyroidism.   The differential for TED includes inflammatory conditions and orbital tumours. Inflammatory orbitopathy, orbital myositis and IgG4 disease are the inflammatory conditions that are in the differential diagnosis. In these inflammatory conditions pain, redness and periorbital oedema would be expected, which is not seen here. Orbital tumours can also present similarly but would most commonly be unilateral and not be associated with symptoms of hyperthyroidism, although they can rarely present bilaterally.   As this patient is presenting with bilateral proptosis without lid retraction on exam, the presentation is atypical and the patient should have a CT scan of the orbits next. As mentioned, orbital tumours can present bilaterally and can present with the findings seen in this patient’s exam. Given the patient's strong family history for malignancy, this is particularly important. Although she has apparent systemic symptoms of hyperthyroidism, her findings are non-specific and can be caused by malignancy. Therefore, a CT scan of the orbits is the most important initial investigation to rule out orbital tumours and malignancy."
Oculoplastics	Morpheaform basal cell carcinoma (BCC) of the eyelid	"A 69-year-old Caucasian female presented to the ophthalmology clinic due to a lesion on her lower left eyelid. She claimed that the lesion had been present for around 9 months. The lesion was painless but was a cosmetic concern. She did not notice any change in the lesion’s size over time. She had no previous medical history.  On examination, the left eye was normal except for a firm, flat, subcutaneous lesion on the lower lid with disruption of the lid margin, indistinct margins and hyperaemia. There was also madarosis in the area of the lesion. There were no abnormalities on examination of the lymph nodes in the head and neck. There were also no detected losses in facial sensation. Her corrected visual acuity was 20/20 OU. Extraocular movements were full and non-painful. Pupil responses were normal OU. Intraocular pressures were 12 mmHg OD and 11 mmHg OS. The eyelid lesion was photographed for documentation."	"The suspected diagnosis is morpheaform basal cell carcinoma (BCC) of the eyelid. The patient presented with an eyelid mass that was asymptomatic and not growing in size. The differential is broad and includes benign and malignant eyelid tumours. Signs seen on external examination provide useful information for differentiating between eyelid lesions.   External examination findings from this patient suggest BCC as the diagnosis. BCC is the most common malignancy of the eyelid. It is also more common on the lower eyelid and in the Caucasian population. There are many forms of BCC such as nodular, morpheaform, superficial and pigmented. The examination findings of a lower lid lesion with lid margin disruption, indistinct margins and associated hyperaemia (most likely due to telangiectatic vessels) are most consistent with a diagnosis of morpheaform BCC.   For confirmation, an incisional biopsy should be performed. On pathology, these tumours have characteristic small islands of tumour within dense fibrous tissue. Morpheaform BCC typically invades the underlying tissue deeper and therefore is more aggressive than the more common nodular subtype. Histopathological confirmation should precede any extensive procedures. "
Oculoplastics	Sebaceous carcinoma	"A 44-year-old female presented to the ophthalmology clinic complaining of swelling of her left upper eyelid. She first noticed the lesion three months ago but did not see an ophthalmologist since she believed it was a common and self-resolving lesion. She has no symptoms but she has noticed that the lesion has gotten bigger. She has a previous history of retinoblastoma in childhood that was treated with chemotherapy and local radiotherapy. She has no other medical history. She has no family history of ocular disease. She also has no previous history of ocular trauma or surgery.   On ocular examination, there was a small, yellow, and firm nodule with irregular borders on the lid margin of her left upper eyelid. Her corrected visual acuities were 20/20 OU. Pupils were 2mm OU and there was no RAPD noted OU. Extraocular movements are full and non-painful OU. Confrontational visual fields were also full OU. On anterior segment exam, there were no other findings noted OU. On posterior segment exam, there were no findings noted OU. Her intraocular pressures were 14 mmHg OD and 17mmHg OS."	"The suspected diagnosis in this patient is sebaceous carcinoma. Sebaceous carcinoma is a highly malignant and potentially lethal neoplasm of the sebaceous glands that should have a high index of suspicion. It is typically seen in elderly females, however it can also be seen in younger individuals if they have a history of receiving periocular radiation, such as in this patient. It is known to be a difficult diagnosis due to how it mimics other conditions. This patient has an atypical presentation of a chalazion with irregular borders and a history of periocular radiation so a sebaceous gland carcinoma should be the suspected diagnosis   The differential diagnosis for sebaceous carcinoma is broad and includes a number of benign and malignant conditions. Benign conditions such as chalazion, blepharitis, and keratoconjunctivitis can mimic sebaceous carcinoma. However, sebaceous carcinoma should be considered when there is recurrent or atypical presentation of any of the mentioned, such as in this case. Histopathologically, the differential diagnosis includes squamous cell carcinoma, basal cell carcinoma, mucoepidermoid carcinoma and melanoma, which will be confirmed through biopsy.   The next step for suspected sebaceous carcinoma is a wide excisional biopsy of the primary lesion and map biopsies of the conjunctiva. This is done as one of the hallmark features of sebaceous carcinoma is pagetoid spread – spreading of malignant cells to epithelium that appears to be separate from the main tumour. The approach for treatment depends on the characteristics of the tumour. Tumours with pagetoid spread are treated with cryotherapy and deep tumours require excision. Orbital exenteration is also often required when orbital invasion is present. Referral to an oncologist or internist for systemic workup and surveillance is important with attention to the lymph nodes, lungs, brain, liver, and bone. "
Oculoplastics	Lagophthalmos secondary to Bell’s palsy	"A 65-year-old man presented to the ophthalmology clinic with a three day history of an inability to fully close his left eyelid. He had been using tape to keep his left eye closed at night. There is no associated pain or diplopia but he does complain of blurry vision in the left eye. He also complained of a left-sided facial droop, slurred speech and diminished taste to the left side of his tongue. He claims that his symptoms came on suddenly but he had the flu before symptom onset. He has no other symptoms and no previous medical history. He also does not have a history of recent head trauma.   On ocular examination, there was left eye redness, left brow ptosis and a left-sided facial droop noted. There was also left lower lid ectropion and left upper lid retraction. When blinking, the right eyelid would close normally but the left eye would make an upward and outward movement instead of the eyelid closing normally. Corrected visual acuity was 20/25 OD and 20/40 OS. Confrontational visual fields were full OU. Extraocular movements were full and non-painful. Pupils were 3 mm in the dark and 2 mm in the light OU. Pupil response was normal bilaterally. On anterior segment examination, there was a rapid tear break-up time OS but there was no anterior chamber involvement. Dilated fundus examination was normal OU. Intraocular pressures were 13 mmHg OD and 14 mmHg OS."	"The suspected diagnosis is lagophthalmos secondary to Bell’s palsy. Cranial nerve VII (CN VII) palsy can be due to many possible causes. History and physical examination are the mainstay for diagnosing Bell’s palsy. The sudden onset of symptoms makes neoplastic and infectious processes less likely to be the cause as these present more gradually. Examination findings are also consistent with an isolated left-sided CN VII palsy. Involvement of other cranial nerves on examination would have made a brainstem lesion more likely. The isolated CN VII palsy, viral prodrome, and lack of other symptoms/recent head trauma makes Bell’s palsy the most likely diagnosis.   However, Bell’s palsy is a diagnosis of exclusion. A complete neurologic and otolaryngological examination should be performed before the diagnosis of Bell’s palsy is made. A comprehensive neurological exam is needed to rule out other cranial nerve involvement, and motor/sensory abnormalities which can suggest a disease process such as a stroke, multiple sclerosis or Guillain Barré syndrome. Otolaryngology examination is necessary to rule out conditions such as acute otitis media, a cholesteatoma, Ramsay-Hunt syndrome and a parotid neoplasm.  "
Oculoplastics	Indirect carotid-cavernous fistula (CCF)	"A 53-year-old female presented to the ophthalmology clinic due to double vision and left eye redness with swelling. The patient has had redness for the last few months. Initially her was diagnosed with viral conjunctivitis but the redness has still not resolved. For the last couple of weeks she has had increased swelling and the development of double vision. She has a history of hypertension that is being controlled with diet and exercise. She has no family history of ocular disease. She has no history of ocular trauma or surgery.   On ocular examination, there was conjunctival injection and eyelid oedema of the left eye noted. Her corrected visual acuity was 20/20 OU. Pupils were 3mm OU and there was no RAPD noted OU. Ocular motility was limited in upgaze OS. Confrontational visual fields were full OU. Exophthalmometer measurements were 22.5mm OS and 19mm OD. There was no orbital bruit heard on auscultation. On anterior segment exam, there were tortuous conjunctival vessels noted OS. On posterior segment exam, there were no findings noted OU. Her intraocular pressures were 16 mmHg OS and 18mmHg OD."	"The suspected diagnosis in this patient is an indirect carotid-cavernous fistula (CCF). Direct CCF’s are direct connections between the internal carotid artery/external carotid artery (ICA/ECA) and cavernous sinus and have a more severe and acute presentation. Indirect CCF’s are connections from meningeal branches and have a more insidious and chronic presentation. This patient most likely had an indirect CCF due to the chronic presentation of eye redness and other signs such as venous engorgement of the eyelids, arterialization of conjunctival veins, proptosis and ophthalmoplegia. An orbital bruit is often not present in CCF. Indirect CCF’s are also more commonly spontaneous and found in post-menopausal women.   Due to the chronic nature of an indirect CCF there are fewer differential to consider compared to a direct CCF. Thyroid eye disease (TED) can be unilateral and present similarly to this patient, however unilateral presentation with chronic conjunctival injection would be uncommon. Orbital tumours can also cause a similar presentation but would also not be expected to cause chronic eye redness. Other conditions such as cavernous sinus thrombosis and retrobulbar haemorrhage would present more acutely/subacutely. Superior orbital fissure syndrome and orbital apex syndrome would be expected to produce cranial nerve, pupil and extraocular findings.   The next step should be to do computed tomography angiography/magnetic resonance angiography (CTA/MRA). CTA/MRA are the first-line investigation for a CCF and can show abnormal flow voids, a dilated superior/inferior orbital vein, orbital congestion, and enlargement of the extraocular muscles which would support the diagnosis. An indirect CCF can close spontaneously and compression therapy is a conservative approach and the least invasive treatment that can be used. A multi-disciplinary approach is often necessary under the care of neurosurgery, oculoplastic surgery and ophthalmology."
Lens and cataract	Posterior capsule tear	"A 72-year-old male is referred for gradual bilateral vision loss. He complains of constant blurry vision and difficulty driving at night over the past few months. He has a history of hypertension, which is well controlled with medication. On examination, he has bilateral nuclear sclerosis cataracts graded as 4+. You propose performing cataract surgery, and the patient agrees. There is no history of ocular trauma or prior surgeries. On the day of the surgery, the patient is positioned and draped adequately. After making your corneal wounds and capsulorrhexis, you begin phacoemulsification using a divide-and-conquer technique. After successfully removing three quadrants, you encounter some difficulty with the last quadrant. As you advance your phacoemulsification probe further into the anterior chamber, you notice a sudden deepening of the anterior chamber. You also observe the formation of some linear folds on the posterior capsule."	"The suspected diagnosis is a posterior capsule tear. The sudden deepening of the anterior chamber and the appearance of lines on the posterior capsule are highly suggestive of this complication. A posterior capsule tear is a complication of cataract surgery, whether by phacoemulsification or extracapsular cataract extraction. Risk factors can be divided into two main categories: lens-related and ocular conditions. Lens-related factors include mature cataract, white cataract, posterior polar cataract, history of ocular trauma, previous vitrectomy, and intravitreal injections. Ocular conditions that increase the risk of rupture include high myopia, high hyperopia, pseudoexfoliation syndrome, floppy iris syndrome, and poor pupil dilation. There are five major signs of posterior capsule rupture: sudden deepening or shallowing of the anterior chamber, momentary pupillary dilation, an occlusion sound without visible phacoemulsification tip occlusion, diminished followability of the cataract fragments (i.e., the fragments do not move as expected), and sudden difficulty rotating the nucleus. Other signs include seeing vitreous strands in the anterior chamber or the appearance of linear folds on the capsule. Once the surgeon identifies that a posterior capsule tear has occurred, the best next step is to stop phacoemulsification and inject dispersive viscoelastic under the cataract fragments to support the remaining pieces and stabilise the chamber. The handpiece can only be removed once the chamber is stable to prevent any depressurization, which can propagate the tear and pull more vitreous. If vitreous is visible in the anterior chamber, an anterior vitrectomy is required. Triamcinolone can be used to stain the vitreous and facilitate its removal. Phacoemulsification can be continued using low vacuum and flow to remove the remaining fragment if it is anterior. Cortex removal can be done using the vitrector. The insertion of the intraocular lens will depend on the size of the tear and the zonular support, but sulcus placement is often a good option. Patients often require glaucoma drops post-operatively, as intraocular pressure tends to be elevated following this complication."
Lens and cataract	Toric IOL rotation	"A 70-year-old man presented with decreased vision in both eyes over the past 6 months. You diagnose him with bilateral nuclear sclerotic cataracts. A decision was made to operate the right eye to start. Pre-operatively, ocular biometry measurements were taken including axial length and keratometry measurments. His pre-operative refraction was -1.00 +2.50 x 95 in the right eye, and his keratometry showed 42/44 at 90. You decide to impant a toric IOL to correct 2 diopetrs of astigmatism at 90 degrees. The surgery is uneventful. On post-operative day 1, the vision is 20/100 and the patient is unhappy. The cornea is clear with mild edema at the incisions, there is 1+ cell and flare. You dilate the patient's pupil and notice that the toric IOL is at the 10 degree axis."	"The diagnosis is postoperative toric IOL rotation. This is known complication following cataract surgery with toric IOL implantation. Toric IOLs are designed to correct pre-existing corneal astigmatism by being precisely aligned along the steep axis of the cornea. However, if the IOL rotates from its intended position, it can lead to residual or induced astigmatism, causing suboptimal visual outcomes. In this case, the patient’s symptoms of decreased visual acuity and misaligned IOL point to the diagnosis. Generally, a rotation of more than 30º negates almost all of the astigmatic correction offered by a toric IOL and in this case explains the decreased vision. The patient must be taken to the operating room to reposition the IOL. Occasionally, repositioning the IOL can be performed after implanting a capsular tension ring (CTR) for increased stability, or exchanging the IOL for a platform with better proven rotational stability."
Lens and cataract	Intraoperative floppy iris syndrome (IFIS).	"An 82-year-old male is seen at the clinic for gradual bilateral vision loss. He’s been having difficulty with his daily activities due to constant blurry vision and has also stopped driving at night. He has a history of hypertension, diabetes, and benign prostatic hypertrophy, all of which are well controlled with metoprolol, metformin, and tamsulosin. On examination, he has bilateral nuclear sclerosis cataracts graded as 2+. You propose performing cataract surgery, and the patient agrees. There is no history of ocular trauma or prior surgeries. On the day of the surgery, the patient is positioned and draped adequately. The pupil is dilated to approximately 6 mm, but no synechiae are observed. You make your corneal incisions and attempt to mechanically stretch the iris, but it keeps returning to its original position. You decide to proceed and complete your capsulorrhexis without any issues. However, during phacoemulsification, you notice that the pupil is gradually constricting and the iris stroma appears flaccid. You successfully remove the cataract, but as you withdraw your phacoemulsification handpiece, the iris prolapses through the main wound. You attempt to push it back into the anterior chamber, but it continues to prolapse, and you notice some loss of iris pigment following this manoeuvre."	"The suspected diagnosis is intraoperative floppy iris syndrome (IFIS). The presence of poor pupillary dilation, flaccid stroma, and iris prolapse are highly suggestive of this condition. IFIS is characterised by the following intraoperative signs: flaccid iris stroma, a tendency for iris prolapse through the corneal wounds (main incision and/or paracentesis), and progressive pupil constriction. Additionally, the iris is elastic and does not respond to mechanical stretching. These findings are due to the antagonism of alpha-1 adrenergic receptors, which prevents the dilation of the iris. These receptors are located within the dilator muscle of the iris and can be blocked by alpha-1 antagonist medications, commonly used for the management of benign prostatic hyperplasia, such as tamsulosin (Flomax®). Other medications with a weaker association with IFIS include finasteride, angiotensin II receptor inhibitors, and certain antipsychotics (e.g., quetiapine, risperidone). The differential diagnosis of IFIS consists of traumatic iris atrophy, pseudoexfolication syndrome and chronic use of pilocarpine."
Lens and cataract	Corneal oedema due to endothelial damage from ultrasound energy	"A 72-year-old lady presents to eye casualty with persistent blurred vision 3 weeks after uncomplicated phacoemulsification cataract surgery in her right eye. There was no associated eye pain or redness. She continues on topical steroids for her postoperative regimen. Her operation note documented a brunescent cataract without any other significant preoperative findings.   Visual acuities were 20/200 OD and 20/30 OS. Intraocular pressures were 12mmHg OD and 15mmHg OS. On examination, both eyes were quiet and the anterior chambers appeared to be deep bilaterally. There was moderate diffuse corneal stromal oedema in the right eye with Descemet membrane folds, no hypopyon, and no obvious anterior chamber activity. No obvious retained lens fragments were seen. The view of the fundus was hazy but the retina appeared grossly flat. The fellow eye featured corneal guttata with a beaten bronze endothelial appearance, a moderate nuclear cataract, and an unremarkable fundus examination."	"The suspected diagnosis is corneal oedema due to endothelial damage from ultrasound energy. Risk factors include a higher ultrasonic energy required to remove the brunescent cataract, and a presumed low endothelial cell count (clues to this would be the guttata and beaten bronze appearance of  Fuchs endothelial dystrophy). Postoperative complications in these patients should be minimised with careful preoperative examination and identification of risk factors. Careful operative planning is crucial here, for example protecting the corneal endothelium with dispersive viscoelastic intraoperatively, using a primary chop technique to reduce ultrasonic energy, or consideration of combined phacoemulsification and partial thickness corneal transplants in patients with more advanced Fuchs dystrophy (e.g. with diurnal variation in symptoms and/or preoperative corneal oedema).  Initial management tends to be medical, and may include hypertonic saline drops and/or ointment to promote corneal deturgescence. If there is progression to pseudophakic bullous keratopathy, the ruptured epithelial bullae should be managed with topical antibiotics, lubricants, and a bandage contact lens as appropriate. If the corneal oedema persists and the visual acuity remains poor, early referral to a corneal surgeon for consideration of partial thickness corneal transplants (e.g. descemet membrane stripping keratoplasty, Descemet membrane endothelial keratoplasty) or full thickness transplants (in the event of corneal scarring) should be considered.  Other causes of postoperative corneal oedema may include intraoperative Descemet membrane detachment and/or damage, toxic anterior segment syndrome (which typically presents within the first 24-48 hours postoperatively), or elevated intraocular pressures. Microcystic corneal oedema would be more suggestive of high intraocular pressures - this could result from retained viscoelastic (which would present more acutely, typically within the first day), secondary acute angle closure (for example iris bombe from a cyclitic membrane resulting from severe postoperative inflammation), or steroid response."
Lens and cataract	Pseudophakic cystoid macular edema (Irvine-Gass syndrome)	"A 65-year-old woman presents to the clinic for a post-operative check-up. She underwent cataract surgery on her right eye 6 weeks ago. She has completed all her post-operative drops. While her vision was initially great following the surgery, it now seems blurrier. She denies any pain, redness, or photophobia and hasn't noticed any metamorphopsia. Her surgery went well without any intraoperative complications, and a monofocal lens was placed. On examination, her visual acuity was 20/40 in the right eye and 20/50 in the left eye. Pupils were symmetric and reactive, without a relative afferent pupillary defect. Intraocular pressure was 15 mm Hg in both eyes. Slit-lamp examination showed a normal anterior segment with a calm chamber and a well-positioned lens. The left eye had a 2+ nuclear sclerosis cataract. The optic nerves were normal without edema. The foveal reflex was diminished in the right eye compared to the left eye. The vitreous was calm, and the peripheral retinal examination was normal bilaterally. A macular optical coherence tomography (OCT) revealed intraretinal cystic hyporeflective areas. Amsler grid testing was positive for metamorphopsia in the right eye."	"The suspected diagnosis is pseudophakic cystoid macular edema, also known as Irvine-Gass syndrome. The presence of cystoid macular edema following cataract surgery in a quiet eye is highly suggestive of this diagnosis. Inflammatory mediators in the aqueous humor and vitreous are upregulated following surgery, leading to the disruption of the blood-retinal barrier and increased vascular permeability. This results in the accumulation of fluid in the inner nuclear and outer plexiform layers, forming large cysts or pockets that appear hyporeflective on OCT. The incidence following cataract surgery depends on the technique used, but is estimated to be around 4-11% after phacoemulsification. Risk factors include surgical complications, diabetic macular edema, uveitis, and glaucoma medications. Pseudophakic cystoid macular edema usually presents 4 to 6 weeks after cataract surgery. Patients may complain of blurry vision, metamorphopsia, or scotoma. Slit-lamp examination can reveal retinal thickening, loss of the foveal depression, or loss of the foveal reflex. OCT is useful for confirming the diagnosis and monitoring the progression of the disease. Fluorescein angiography can also be performed to rule out other causes of cystoid macular edema, such as diabetic macular edema, uveitidies or vein occlusions. In this patient, the best next step would be to start topical medication to help reduce the inflammatory cascade and resolve the fluid leakage. Options include nonsteroidal anti-inflammatory drugs (NSAIDs), corticosteroids, carbonic anhydrase inhibitors, or a combination of these. Intravitreal injections or drug-delivery systems of corticosteroids can also be considered. Finally, vitrectomy can be performed in persistent cases."
External eye disease and cornea	Limbal stem cell deficiency (LCSD)	"A 40-year-old woman was referred to the eye clinic with a 12-month history of persistently blurred vision and foreign body sensation in her left eye. She describes being troubled by episodes of severe eye pain and blurred vision that improve temporarily before recurring again. These symptoms started soon after she was treated for a severe chemical injury to her left eye. Past medical and ocular history was otherwise unremarkable.  Visual acuities were 20/25 OD and 20/80 OS. Intraocular pressures were 18mmHg OD and 16mmHg OS. On examination, the right eye was unremarkable. The left featured diffuse punctate epitheliopathy most prominent in the perilimbal region and extending centrally in a whorl-like pattern. The corneal epithelium appeared slightly hazy, and reduced tear break-up time was present as well."	"The suspected diagnosis is limbal stem cell deficiency (LCSD). Limbal stem cells are responsible for regenerating the corneal epithelium, and any loss or damage of the limbal stem cells can impair corneal surface maintenance.  The most likely precipitant of the LCSD in this case would be the previous chemical injury. However, there are many causes of LCSD. LCSD can be congenital (e.g. PAX6 mutation) or acquired secondary to inflammatory causes (e.g. Stevens-Johnson syndrome, graft-versus-host disease), infectious causes (e.g. recurrent herpes simplex keratitis), traumatic causes (e.g. thermal or chemical injuries), or iatrogenic causes (e.g. topical glaucoma medications, radiotherapy, and topical or systemic chemotherapeutic agents), amongst many others.  Impression cytology is needed to confirm the diagnosis. Management often takes a step-wise approach in the earlier stages of LCSD. To start with, medical management comprises intensive preservative-free lubrication, topical corticosteroids (and/or topical cyclosporine drops) to treat any ocular surface inflammation, and/or autologous serum eye drops. Bandage contact lenses may be helpful in managing the recurrent corneal erosions; however care must be taken to reduce the risk of infection. Surgery is indicated if the disease progresses despite adequate medical treatment. Limbal stem cell transplantation is the mainstay of surgical treatment, and may take the form of autologous transplantation from the fellow eye in cases of unilateral LCSD, or allogeneic transplantation in bilateral disease. This requires intensive postoperative monitoring and immunosuppression due to the risk of graft rejection."
External eye disease and cornea	Pterygium	"A 55-year-old woman was referred to the eye clinic with a 3-year history of a progressively enlarging white growth in her right eye. This is becoming more noticeable and the appearance is bothering her. She also reports noticing slightly blurred vision and a foreign body sensation in the last few months. Past medical history includes three previous basal cell carcinomas on her cheek and back region which were successfully treated with surgery. Past ocular history were otherwise unremarkable. She has a long history of sun exposure from sunbathing and frequently visits tanning salons.   Visual acuities were 20/25 OD and 20/80 OS. Intraocular pressures were 18mmHg OD and 16mmHg OS. On examination, the right eye featured a wing-shaped and slightly elevated fibrovascular lesion extending from the nasal conjunctiva towards the pupillary margin. There is a smaller growth in the left which only extends 0.5mm into the cornea. No obvious feeder vessels were observed. Dilated fundus examination was unremarkable."	"The suspected diagnosis is pterygium, with characteristics appearances as described above. This is a clinical diagnosis. Important differentials for a thickened conjunctival growth include ocular surface squamous neoplasia - although the clinical presentation can vary, this typically presents as a gelatinous, papillomatous, or leukoplakic lesion with an irregular surface, feeder vessels, with varying degrees of keratinization and vascularization. Importantly, the bilateral (albeit asymmetric) changes and slow progression are more in keeping with a pterygium.  Conservative management with topical lubricants to alleviate any discomfort is often appropriate, particularly in asymptomatic patients with stable pterygia. However, surgical excision with conjunctival autograft is indicated if the patient experiences reduced vision from pterygium-induced astigmatism, if the pterygium progresses towards the visual axis, or if the patients experiences recurrent inflammation or significant discomfort from a bulky pterygium. Surgery carries a risk of recurrence rate of up to 10%. The excised pterygium tissue can be sent for histopathological testing to exclude any occult neoplasia, although the risk of this is very low."
External eye disease and cornea	Keratoconus	"A 22-year-old man was referred by his optometrist who had noted ‘scissoring reflex’ on retinoscopy at a routine sight test. He had presented due to increasingly blurred vision in his right eye. Past medical history was significant for eczema and hayfever. Past ocular history was unremarkable apart from seasonal ‘itchy and watery eyes’ during hayfever season. ’  Best corrected visual acuity was 20/80 OD and 20/40 OS. Subjective refraction was -1/-4.75 x 70 OD and 1/-3.25 x 155 OS. Intraocular pressures were 15mmHg in both eyes. Anterior segment examination was quiet. The bulbar conjunctiva was white with few papillae present on the tarsal conjunctiva, and mild apical scarring was present in the right cornea. Corneal topography showed steepening in the inferior corneas bilaterally and an asymmetric bow-tie pattern."	"The most likely diagnosis is keratoconus. Risk factors in this patient include the history of atopy and eye rubbing. Other risk factors include a positive family history and conditions such as Down’s syndrome.   Corneal tomography (e.g. Pentacam with Belin/Ambrósio Enhanced Ectasia display) can be helpful in mapping the corneal thickness and elevation to confirm the diagnosis and rule out other ectatic disorders. The patient should be counselled regarding the condition, natural history, and management options. The next step would be to monitor the patient closely with the use of serial imaging - corneal topography or tomography is helpful in identifying patients at risk of progression, in particular young patients under the age of 35.   Management depends on the patient’s vision, progression, and visual needs. Key principles of management comprise 1) managing risk factors such as treating any allergic eye disease and minimising eye rubbing ; 2) optimising vision/ visual rehabilitation, for example with spectacles or rigid gas permeable contact lenses, or corneal transplant in advanced cases; and 3) reducing the risk of progression and stabilising disease, such as with corneal cross-linking. Cross-linking is indicated in patients with progressive disease. The thresholds for progression and minimum corneal thickness may vary according to local treatment protocols."
External eye disease and cornea	Infectious keratitis	"A 71 year old woman presents to the emergency department with left eye pain and redness. Yesterday, she felt well, but after sleeping in her contact lenses, she woke up this morning with worsened redness and pain of the left eye. Additionally, she is reporting some blurred vision. The patient has no known past ocular history other than contact lens wear. She uses no eyedrops regularly.   On exam, vision is 20/20 with correction in the right eye and 20/80 with correction in the left eye. Intraocular pressures are 17 in the right eye and 18 in the left eye. Pupils are equal, round, and reactive to light with no afferent pupillary defect. Extraocular movements are intact, and visual fields are full to confrontation.   On slit lamp exam, the right eye is normal. The left eye displays normal lids and lashes, diffuse 2+ conjunctival injection, and 1+ corneal haze. There is a 2.5 mm by 3 mm stromal infiltrate with overlying epithelial defect as seen on fluorescein staining; the infiltrate is approximately 3 mm from the central visual axis. The anterior chamber is deep with no cell visualized. The iris is round and reactive, and the lens has 1+ nuclear sclerosis."	"The most likely diagnosis is infectious keratitis. The examination suggests a corneal ulcer given the stromal infiltrate and overlying epithelial defect. The most likely etiology is bacterial infection, but fungal and protozoal (such as acanthamoeba) should be considered in this contact-lens wearer. Corneal ulcers are often cultured if larger than 1-2 mm in size (such as in this case), are sight threatening, near the visual axis (often defined as within 3 mm), chronic, or unresponsive to treatment. For treatment of small, peripheral ulcers, a fluoroquinolone +/- polymyxin B/trimethoprim drops q1-2 hours while awake can be used. If vision threatening or unresponsive to initial treatment, fortified antibiotics (such as vancomycin 25 mg/mL and tobramycin or gentamicin 15 mg/mL) are often used, each q1h, alternating.   Additionally, the patient should refrain from contact lens wear while the infection is being treatment. Sometimes systemic antibiotics, such as an oral fluoroquinolone, can be used for patients with scleral extension of the infiltrate. If the patient has difficulty administering the necessary eyedrops at the required frequency, admission to the hospital can be considered. Additionally, these patients should be followed closely pending the results of the cultures and antibiotics should be tailored according to sensitivies."
External eye disease and cornea	Acanthomoeba keratitis	"A 33-year-old woman was referred to the ophthalmology clinic with blurry vision, photophobia and a red, painful right eye. She is a contact lens wearer and was wearing disposable soft contact lenses for the last four months. Three weeks before presentation she developed blurry vision, photophobia and increasingly severe pain in the right eye. The patient was treated with topical antibiotics at another facility however her symptoms worsened. When presenting to the ophthalmology clinic she was on topical antibiotic drops every hour and cyclopentolate 1% twice daily OD. She has no previous medical history. She also has no history of ocular trauma.   On examination, there was significant eye redness noted in the right eye. Corrected visual acuity was counting fingers at 6 inches OD and 20/20 OS. Extraocular movements were full OU. Both pupils were 4 mm in the dark and 2 mm in light. Pupil response was normal OU. Corneal sensation was reduced in the right eye and normal in the left eye. On anterior segment examination, the left eye was normal but there was conjunctival injection, multifocal stromal infiltrates, keratic precipitates and hypopyon noted in the right eye. Small epithelial defects, anterior chamber cells and flare were also noted OD. On dilated fundus examination, there was a hazy posterior view OD and a normal fundus seen OS. Intraocular pressures were 17 mmHg OD and 18 mmHg OS."	"The suspected diagnosis is acanthamoeba keratitis. Acanthamoeba keratitis classically presents in contact lens wearers with severe ocular pain that is out of proportion with early clinical signs. However, early clinical signs such as epithelial irregularities, pseudodendrites/true dendrites and epithelial/stromal infiltrates can be mild and non-specific, which broadens the differential diagnosis. The differential includes herpes simplex keratitis, bacterial keratitis, acanthamoeba keratitis and fungal keratitis. Herpes simplex keratitis is the first differential that should be considered due to its similar early presentation.   This patient was most likely diagnosed with contact lens associated bacterial keratitis. She was treated with antibiotics but her symptoms have worsened. This suggests that her keratitis is most likely due to acanthamoeba or fungal infection. On examination the patient has multifocal stromal infiltrates which can be seen in both infections (typically a stromal ring infiltrate in acanthamoeba infection and satellite lesions in fungal infection). Conjunctival injection, keratic precipitates and hypopyon with an anterior chamber reaction can also be seen in both disease processes. However, contact lens use is a major risk factor for acanthamoeba infection. Risk factors for fungal keratitis include a history of ocular trauma, ocular surface disease and topical steroid use. These risk factors are not present in this patient’s history, making acanthamoeba keratitis the more likely diagnosis.  For confirmation, the patient should have corneal scrapings sent for acanthamoeba evaluation which would most likely show double-walled cyst structures. If available confocal microscopy can also be performed which can show acanthamoeba trophozoites/cysts and inflamed corneal nerves in acanthamoeba keratitis."
External eye disease and cornea	HSV keratitis	"A 33-year-old male presented to the ophthalmology clinic with a two-day history of blurred vision, photophobia and pain in his left eye. His symptoms developed gradually and were not associated with trauma. The patient is known for -3.50 myopia in his left eye and -4.00 myopia in his right eye. He wears soft contact lenses daily. He has a medical history of recurrent genital lesions. He has no personal or family history of ocular disease.   On ocular examination, there was eye redness noted OS. Corrected visual acuity was 20/60 OS and 20/20 OD. Extraocular movements were full and non-painful OU. The left pupil was 2mm and the right pupil was 3mm in diameter. Pupil responses were normal OU. There were no findings on external exam OU. On anterior segment exam, there was conjunctival injection OS. On examination of the cornea with fluorescein staining, two linear and branching epithelial ulcerations with bulbs at the end of each branch were noted OS. Corneal sensation was normal OU. Dilated fundus exam was normal OU. Intraocular pressures were 17mmHg OS and 18mmHg OD."	"The suspected diagnosis is HSV keratitis. HSV keratitis involves the corneal epithelium, stroma and endothelium. HSV epithelial keratitis can cause classic dendritic lesions which are pathognomonic for HSV keratitis. These dendritic lesions are described as “thin, linear and branching epithelial ulcerations with club-shaped terminal bulbs”. In addition, the patients history of recurrent genital lesions, which are commonly caused by HSV, also supports the diagnosis.   The differential diagnosis for HSV epithelial keratitis is broad and can include acanthamoeba keratitis, herpes zoster ophthalmicus, recurrent corneal erosions and corneal abrasions. Acanthamoeba can present similarly to HSV keratitis but is less likely due to the presence of dendritic lesions and history of genital lesions. Herpes zoster ophthalmicus causes pseudodendrite formation and commonly causes associated skin lesions in the distribution of the V1 dermatome. Recurrent corneal erosions and corneal abrasions are also unlikely due to the lack of ocular trauma and personal/family history of a corneal dystrophy or other ocular disease.   The diagnosis of HSV keratitis is clinical and highly suggested in this patient so the next step is to initiate treatment. Systemic or topical antiviral therapy (acyclovir, valacyclovir, famciclovir or ganciclovir) can be used and should be continued for up to 14 days. Importantly, topical steroids are contraindicated but cycloplegic agents and topical antibiotics to prevent bacterial superinfection can also be given. For epithelial defects that do not resolve in 1-2 weeks, bacterial co-infection or acanthamoeba keratitis should be considered and managed accordingly."
External eye disease and cornea	HZO with acute retinal necrosis	"A 77-year-old male was referred by the emergency physician for conjunctival hyperemia and skin lesions on the eyebrow and eyelids. Four days ago, he started experiencing pain and photophobia in his right eye. He also noticed redness and tearing. Yesterday, multiple vesicular skin lesions appeared on the scalp, eyebrow, and eyelids on the right side, respecting the vertical meridian. He feels that his vision is blurrier and has been seeing lots of new floaters. He denies flashes of light or the presence of a black curtain.  On examination, his visual acuity was 20/50 in the right eye and 20/25 in the left eye. Pupils were reactive, symmetric, and there was no relative afferent pupillary defect. Intraocular pressure was 20 mm Hg in both eyes. Eye movements and visual fields were full bilaterally. On the right side, there was 1+ conjunctival hyperemia with follicles and a large central pseudodendrite that stained positively with fluorescein. There were 1+ of cells in the anterior chamber. On dilated fundus examination, there were 2+ of cells in the vitreous. From 3 to 9 o’clock, there were multiple peripheral patches of retinal whitening and opacification with rare hemorrhages. Perivascular sheathing was noted in the arteries in that region. You decide to perform a vitreous tap."	"The suspected diagnosis is herpes zoster ophthalmicus with acute retinal necrosis. The presence of vesicular skin lesions, along with vitritis, arteriolitis, and retinitis, is highly suggestive of this condition.  Herpes zoster ophthalmicus, also known as shingles, is a disease characterised by a painful skin rash in one or multiple dermatomes in the distribution of the trigeminal nerve. It is caused by the varicella-zoster virus (VZV), which has reactivated from its dormant state. The skin rash presents as an acute eruption of multiple vesicles that crust and heal within a few weeks. In addition to the rash, the virus can have multiple ocular manifestations, including conjunctivitis, keratitis, uveitis, retinitis, optic neuritis, and many more.  One of the most severe complications of zoster ophthalmicus is acute retinal necrosis (ARN). Patients typically present with acute vision loss in one eye, which may be associated with redness, photophobia, floaters, and flashes. The majority of cases are unilateral, but about a third will present with bilateral ARN. On clinical examination, ARN classically includes vitreous inflammation, retinal vascular arteriolitis, and peripheral retinitis. The vasculitis primarily involves the arterioles, which can help differentiate ARN from other vasculitides. Progressive outer retinal necrosis (PORN) has a more extensive and severe course and is usually seen in immunocompromised patients, especially those with HIV/AIDS. One key differentiating feature of PORN is the absence of vitreous inflammation.  The diagnosis of ARN is clinical, but it is appropriate to analyse an aqueous or vitreous sample for viral PCR and/or culture. The next step would be to start antiviral therapy. Antiviral therapy can be administered systemically (intravenously or orally) as well as via intravitreal injections. Commonly used agents include acyclovir, valacyclovir, and ganciclovir. Intravitreal injection of antiviral medication is beneficial in cases where the retinitis is threatening the macula or optic disc. Systemic corticosteroids can also be beneficial but should only be started 48 hours after the initiation of antiviral therapy."
External eye disease and cornea	Cicatrizing conjunctivitis	"A 60-year-old woman is referred to the eye clinic with a 6-month history of persistent bilateral eye redness, discomfort, light sensitivity, and slightly blurred vision. Her past medical history and past ocular history was unremarkable, and she denies any history of autoimmune conditions. On systems review, she reported a recent onset of mouth ulcers and blisters. She denies the use of any topical eyedrops or antecedent trauma or infections.   Visual acuities were 20/40 OD and 20/50 OS. Intraocular pressures were 15mmHg OD and 18mmHg OS. On examination, her eyelids were erythematous and oedematous with signs of blepharitis and a mild entropion of the left lower lid. The conjunctiva in both eyes was diffusely injected, with areas of subconjunctival and caruncular fibrosis, superior and inferior symblephara, and shortening of the inferior fornix bilaterally. Fluorescein staining revealed diffuse punctate epitheliopathy and reduced tear break-up time in both eyes. Dilated fundus examination featured a hazy view of the fundus due to the ocular surface disease, but was grossly unremarkable."	"The suspected diagnosis is cicatrizing conjunctivitis. Possible causes of this include oculocutaneous conditions (most commonly ocular mucous membrane pemphigoid, but also rosacea and atopic keratoconjunctivitis), iatrogenic causes (e.g. Stevens-Johnson syndrome, toxic epidermal necrolysis, topical anti-glaucoma medications, radiotherapy), infections (e.g. chlamydia trachomatis, epidemic keratoconjunctivitis from adenovirus), systemic autoimmune conditions (e.g. scleroderma, systemic lupus erythematosus), or traumatic causes (e.g. thermal or chemical injuries), amongst many others.  The cicatrising conjunctivitis in this patient is most likely to be related to ocular pemphigoid, given the mouth ulcers and absence of any precipitating agents or events. This is a clinical diagnosis supplemented by immunohistopathologic evidence from direct immunofluorescence testing on conjunctival biopsy; it is important to note that while a positive test clinches the diagnosis, a negative test does not necessarily rule this out. As sampling severely inflamed conjunctiva can produce extensive scarring while also reducing the diagnostic yield, it is generally advised that the biopsy should be obtained from tissue adjacent to an inflamed site where possible. The sample should be transported to the lab in Michel medium for analysis.  Systemic immunosuppression is the mainstay of treatment. This may include systemic corticosteroids to treat acute exacerbations, and maintenance therapy with steroid-sparing agents such as oral dapsone for milder disease, or azathioprine, mycophenolate mofetil, or methotrexate for more severe cases. The ocular surface should be treated with intensive topical lubricants, and topical steroids and/or cyclosporine may be considered as well. Surgical intervention may be required to address complications such as persistent epithelial defects, corneal scarring, entropion, or trichiasis."
External eye disease and cornea	Peripheral ulcerative keratitis (PUK)	"An 86-year-old woman was referred to the ophthalmology clinic following the development of corneal ulcers in both eyes. She has had pain, tearing and blurry vision bilaterally for the last 4 weeks. The patient has a history of “some kind of arthritis” for which she receives prednisolone 40 mg once daily from her primary care physician for treatment. Her past medical history also includes hypertension and Alzheimer’s dementia. She had cataract surgery on both eyes five years ago.   On ocular examination, there were crescent-shaped corneal ulcers noted bilaterally in the periphery of the patients corneas. Her corrected visual acuity was 20/400 OS and 20/200 OD. Extraocular movements were full and non-painful OU. Confrontational visual fields were full OU. Pupils were 2mm OU and had normal pupil responses. On anterior segment exam, crescent-shaped corneal ulcers were noted in the superior corneas and were associated with epithelial defects and stromal thinning OU. Dilated fundus exam was normal OU. Intraocular pressures were 14mmHg OU."	"The suspected diagnosis is peripheral ulcerative keratitis (PUK). PUK is a group of inflammatory diseases that share the final common pathway of peripheral corneal thinning. Although more commonly unilateral, PUK can be bilateral. The patient's history of “some kind of arthritis” that is being treated with steroids raises the possibility of systemic inflammatory arthritis such as rheumatoid arthritis, a common association with PUK. Other systemic inflammatory conditions such as Granulomatosis with polyangiitis, Polyarteritis nodosa and Systemic Lupus Erythematosus are also associated.   Mooren ulcers can also present as bilateral crescent-shaped corneal ulcers but is a diagnosis of exclusion, and given the patient’s history of arthritis, PUK is the suspected diagnosis. Other conditions that can cause bilateral and peripheral corneal thinning such as Terrien marginal degeneration, Pellucid marginal degeneration and Furrow degeneration are usually painless and have an intact epithelium.   The most appropriate next step in this patient is evaluation for an underlying systemic autoimmune condition by ordering lab tests such as a complete blood count (CBC), erythrocyte sedimentation rate (ESR), rheumatoid factor (RF), antinuclear antibodies (ANA), anti-neutrophil cytoplasmic antibodies (ANCA), anti-Ro and anti-La antibodies, and angiotensin-converting enzyme (ACE). The patient should also be referred to rheumatology for further evaluation and long-term care. As a large number of patients with PUK also present with scleritis, the patient can also have a B-scan ultrasound performed. The patient’s primary care physician can also be contacted for information but blood tests should still be done as the patient may have an undiagnosed autoimmune condition that provoked the development of PUK."
External eye disease and cornea	Primary acquired melanosis (PAM)	"A 45-year-old Caucasian female presented to the ophthalmology clinic after noticing a pigmented lesion in her right eye last week. The patient is not experiencing any pain or visual changes but is worried about having a malignancy as her close friend, who she spent many summers holidays with, was recently diagnosed with basal cell carcinoma. She has no previous medical history. Her mother was diagnosed with breast cancer at 70 that is currently in remission and her father has hypertension, diabetes and dementia.   On ocular examination, there was an area of conjunctival pigmentation noted OD. Uncorrected visual acuities were 20/20 OU. Extraocular movements were full and non-painful OU. Confrontational visual fields were also full OU. Pupils were 2mm OU and had normal pupil responses. On anterior segment exam, a flat, light-tan patchy area of pigmentation was noted in the conjunctiva of the right eye. The lesion measured between 1-2 clock hours, was asymmetrical and poorly circumscribed with no associated vasculature. There were no corneal abnormalities noted OU. Dilated fundus exam was normal OU. Intraocular pressures were 18 mmHg OD and 16 mmHg OS."	"The suspected diagnosis in this patient is primary acquired melanosis (PAM). PAM is an acquired pigmentation in the conjunctival epithelium. It is typically a benign and asymptomatic condition but a substantial minority of patients can progress to a melanoma. PAM is suspected due to the classical exam findings of a flat, patchy and tan coloured area of pigmentation. PAM is also most common in middle-aged and elderly Caucasians.    There are many differential diagnoses for conjunctival pigmentation. The most important differential to exclude is a conjunctival melanoma. However, conjunctival melanomas are more commonly elevated lesions that are well-vascularised so is less likely in this patient. Naevi are a common benign differential diagnosis and characteristically present with small cysts in the lesion and are well-circumscribed, which is not seen with this patient’s lesion.   This patient’s PAM measures between 1-2 clock hours, so the next step in the management of this patient is to discharge the patient and inform her of the careful and regular observation of the lesion that is advised for this condition. This is because the management of PAM is defined by the size of the lesion. A lesion larger than 2 clock-hours is an indication for biopsy/removal of the lesion. If there are any developments in the lesion on follow-up then further management may be indicated."
External eye disease and cornea	Dry eye syndrome	"A 73-year-old man presented to the ophthalmology clinic complaining of foreign body sensation and decreased vision in both eyes. He has had these symptoms for the last 4 weeks. He is known for -4.50 myopia in both eyes and wears glasses. He was diagnosed with Parkinson’s disease one year ago and is being treated with carvidopa-levodopa. He also has hypertension and diabetes mellitus that is treated with daily medication. He had uncomplicated cataract surgery on both eyes 4 years ago. He has no other history of ocular disease or surgery.  On examination, there was reduced facial expression noted. Corrected visual acuities were 20/40 OU. Extraocular movements were full and non-painful OU. Confrontational visual fields were full OU. Pupils were 3mm OU and had normal pupil responses. On anterior segment exam with fluorescein staining there were a few, scattered punctate epithelial erosions and a decreased tear break-up time noted OU. Schirmer test without anaesthesia showed 10mm of moisture on the filter paper. Dilated fundus exam was normal OU. Intraocular pressures were 15 mmHg OU."	"The suspected diagnosis is dry eye syndrome secondary to Parkinson’s disease. Parkinson’s disease can cause infrequent blinking in patients which can result in the development of dry eye syndrome. Dry eye syndrome classically presents with burning, foreign body sensation and decreased vision. It can also cause ocular surface defects such as punctate epithelial erosions and a significant decrease in the tear break-up time. Schirmer test without anaesthesia is considered normal with moisture between 10-15mm on the filter paper but can be insensitive for mild dry eye syndrome.  Given the patient's symptoms, history of Parkinson’s disease and decreased tear break-up time, dry eye syndrome is highly suspected. The next step should be to begin treatment with preservative free artificial tears q.i.d. If the patient's symptoms do not improve adequately then the frequency can be increased to q1-2h or the artificial tears can be supplemented with lubricating gel/ointment."
External eye disease and cornea	Chemical eye injury	"A 46-year-old female presents to the emergency department of the eye hospital following exposure to liquid laundry detergent. An hour ago she was preparing to do the laundry, however when reaching for the laundry detergent on the upper shelf, it tipped over and splashed directly into her face and both eyes. She instantly experienced tearing, decreased vision and intense pain bilaterally. She rinsed her eyes with tap water for 2-3 minutes afterwards but still experienced symptoms. She does not wear glasses and was not wearing any eye protection. She was diagnosed with multiple sclerosis eight years ago and receives regular care from neurology and neuro-ophthalmology. On ocular examination, there was erythema and mild swelling of the periorbital skin bilaterally. Her vision was 20/40 in the right eye and 20/80 in the left. Pupils were equal, round, and reactive to light with no afferent pupillary defect noted. On anterior segment examination, marked conjunctival injection and chemosis were observed in both eyes. Mild subconjunctival hemorrhage was noted in the right eye. There was diffuse punctate keratopathy in both eyes but no limbal ischemia. The anterior chambers were deep and quiet and the rest of the ocular examinatio nwas normal."	"The suspected diagnosis is a chemical eye injury. Chemical eye injury cases that are not associated with occupational exposure, commonly occur at home and during house chores. The liquid laundry detergent that this patient’s eyes were exposed to is most likely an alkali solution, which are lipophilic and penetrate tissues more rapidly than acidic solutions.   Before even performing a full ophthalmological exam, the patient’s ocular pH should be checked and if abnormal, emergency treatment should be given. Measuring the ocular pH would also support the suspected diagnosis. Emergency treatment includes copious irrigation with normal saline or ringer lactate solution. An eyelid speculum and topical anaesthesia may be placed before irrigation. Irrigation is continued until neutral pH is achieved. The pH should be measured at least five minutes after irrigation is stopped to ensure that the pH does not rise or fall secondary to retained particulate matter, which should be flushed or manually removed if present. Conjunctival fornices should be swept with a moistened cotton-tipped applicator to remove any sequestered particles of caustic material and necrotic conjunctiva, especially in the case of abnormal pH. Following emergency treatment, this patient should receive further treatment including frequent preservative-free artificial tears and cycloplegia. Given that the laundry detergent is alkali, topical steroids drops are indicated with adequate antibiotic coverage. "
External eye disease and cornea	Corneal endothelial rejection	"A 42 year-old man presents with a 3 day history of reduced vision in his left eye associated with eye redness and pain. These symptoms were preceded by a 1-week history of foreign body sensation. Past ocular history is significant for corneal perforation secondary to herpetic eye disease, for which he had a penetrating keratoplasty in his left eye 10 months before. The postoperative course had been uneventful till date, and he remains on once daily topical steroids in this eye as well as prophylactic oral Aciclovir treatment. Past medical history was unremarkable.  On examination, his best corrected visual acuity was 20/20 OD (with rigid gas-permeable contact lens) and 20/200 OS. Intraocular pressures were 15mmHg in both eyes. The right eye was quiet. The left eye was injected, the cornea was diffusely oedematous, and there were 3 clock hours of stromal neovascularisation extending to the host cornea temporally. Fluorescein staining revealed a loose suture temporally and an intact corneal epithelium. There were cells in the anterior chamber and keratic precipitates on the donor endothelium."	"The most likely diagnosis is corneal endothelial rejection. This is an immunological response of the host cornea to the donor corneal tissue. Risk factors include his young age, stromal neovascularisation of the host cornea (>2 clock hours of involvement), history of herpetic eye disease, and the loose corneal suture. The risk is highest in the first 18 months postoperatively.  The next step would be to remove the loose corneal suture and start intensive topical steroid therapy. Some clinicians would consider systemic steroid treatment (oral or intravenous) in cases of severe rejection. Close monitoring is warranted for signs of recurrent herpetic disease and complications of topical treatment such as steroid response/ glaucoma. Repeat corneal transplant may be necessary if the rejection episode does not resolve and leads to graft failure.   Other suture-related complications may include microbial keratitis, which was not present in this instance. This would typically be associated with an infiltrate and should prompt microbiological testing of the suture as well as topical antibiotic treatment."
External eye disease and cornea	Fuch’s endothelial dystrophy (FED)	"A 41-year-old female presented to the ophthalmology clinic with a four-day history of painless intermittent blurry vision and foreign body sensation in both eyes. She noticed that her vision is worse upon awakening and improves as the day progresses. She is known for -2.50 myopia OU and wears glasses, with occasional use of contact lenses at social gathering. She has no previous medical history. She also has no previous history of ocular trauma or surgery.   On ocular examination, there were no findings noted on external exam in both eyes. Corrected visual acuities were 20/25 OU at distance and near. Extraocular movements were full and non-painful OU. Confrontational visual fields were full OU. Pupils were 3mm OU and had normal pupil responses. On anterior segment exam, a “beaten metal” endothelial appearance was noted with stromal oedema OU. There were no iris abnormalities noted OU. Dilated fundus exam was normal OU. Intraocular pressures were 16 mmHg OU."	"The suspected diagnosis is Fuch’s endothelial dystrophy (FED). FED is a non-inflammatory dystrophy involving the corneal endothelium. This patient has the typical presentation of bilateral blurry vision that is worse upon awakening. In addition, corneal oedema and corneal guttae are highly associated, corneal guttae can be seen as a “beaten metal” appearance.   The differential for FED includes many similar conditions so it can be challenging to make the diagnosis. The two main differential diagnoses in this case are iridocorneal endothelial (ICE) syndrome and Posterior Polymorphous Corneal Dystrophy (PPCD). Although ICE syndrome is classically associated with a “beaten metal” endothelial appearance,  ICE syndrome is typically unilateral, causes iris abnormalities and is associated with elevated intraocular pressures. PPCD typically also causes elevated intraocular pressures, in addition it also causes several corneal changes such as corneal thickening, opacities, grouped vesicles and band lesions, which are not seen in this patient.   As the diagnosis of FED is clinical, next the patient should have FED progression monitored with pachymetry or specular microscopy. Central corneal thickness (CCT) measurement with pachymetry at diagnosis is important as FED progression results in an increase in CCT over time. The rate at which corneal thickness increases will be important for counselling the patient. Similarly, specular microscopy can be done to measure the endothelial cell count and determine the severity of disease and the rate of progression in follow-up appointments. In addition to pachymetry and specular microscopy, corneal tomography can also be considered to monitor disease progression."
Uveitis and ocular inflammation	Multiple evanescent white dot syndrome (MEWDS)	"A 30-year old woman was referred to the eye emergency clinic with a 3-day history of flashing lights. On further questioning, she reports blurry central vision and shimmering lights in her left eye. She is otherwise fit and well and denies any past medical history or antecedent trauma. Past ocular history was only significant for moderate myopia (-4DS right and -4.25DS left).   On examination, her best corrected visual acuity was 20/25 OD and 20/30 OS. Pupils were equal and reactive to light. No relative afferent pupillary defect was detected. Both anterior segments were quiet with no anterior chamber activity present; however occasional cells were noted in the vitreous. No posterior vitreous detachment was evident and the retinas were flat bilaterally. Numerous faint white dots were visible in the mid-peripheral retina, centred on the optic disc."	"The suspected diagnosis is multiple evanescent white dot syndrome (MEWDS). Differentials for white dots in the fundus include other white dot syndromes. This is an umbrella term for a heterogenous group of inflammatory chorioretinopathies of unknown aetiology, such as birdshot chorioretinopathy, punctate inner choroiditis, or multifocal choroiditis. However, these tend to be bilateral and phenotypically distinct (albeit with some overlap in their signs and symptoms). As above, multimodal retinal imaging will help in distinguishing these conditions.  Retinal imaging should be performed next to confirm the diagnosis. Multimodal imaging is particularly helpful in clinching the diagnosis in patients with myopic fundi, in whom the white dots may not always be easily visualised, and whose symptoms may often be erroneously attributed to posterior vitreous detachment. Fundus autofluorescence and indocyanine green angiography are the most helpful ancillary tests to consider. Fundus autofluorescence reveals hyperautofluorescent dots corresponding to the white dots. Late-phase indocyanine green angiography demonstrates corresponding hypocyanescent dots that can often outnumber the lesions seen fundoscopically. Optical coherence tomography of the macula typically shows discontinuities in the inner segment/ outer segment junction in the acute phase.   As MEWDS is a self-limiting condition, the patient can be monitored closely until her symptoms and retinal changes resolve. This typically occurs by 3 months after symptom onset."
Uveitis and ocular inflammation	Vogt-Koyanagi-Harada (VKH) disease	"A 38-year-old woman presented to the ophthalmology clinic with a progressive decrease in vision, photophobia and painful eyes. She believed that she had a viral illness because six days before presentation she developed a fever, headache, fatigue and ear ringing that she was treating at home with paracetamol. The patient has no previous medical history. She is known for -2.50 myopia and wears glasses. She has no history of ocular trauma or surgery.   On examination, there was bilateral eye redness. Her corrected visual acuity was 20/400 OD and 20/800 OS. Extraocular movements were full. Her pupils were 4mm bilaterally and the pupil response was normal OU. On anterior segment examination, anterior chamber flare and keratic precipitates were noted OU. On dilated fundus examination, optic disc oedema and hyperaemia along with subretinal fluid and multiple serous retinal detachments were noted in both eyes. Intraocular pressures were 14 mmHg OD and 16 mmHg OS."	"The suspected diagnosis is the acute uveitic stage of Vogt-Koyanagi-Harada (VKH) disease. The diagnostic criteria for VKH disease includes the following: (1) no history of ocular trauma, (2) no evidence of other disease processes, (3) bilateral anterior or panuveitis, (4) neurologic/auditory findings and (5) dermatologic findings.   The patient has an acute presentation of bilateral panuveitis. The differential for bilateral panuveitis includes inflammatory processes and infectious diseases. The most important autoimmune differential is sympathetic ophthalmia. This condition is histopathologically identical to VKH disease but is usually preceded by ocular trauma or surgery. The viral illness-like prodromal stage and the neurological/auditory symptom of tinnitus are suggestive of VKH disease.   A diagnosis of incomplete VKH disease (4/5 criteria) would require an exclusion of other disease processes that can cause bilateral panuveitis. Initial blood tests and investigations include a Full Blood Count (FBC), Syphilis testing (VDRL/RPR or FTA-ABS), Tuberculosis testing (PPD/IGRA), and Sarcoidosis testing (ACE levels). Once other disease processes have been excluded in this patient, she is considered to have incomplete VKH disease and should receive high dose systemic corticosteroids to treat the acute stage of the disease."
Uveitis and ocular inflammation	Ocular syphilis	"A 60 year old man presents with slowly decreased vision in the right eye over the past 2 weeks. He has no pain in the eye and denies flashes and floaters. There are no symptoms in the left eye. He reports a new recent female sexual partner.   On exam, visual acuity is 20/60 in the right eye and 20/20 in the left eye. Intraocular pressure is 16 and 15 in the right and left eye, respectively. Pupils are equal, round, and reactive to light without afferent pupillary defect. Extraocular movements are intact, and visual fields are full to confrontation bilaterally. Slit lamp examination of the lids, lashes, cornea, anterior chamber, iris, and lens of both eyes is normal. There is 2+ vitreous cell in the right eye and no cell or infiltrates in the vitreous of the left eye.   On dilated fundus exam of the right eye, there is no optic disc edema or pallor. There is macular edema. The vessels are normal. There are yellow confluent placoid lesions in the inferotemporal periphery. Dilated fundus exam of the left eye is normal without optic disc edema/pallor, vascular attenuation, or macular edema or lesions. There are no lesions in the periphery of the left eye."	"The differential diagnosis is very broad and dependent on presentation (anterior uveitis vs posterior uveitis vs panuveitis, age of patient, and risk factors). The differential diagnosis may include sarcoidosis, tuberculosis, vascular disease, ocular herpes, and lymphoma. In this patient’s case, the placoid lesions in the retina (posterior uveitis), evidence of vitritis, and new sexual partner suggest that syphilis should remain high on the differential diagnosis for this patient. Further testing should be performed.   Ocular syphilis is known to present in many varied ways, most commonly as posterior uveitis or panuveitis. In the retina, syphilis often presents with placoid confluent lesions, and there can be necrotizing or non-necrotizing lesions. Retinal edema and vasculitis can also be present. Syphilis can also present with an optic neuritis or perineuritis, involving optic disc pallor and edema with relative afferent pupillary defect and impaired color vision. Classically, syphilis may present with an Argyll Robertson pupil, or a miotic pupil that constricts in response to near vision but not to light (light-near dissociation); however, syphilis may also present with tonic pupils or Horner syndrome.   OCT imaging in syphilis often include outer retinal abnormalities such as thickening/irregularities of the retinal pigment epithelium or ellipsoid zone disruption or loss. Inner retinal findings may also be present, such as hyperreflective lesions. Intraretinal and/or subretinal fluid may be present. OCT is often helpful in leading toward a diagnosis of syphilis but has not yet been found to have prognostic value.   Diagnosis of syphilis is confirmed via treponemal and non-treponemal testing, and treatment with IV penicillin is effective."
Uveitis and ocular inflammation	Neuroretinitis	"A 47-year-old woman presents to the ophthalmology clinic with gradual loss of vision in her left eye over the past 2 weeks. She also feels that her vision is distorted as straight lines are now wavy with the left eye. There is no pain, redness or photophobia. She denies any headache or neurological symptoms. Her past medical history is only positive for mild depression treated with a selective serotonin reuptake inhibitor. She has not traveled recently and hasn’t been in contact with sick individuals. She has one adult cat and recently adopted a kitten.    On examination, visual acuity was 20/20 in the right eye and 20/100 in the left eye. Pupils were symmetric and reactive with a relative afferent pupillary defect in the left eye. Extraocular movements were full. Metamorphopsia paracentrally was objectified using the Amsler grid. Anterior chamber and vitreous were deep and calm in both eyes. Fundus examination revealed a diffuse optic nerve head edema with exudates distributed around the macula in a star pattern. Optical coherence tomography showed subretinal fluid in the macular region of the left eye with exudates within the outer plexiform layer."	"The suspected diagnosis is neuroretinitis. The patient presented with unilateral optic nerve head edema accompanied by exudates in the outer plexiform layer, which is highly suggestive of this pathology.  Neuroretinitis is an inflammation of the optic disc vasculature with exudation of fluid into the peripapillary retina. The exudates are usually found in the outer plexiform layer and are distributed in a star pattern around the fovea, creating the “macular star.” The macular star usually appears 1 to 2 weeks following the onset of symptoms. It is divided into two main categories: infectious and idiopathic. The most common agent of infectious neuroretinitis is cat-scratch disease, caused by Bartonella species. Other infectious causes include syphilis, Lyme, and tuberculosis. These are also part of the differential diagnosis of neuroretinitis, which also includes hypertensive retinopathy, diabetic retinopathy, anterior ischemic optic neuropathy, and papilledema.  It is important to perform a complete ophthalmological exam, including blood pressure measurement, to rule out hypertensive retinopathy. Fluorescein angiography can be performed to assess for disc leakage or vascular pathology. In this case, the best next step would be ordering blood work-up and starting antibiotics.   Laboratory testing is guided by clinical suspicion but should always include Bartonella titers. Other useful tests include interferon-gamma assay, Lyme serology, angiotensin-converting enzyme, and chest radiography. Neuroimaging is usually not required as the diagnosis is clinical, and imaging tends to be normal. In contrast to optic neuritis, neuroretinitis is not considered a risk factor for the future development of multiple sclerosis. Neuroretinitis is typically self-limited, and the prognosis is generally good. Treatment is focused on the underlying cause, with antibiotics being the mainstay for infectious etiologies, such as those caused by Bartonella. Antibiotics commonly prescribed include azithromycin, ciprofloxacin, or doxycycline in combination with rifampin. In some cases, corticosteroids may be used to reduce inflammation."
Uveitis and ocular inflammation	Progressive outer retinal necrosis (PORN)	"A 35-year-old male presents to the clinic with bilateral progressive vision loss. In addition to blurry central vision for the past 2 weeks, he also reports difficulty seeing on the sides, as if his visual fields were constricted. There is no pain, redness, or photophobia. He hasn’t noticed any floaters but mentions seeing occasional flashes of light. He saw a primary care physician last week who ordered some blood tests, but he hasn’t had a follow-up yet. He mentions having high-risk sexual behaviors, including unprotected intercourse and multiple partners. He had an episode of zoster ophthalmicus a few years ago that was treated with oral antiviral medication.  On examination, his visual acuity was 20/40 in both eyes. Intraocular pressure was 13 mm Hg bilaterally. Pupils were reactive and symmetric without a relative afferent pupillary defect. Visual fields by confrontation confirmed constriction in both eyes. The anterior segment exam was normal in both eyes with calm and deep chambers. The optic nerve and macula were normal bilaterally. Posterior pole and peripheral retina examination revealed multiple patches of yellow opacification with associated vascular sheathing. No cells were present in the vitreous. The blood workup ordered by the primary care physician showed a decreased CD-4 count."	"The suspected diagnosis is progressive outer retinal necrosis (PORN). The presence of bilateral retinal necrosis without vitreous inflammation in a patient with a low CD-4 count is highly suggestive of this condition.  Progressive outer retinal necrosis (PORN) is a syndrome characterized by necrosis of all retinal layers with associated vasculitis, primarily seen in immunocompromised individuals, especially patients with HIV/AIDS. The majority of cases are attributed to infection by varicella-zoster virus, but herpes simplex virus and cytomegalovirus can also be causative agents. These retinal lesions can appear in both the posterior pole and the peripheral retina. A hallmark feature of this diagnosis is the absence of intraocular inflammation. Retinal detachments can occur in approximately half of the patients, either serous due to fluid leakage from the necrosis or rhegmatogenous due to holes secondary to full-thickness necrosis.  Diagnosis is primarily clinical, but testing modalities such as fluorescein angiography and optical coherence tomography (OCT) can be useful. A similar condition, acute retinal necrosis (ARN), has a similar pathophysiology but is seen in immunocompetent patients. ARN tends to be unilateral and presents with significant vitritis, which is the key differentiating feature. The patches of retinitis usually develop in the periphery, with the posterior pole being spared until late in the disease course. The differential diagnosis of ARN/PORN is broad and includes other infectious etiologies (e.g., syphilis, tuberculosis), inflammatory conditions (e.g., sarcoidosis, Behcet’s disease), and neoplasms.  The best next step for this patient would be to obtain a vitreous tap with viral PCR (VZV, HSV, CMV) to identify the pathogen. This patient should also be referred to an infectious disease specialist to confirm the diagnosis of HIV/AIDS and initiate appropriate therapy. Treatment of PORN typically involves a combination of systemic and intravitreal antiviral medications."
Uveitis and ocular inflammation	Toxoplasmosis chorioretinitis	"A 38-year-old woman is referred by her optometrist for unilateral uveitis. She complains of redness, photophobia, and pain in her left eye for the past 3 days. She also noticed a significant increase in floaters and feels that her vision is blurry. She believes she had a similar episode a few years ago but didn’t seek treatment. She immigrated from Brazil more than 15 years ago. Her past ocular and medical history is non-contributory.  On examination, her corrected visual acuity was 20/20 in the right eye and 20/50 in the left eye. Pupils were symmetric and reactive to light, without a relative afferent pupillary defect. Intraocular pressure was 18 mm Hg in the right eye and 12 mm Hg in the left eye. Slit-lamp examination revealed 2+ cells and flare in the anterior chamber of the left eye with a few large yellowish keratic precipitates. No nodules were present on the iris. Dilated fundus examination showed 3+ cells in the vitreous. The optic nerve and macula were normal in both eyes. In the left eye, there was a pigmented chorioretinal scar at the 2 o’clock position peripherally. Adjacent to the scar, there was a focal area of retinitis and vasculitis."	"The suspected diagnosis is toxoplasmosis chorioretinitis. The presence of granulomatous uveitis and vitritis with retinitis adjacent to a chorioretinal scar is highly suggestive of this diagnosis.  Ocular toxoplasmosis is caused by the intracellular parasite Toxoplasma gondii. It is the leading cause of posterior uveitis worldwide, with a particularly high prevalence in Brazil and Africa. It classically presents with white focal retinitis adjacent to a pigmented chorioretinal scar. Vitreous inflammation overlaying the retinitis is often seen, giving the appearance of a “headlight in the fog.”   Granulomatous keratic precipitates and uveitis can also be seen. Rarer ocular manifestations include papillitis, neuroretinitis, and scleritis, among others. Given the high prevalence of positive toxoplasma titers in many populations, serology is mainly used to exclude the pathology. If titers are positive, it indicates that toxoplasmosis should remain in the differential diagnosis. The differential diagnosis of vitritis with chorioretinitis includes tuberculosis, viral retinitis, syphilis, and some autoimmune conditions such as Behcet’s disease or sarcoidosis.  The next step would be to perform a granulomatous uveitis workup to rule out other causes of retinitis, as listed above. Topical steroids and cycloplegics would be used to treat the anterior inflammation. Small peripheral retinal lesions are often observed, as they typically heal spontaneously. Indications for systemic treatment include optic nerve or macula-threatening lesions, dense vitritis, significant visual impairment, pregnancy, and immunodeficiency. If a decision is made to treat systemically, the classic triple therapy consists of pyrimethamine, sulfadiazine, and corticosteroids. "
Uveitis and ocular inflammation	CMV retinitis	"A 48-year-old male presented to the ophthalmology clinic with decreased vision and floaters in his right eye. These symptoms have been progressively worsening and he began experiencing them many weeks ago. The patient has Acquired Immunodeficiency Syndrome (AIDS) with Human Immunodeficiency Virus (HIV) transmitted through intercourse with a previous partner. His last known CD4 count was 29 cells/mm3. He is inconsistent with his anti-retroviral treatment regimen. He has no personal or family history of ocular disease. He also has no history of ocular trauma or surgery.   On examination, there were no abnormalities noted on external exam. His corrected visual acuity was 20/200 OD and 20/20 OS. Extraocular movements were full and non-painful OU. Confrontational visual fields were full OU. Pupils were 3mm OU and there was a RAPD noted in the right eye. On anterior segment examination, there were keratic precipitates and aqueous cells with flare noted OD. On posterior segment examination, some vitreous cells and large yellow-white retinal patches with haemorrhages were noted in the peripheral retina OD. His intraocular pressures were 15 mmHg OU."	"The suspected diagnosis is CMV retinitis. CMV retinitis is a common ocular opportunistic infection in patients with AIDS. Fulminant CMV retinitis classically presents with haemorrhagic necrosis on white/yellow cloudy retinal lesions. Signs of inflammation in the anterior and posterior segment can also commonly be seen in CMV retinitis but are typically minimal due to the chronic immunosuppressed states in these patients.   The two main differential diagnoses in this patient are HIV retinopathy and Acute Retinal Necrosis (ARN). HIV retinopathy is characterised by retinal haemorrhages, cotton wool spots and microaneurysms, from which only the haemorrhages are seen in this patient. ARN is a pan-uveitis that presents with significant vitreous inflammation and pain so is less likely in this case. Retinal haemorrhages can be found but are also uncommon in ARN.   Next, the patient should be treated with antiviral therapy. First line therapy for CMV retinitis is oral valganciclovir. The induction dose is 900 mg twice daily for 21 days followed by a maintenance of 900 mg daily. Alternatively, intravenous ganciclovir or foscarnet can also be given. Patients that have progression of retinitis despite induction can be given adjuvant intravitreal antiviral injections. The patient should also be referred to his infectious disease specialist for evaluation and optimization of his anti-retroviral treatment regimen."
Uveitis and ocular inflammation	Tubulointerstitial Nephritis and Uveitis (TINU) syndrome	"A 22-year-old female patient presented to the ophthalmology department in the hospital with a two day history of painless eye redness and blurry vision OU. Two weeks before presentation she was discharged from hospital. She was admitted for non-specific symptoms that included a fever, unexplained weight loss, a reduced appetite and abdominal pain. She underwent intensive investigation but no diagnosis was made. Testing for sarcoidosis was negative. ANA and ANCA testing was also negative.  Her laboratory studies showed a normal full blood count. Her urinalysis was positive for trace protein, leukocyte esterase, 9 white blood cells and 15 red blood cells per high-power field. The urine protein-to-creatinine ratio was normal. Beta-2 microglobulin was 3521 mcg/L in urine and 1242 mcg/L in serum.  On ocular examination, the patient had bilateral eye redness. Corrected visual acuity was 20/25 OD and 20/30 OS. Extraocular movements were full. The left pupil had an irregular shape and the right pupil was normal. The pupil response was normal OU. On anterior segment examination, there were keratic precipitates and anterior chamber cells and flare noted OU. On dilated fundus examination, there were anterior vitreous cells noted OU. Intraocular pressures were 13 mmHg OU."	"The suspected diagnosis is Tubulointerstitial Nephritis and Uveitis (TINU) syndrome. She presents with acute bilateral uveitis on examination. Acute bilateral uveitis has a broad differential diagnosis which includes several infectious and autoimmune diseases. The differentiating factor is that the patient was recently admitted to hospital with non-specific constitutional symptoms and was undiagnosed. Sarcoidosis, ANA and ANCA testing was negative which rules out several important autoimmune causes of the patient’s preceding symptoms and acute bilateral uveitis.   TINU syndrome is a rare condition that commonly presents with vague and non-specific constitutional symptoms before acute bilateral uveitis develops, as was seen in this patient. Ocular symptoms occur most commonly following constitutional symptoms, however they can also precede or occur concurrently with constitutional symptoms. Urinalysis findings also support the diagnosis. She has proteinuria, pyuria and haematuria. The most specific finding for TINU syndrome is the elevation in urinary and serum levels of Beta-2 microglobulin.   Following her sudden-onset bilateral uveitis and urinalysis findings, TINU syndrome is highly suspected and a renal biopsy should be performed to confirm the diagnosis. Renal biopsy would most likely show eosinophilic or mononuclear cellular infiltrates with glomerular sparing in TINU syndrome. In a patient with isolated acute bilateral uveitis, infectious disease testing could be performed instead to rule out pathogens such as tuberculosis, toxoplasma, bartonella, lyme, and syphilis. But in this case, TINU is strongly suggested and referral to a nephrologist for a renal biopsy is the most appropriate next step."
Uveitis and ocular inflammation	Endophthalmitis	"An 82-year-old lady presents to eye casualty with a 1-day history of a red, painful, and photophobic left eye associated with reduced vision. Past medical history is significant for hypertension and type 2 diabetes. Past ocular history includes cataract surgery in this eye 3 days ago, as well as bilateral diabetic macular oedema for which she is receiving intravitreal anti-VEGF injections, with the last set taking place 8 weeks before. She is systemically well and apyrexial.   Her best corrected visual acuity was counting fingers in the left eye and 20/30 in the right eye. Intraocular pressures were 12 and 14 respectively. On examination, the left eye was very injected with mild lid oedema and sticky yellow discharge. There was mild corneal oedema with Descemet folds, 4+ cells and fibrin with a slim hypopyon in the anterior chamber, as well as dense vitritris with no fundal view. The right eye featured a moderate cataract and mild non-proliferative diabetic retinopathy only."	"The suspected diagnosis is acute postoperative endophthalmitis. This is an ocular emergency and the key condition to exclude in patients with such presentations. Other differentials may include toxic anterior segment syndrome (which typically occurs within the first 24 hours and predominantly affects the anterior segment), retained lens fragment (which typically presents after the patient has completed their postoperative regimen), or reactivation of previous uveitis.  During the initial evaluation, ultrasound B-scan should be performed to evaluate the level of vitreous inflammation (which can be helpful for monitoring the condition in the absence of a fundal view) and to examine for any retinal detachment.   Initial management steps will depend on the local treatment protocol in the absence of a global consensus. ‘Tap and inject’ is commonly performed in many institutions - this comprises an intravitreal tap (either through needle aspiration or mechanical cutter biopsy), immediately followed by intravitreal injection of empirical antibiotics (typically ceftazidime and vancomycin according to local treatment protocols and microbiological spectrum). Anterior chamber tap is occasionally performed. Intraocular fluid samples are sent to the microbiological lab urgently for gram staining and culture and sensitivities.   Much of the evidence on acute postoperative endophthalmitis stems from the Endophthalmitis Vitrectomy Study (EVS), a landmark multicentre randomised controlled trial conducted between 1990-1994 comparing ‘tap and inject’ versus early pars plana vitrectomy (PPV) for post-cataract surgery endophthalmitis. A recent review has linked the general consensus on vitreous sampling for microbiological testing to the EVS. The EVS recommends early PPV for cases of post-cataract surgery endophthalmitis with a visual acuity of light perception or worse at presentation. However, some centres opt to perform an early pars plana vitrectomy (PPV) with vitreous biopsy and intravitreal antibiotics instead of ‘tap and inject’ regardless of presenting visual acuity or the type of surgery precipitating the endophthalmitis.   Postoperative treatment regimen also varies according to local treatment protocols, but may include topical and oral antibiotics, topical steroids, topical cycloplegics agents, and intravitreal or systemic steroids at a later stage. Close monitoring is key as the clinical course often guides subsequent management."
Glaucoma	Normal Tension Glaucoma (NTG)	"A 63-year-old female presented to the ophthalmology clinic complaining of progressive vision loss. She feels that she regularly misses things in her line of sight and believes that it’s been getting worse over time. She has a 15-year history of migraine with aura that is treated with propranolol for prevention and sumatriptan for abortive therapy. There is no personal or family history of ocular disease. She also has no history of ocular trauma or surgery.   On examination, there were no findings noted on external exam OU. Her corrected visual acuities were 20/20 OU. Extraocular movements were full and non-painful OU. Pupils were 3mm OU and there were normal pupil responses noted OU. On anterior segment examination, there were no abnormalities noted OU. On posterior segment exam, there was a 0.6 cup-to-disc ratio noted OD and a 0.7 cup-to-disc ratio noted OS. There was also focal thinning of the optic nerve head rim noted OU. Her intraocular pressures were 14mmHg OU."	"The suspected diagnosis in this patient is Normal Tension Glaucoma (NTG). NTG involves glaucomatous optic neuropathy in the setting of an open anterior chamber angle and a range of normal, supranormal and subnormal intraocular pressures (IOP). This patient has evidence of glaucoma as seen by the optic disc cupping and the peripapillary atrophy in both eyes. NTG is also believed to be associated with vasospastic conditions, with the most common association being with migraine.   The differential diagnosis includes all the conditions that can cause glaucomatous optic nerve damage. Due to its similar clinical presentation an important differential to exclude is primary open-angle glaucoma (POAG). By definition the only difference between NTG and POAG is the elevated IOP seen in POAG. On examination, the IOP was in a normal range which suggests that the patient has NTG, however continuous and diurnal measurements of IOP are needed to confirm the diagnosis.   Therefore, the next step in this patient should be to have a baseline glaucoma evaluation. This includes formal visual field testing, optic disc imaging, pachymetry, colour vision testing, gonioscopy and diurnal IOP measurements. This is important to detect and monitor the visual loss seen in this patient’s condition. In addition, the patient should be advised to avoid using propranolol at night-time and to have it in the morning, as night-time use of beta-blockers is an independent risk factor for visual field loss progression in NTG."
Glaucoma	Angle recession glaucoma	"A 44-year-old man presented to the ophthalmology clinic complaining of painless and gradual visual loss in his right eye. He noticed a few weeks ago that he can have difficulty seeing things from his right eye. He has no refractive error and does not wear glasses. He has no chronic medical conditions. He is a baseball coach and five years ago he sustained a direct blow to the right orbit from a baseball bat. He suffered from a fracture of the orbital floor that we repaired operatively with no complications. He is currently a smoker and has a 5 pack-year history of cigarette smoking.  On ocular examination, there was no scarring noted around the right eye. Uncorrected visual acuities are 20/20 OS and 20/25 OD. Extraocular movements were full and non-painful OU. Humphrey visual fields were normal OS and showed an inferior arcuate defect with an inferior nasal step OD. Pupils were 3mm OU and there was no RAPD noted OU. On anterior segment examination, gonioscopy showed a broad white band at the junction of the ciliary body and scleral spur that was only seen in the right eye. On posterior segment exam, there was a 0.4 cup-to-disc ratio noted OS and a 0.7 cup-to-disc ratio noted OD. Her intraocular pressures were 17mmHg OS and 28mmHg OD."	"The suspected diagnosis is angle recession glaucoma. Angle recession is defined as a tear that occurs between the longitudinal and circular layers of the ciliary muscles leading to a widened ciliary body band and a decrease in the outflow of aqueous. It is a late onset traumatic glaucoma and occurs many years following trauma to the eye. The key exam finding in angle recession is widening of the ciliary body band which is seen on gonioscopy in this patient. This exam finding in the presence of elevated pressure and nerve damage, leads to the diagnosis of angle recession glaucoma.  Given that the diagnosis of angle recession glaucoma is highly likely in this patient, next he should be treated for glaucoma. Topical aqueous suppressants such as beta-blockers, alpha agonists and carbonic anhydrase inhibitors are effective. Importantly, miotics such as pilocarpine should be avoided as they can exacerbate angle recession. Surgical therapy may be needed if the IOP is not controlled effectively. "
Glaucoma	Pigmentary dispersion glaucoma	"A 39-year-old male is referred by his optometrist for suspicion of glaucoma. He went to see his optometrist to adjust his prescription glasses, and his intraocular pressure was found to be elevated to 26 mmHg in the right eye and 27 mmHg in the left eye. Besides myopia of -6.0 diopters, he has no medical or ocular history. There is no family history of glaucoma or blindness. He doesn’t take any regular medication and denies any history of ocular trauma. He has never had eye surgery.  On examination, his corrected visual acuity with his new glasses was 20/20 in both eyes. Pupils were symmetric and reactive to light, without a relative afferent pupillary defect. Intraocular pressure was elevated to 27 mmHg in both eyes. Slit-lamp examination revealed pigment on the endothelium distributed along a vertical line, more prominent in the right eye. Some mild transillumination defects were seen in the mid-periphery of the iris bilaterally. Gonioscopy revealed open angles but with a heavily pigmented trabecular meshwork. The cup-to-disc ratio was 0.6 in the right eye and 0.7 in the left eye, without peripapillary haemorrhages. Optical coherence tomography revealed thinning of the retinal nerve fiber layer in the superior and inferior quadrants. Automated visual field testing showed mild inferior arcuate defects in both eyes."	"The suspected diagnosis is pigmentary dispersion glaucoma. The presence of pigment on the endothelium, transillumination defects, and pigmented trabecular meshwork with signs of glaucomatous neuropathy are highly suggestive of this diagnosis.  Pigmentary glaucoma is a type of secondary open-angle glaucoma, seen most often in young myopic males. The iris has a concave configuration and rubs against the anterior portion of the lens zonules, leading to the dispersion of pigment granules. These can accumulate in the trabecular meshwork, reducing aqueous humour outflow and causing elevation in intraocular pressure (IOP). Following the flow of aqueous humour, these pigments also deposit on the endothelium in a vertical fashion, known as the Krukenberg spindle. The contact between the iris pigment epithelium and anterior lens zonules is maximal in the mid-peripheral iris, leading to the formation of spoke-like transillumination defects.  Another condition that can present with a heavily pigmented trabecular meshwork and elevated IOP is pseudoexfoliation syndrome. It is characterised by the deposition of fibrillar material in the anterior segment of the eye. It has a higher prevalence in the elderly and those of Scandinavian descent. Fibrillar white flaky material can be seen on the anterior lens capsule or at the pupillary border. Transillumination defects are also seen in this condition, but they are peripupillary instead.  Given the presence of cupping, OCT thinning, and visual field defects, this patient requires treatment to reduce his elevated IOP. The next step would be to initiate glaucoma drops such as prostaglandin analogues. Selective laser trabeculoplasty can also be considered as a first-line treatment to reduce the pressure. The patient would require periodic follow-up to ensure adequate control of the pressure. Glaucoma surgery can be considered in severe cases. "
Glaucoma	Phacomorphic glaucoma	"A 78-year-old male presents to the clinic with blurry vision, redness, and severe pain in his right eye that started a few hours ago while he was watching television. He is also complaining of headaches and nausea, which have caused him to vomit three times since the onset of symptoms. This is the first time he has experienced such symptoms. He is known to have diabetes, which isn’t well controlled. He hasn’t seen an optometrist or ophthalmologist in the past decade.  On examination, visual acuity is counting fingers in the right eye and 20/100 in the left eye. The right pupil is mid-dilated and doesn’t react to light. Intraocular pressure measured by applanation tonometry is 54 mm Hg in the right eye and 20 mm Hg in the left eye. Slit-lamp exam reveals diffuse microcystic edema with Descemet folds OD. The chamber is very narrow with peripheral iridocorneal touch. There is 1+ cells and flare and a very dense brown cataract. There is no view of the fundus, but a B-scan is performed and doesn’t show any mass or hemorrhage. The left eye has a 3+ nuclear sclerosis cataract with a normal optic nerve and macula. Gonioscopy shows a completely closed angle in the right eye, while it is open in the left eye."	"The suspected diagnosis is phacomorphic glaucoma. The presence of acute angle closure in an eye with an advanced cataract, in the setting of an open angle in the contralateral eye, is highly suggestive of this diagnosis. Phacomorphic glaucoma is one of the main four types of lens-induced glaucomas.  Lens-induced glaucomas describe a group of pathologies that lead to open or closed-angle glaucoma, where the lens plays a major role in the pathophysiology. All these conditions present with decreased visual acuity, pain, redness, and elevated intraocular pressure (IOP).  First, phacomorphic glaucoma occurs when an advanced or intumescent cataract leads to pupillary block and the obstruction of aqueous outflow. It presents with elevated IOP and a narrow chamber with mild cells and flare. Usually, the angle is open in the contralateral eye, which can help differentiate it from primary angle-closure glaucoma. Second, phacolytic glaucoma occurs when there is material leakage from a mature cataract that clogs the trabecular meshwork. On examination, particles or white material can be seen in the anterior chamber without keratic precipitates. Gonioscopy shows an open angle. Third, lens-particle glaucoma occurs when the lens capsule is disrupted, and lens material enters the anterior chamber and blocks the trabecular meshwork. Potential causes of this disruption include cataract surgery or penetrating lens injury. The presentation can follow the precipitating event by weeks, months, or years. The exam will show similar findings to phacolytic glaucoma. Finally, phacoantigenic glaucoma is a granulomatous inflammatory reaction directed against the lens antigens that are liberated after surgery or trauma. A key differentiating feature of this condition is the presence of keratic precipitates and a severe anterior chamber cell/flare reaction. Residual lens material can often be identified.  For all these conditions, the initial management is to start medication to lower the intraocular pressure. For this patient with phacomorphic glaucoma, the next step would be to start topical and oral glaucoma medications (e.g., beta blockers, carbonic anhydrase inhibitors) and hyperosmotic agents such as intravenous mannitol. A laser peripheral iridotomy can be performed to lower the pressure, as well as prophylactically in the other eye if it seems predisposed to the same condition. Definitive treatment, however, consists of cataract extraction. "
Glaucoma	Steroid-induced ocular hypertension	"A 14-year-old boy is reviewed routinely in the paediatric ophthalmology clinic following a recent diagnosis of atopic keratoconjunctivitis. He is currently on a tapering regimen of topical corticosteroids, in addition to topical olopatadine 0.1% and systemic antihistamines.  He has a background of periocular eczema which is being treated with hydrocortisone cream by his primary care physician. He is otherwise well and denies any visual symptoms.   Visual acuities were 20/20 OD and 20/25 OS. Intraocular pressures were 48mmHg OD and 42mmHg OS. Gonioscopy revealed open iridocorneal angles and central corneal thickness was 540 OU. Anterior segment examination demonstrated resolution of the ocular surface inflammation and clear crystalline lenses. Posterior segment examination was unremarkable - specifically both optic discs were within normal limits."	"The most likely diagnosis is steroid-induced ocular hypertension. If this is not adequately treated, it may lead to irreversible optic nerve damage and hence glaucoma. Risk factors include patients at the extremes of age, primary open angle glaucoma or a family history of glaucoma, high myopia, and Afro-Caribbean ethnicity.   The precipitating agents (topical and periocular steroids in this case) should be discontinued where possible. Topical steroid drops should be tapered rapidly, and consideration of steroid-sparing agents, such as topical ciclosporin for patients with atopic keratoconjunctivitis, should be considered if the patient is likely to require recurrent or chronic treatment. This applies to periocular steroid creams as well, and the input of the primary care physician and/or dermatologist should be sought.   The patient should be carefully monitored for resolution, and intraocular pressure (IOP) lowering therapy, typically topical agents, should be considered as well. In acute cases of steroid-induced IOP elevation, IOP should normalise within days of discontinuing steroids. The chronic forms may take weeks to resolve. In a small proportion of patients, IOP remains persistently elevated and/or refractory to medical treatment, and IOP-lowering surgeries may occasionally need to be considered."
Glaucoma	Primary angle closure (PAC)	"A 52-year-old woman is referred to an ophthalmologist by her optometrist for suspicion of glaucoma. She doesn’t have any visual symptoms and initially visited her optometrist for a routine exam, during which elevated intraocular pressure was noted bilaterally. Her family history is positive for glaucoma in her father and aunts. She is followed by her primary care physician for pre-diabetes and doesn’t take any regular medication.  On examination, her visual acuity is 20/30 in both eyes, improving to 20/20 with pinhole. Intraocular pressure is 25 mm Hg in the right eye and 26 mm Hg in the left eye. Pupils are reactive and symmetric, without a relative afferent pupillary defect. The anterior segment is normal except for a 1+ nuclear sclerosis cataract bilaterally. The optic nerves have a cup-to-disc ratio of 0.3, without notching or hemorrhages. On gonioscopy, only the trabecular meshwork is visible in all four quadrants, with no peripheral anterior synechiae. Optical coherence tomography (OCT) shows normal retinal nerve fiber layer (RNFL) thickness in both eyes, as well as normal ganglion cell layer (GCC) thickness. Automated visual fields were performed and showed no defects."	"The suspected diagnosis is primary angle closure (PAC). The presence of a narrow iridocorneal angle and elevated intraocular pressure without optic nerve damage are characteristic of this condition.  Primary angle closure can be classified as primary angle-closure suspect (PACS), primary angle closure (PAC), and primary angle-closure glaucoma (PACG). These conditions exist on a spectrum, with the common finding being iridotrabecular contact over more than 180 degrees. PACS is defined as 180 degrees of iridotrabecular contact without trabecular meshwork or optic nerve damage, resulting in normal intraocular pressure (IOP) and the absence of peripheral anterior synechiae (PAS). In PAC, in addition to iridotrabecular contact, there is evidence of trabecular meshwork damage, which presents with elevated IOP and/or PAS, but without optic nerve damage. Finally, PACG is characterised by iridotrabecular contact, elevated IOP or presence of PAS, and glaucomatous optic neuropathy. Optic nerve damage can be identified by an increased cup-to-disc ratio, thinning of the RNFL and/or GCC, and visual field defects.  The best next step would be to perform a peripheral laser iridotomy given the narrow angles with elevated intraocular pressure. This procedure is done to prevent acute angle-closure glaucoma crises and to relieve the pupillary block component, which can help lower IOP. If the IOP remains high or there are signs of glaucomatous neuropathy, then topical therapy or glaucoma surgery should be considered. Based on the EAGLE study, many ophthalmologists consider performing clear lens extraction in patients with PAC with IOP above 30 mm Hg, or in those with PACG with IOP above 21 mm Hg."
Glaucoma	Neovascular glaucoma (NVG)	"A 66-year-old man presented to the ophthalmology clinic complaining of a gradual and painful loss of vision in both eyes. He had a 15-year history of diabetes mellitus that was being treated with metformin. He said that he finds it difficult to control his blood glucose levels. He had been receiving regular intravitreal anti-VEGF injections but stopped receiving them three years ago because he believed they were providing no benefit. He has no other medical history.   On examination, corrected visual acuity was light perception in the right eye and 20/400 in the left eye. Extraocular movements were full. The left pupil was normal but the right pupil was mid-dilated and had a RAPD. On anterior segment examination, there was corneal oedema noted OD. Dilated funduscopic examination of the right eye showed a cup-disk ratio of 0.6 with neovascularisation on the disk, dot blot haemorrhages, and exudates in all quadrants of the retina. Gonioscopic findings in the right eye showed a closed angle with neovascularisation of the angle in all quadrants, while the angles of the left eye were open with neovascularization of the angle in three quadrants. Intraocular pressures were 57 mmHg OD and 50 mmHg OS."	"The suspected diagnosis is neovascular glaucoma (NVG) secondary to proliferative diabetic retinopathy (PDR). The most common causes of NVG are diabetes mellitus, ischaemic central retinal vein occlusion and carotid occlusive disease. Diabetic retinopathy can progress to PDR that can result in neovascularisation of the iris and the iridocorneal drainage angle, leading to the development of NVG.  In this patient, there is a long history of poorly controlled diabetes mellitus. In addition, he has a history of receiving anti-VEGF injections which suggests that he had progressed to an advanced stage of diabetic retinopathy. On examination, there is evidence of PDR and an increased cup-to-disc ratio OD and neovascularisation of the iridocorneal drainage angle present OU. There is also a remarkable increase in intraocular pressures (IOP) that is ≥50 mmHg in both eyes. These examination findings are strongly suggestive of NVG.  Ocular treatment modalities in NVG that should be employed immediately include IOP-lowering agents, anti-VEGF, pan-retinal photocoagulation (PRP) and surgical intervention. Aqueous suppressants (beta-blockers, alpha-agonists and carbonic anhydrase inhibitors) should be used to lower IOP. Prostaglandin analogues should be used if IOP remains high (promotes uveoscleral outflow). Topical steroids and cycloplegics should also be used to control inflammation and improve the patient's pain. Osmotic agents should be used to treat the corneal oedema seen in the right eye. If IOP cannot be adequately controlled medically, glaucoma filtration surgery should be performed. Trabeculectomy is avoided in NVG due to the high-rates of bleb failure. The main surgical treatment used are glaucoma drainage devices. After the IOP is controlled, anti-VEGF should be given as an adjuvant treatment to medical/surgical treatment to prevent NVG progression. Because of the temporary nature of anti-VEGF, PRP is considered to be the more definitive treatment and should also be given as an immediate treatment."
Glaucoma	Glaucomatous optic neuropathy	"A 55 year old man presents to ophthalmology clinic for an annual eye examination. He has no complaints. He has a history of myopic LASIK in both eyes, done 10 years prior. His visual acuity is 20/20 in each eye, and his intraocular pressures (IOP) are measured as 16 and 18 in the right and left eye, respectively, as measured by Goldmann applanation. Pupils are equal, round, and reactive to light in both eyes without afferent pupillary defect. Anterior segment examination, gonioscopy, and posterior pole examination are normal. The cup-to-disc ratios are 0.8 in both eyes with inferior rim thinning. Humphrey visual fields demonstrate bilateral superior arcuate defects."	"This patient likely has glaucoma as evidenced by glaucomatous optic neuropathy given the neuroretinal rim thinning and corresponding visual field defects. The patient likely has open-angle glaucoma given the normal gonioscopy but the exact subtype of glaucoma cannot yet be determined. Primary open-angle glaucoma is a possible diagnosis and the normal intraocular pressures does not exclude this diagnosis. The normal IOP can be due to understimation of the IOP given the history of myopic LASIK, or can be due to diurnal IOP variations that can be up to 10 mmHg in patients with glaucoma. Normal tension glaucoma can be on the differential diagnosis. The patient must be started on IOP lowering medications such as topical prostaglandin agnosists and followed closely to monitor the IOP reduction and repeat the visual field testing."
Glaucoma	Primary open-angle glaucoma (POAG)	"A 72-year-old male is referred by his optometrist for suspicion of glaucoma. He visited his optometrist for a routine exam, during which his intraocular pressure was found to be elevated in both eyes. He has no visual symptoms and feels that his vision is stable. His medical history is significant for hypertension and diabetes mellitus, both well-controlled with medication. He had cataract surgery about 5 years ago and does not recall any mention of glaucoma at that time.  On examination, his visual acuity was 20/20 in both eyes. Intraocular pressure was 27 mm Hg in the right eye and 34 mm Hg in the left eye. The anterior segment examination was normal with a posterior chamber intraocular lens in place. The optic nerves appeared pink, with a cup-to-disc ratio of 0.6 in the right eye and 0.8 in the left eye, with an inferior notch in both eyes. There was no peripapillary hemorrhage. On gonioscopy, the ciliary body was visible in all four quadrants in both eyes. Optical coherence tomography (OCT) showed moderate inferior and superior thinning of the retinal nerve fiber layer (RNFL) in both eyes, more pronounced in the left eye. Automated visual field testing revealed a nasal step in the right eye and a superior altitudinal defect in the left eye."	"The suspected diagnosis is primary open-angle glaucoma (POAG). The presence of elevated intraocular pressure (IOP), optic nerve excavation, RNFL thinning, and corresponding visual field defects strongly suggests this diagnosis.  Glaucoma is a neurodegenerative disease of the optic nerve and is the most common optic neuropathy. However, many other causes of optic neuropathy can be mistaken for glaucoma. Therefore, the best next step would be to complete the examination by evaluating pupillary reactions and testing color vision. These features can help differentiate glaucoma from other pathologies. In glaucoma, pupils are typically symmetric and reactive, and a relative afferent pupillary defect (RAPD) is usually only present in very severe and asymmetric cases, which are rare. Color vision remains intact until very late in the disease, and in this patient with moderate RNFL thinning and visual field defects, color vision impairment would not be expected. The appearance of the optic nerve can also be quite helpful, as glaucomatous optic nerves retain their natural orange-pink color until the disease reaches an advanced stage.  Other causes of optic neuropathy include damage from previous episodes of optic neuritis, ischemic optic neuropathy, compressive optic neuropathy, nutritional deficiencies, toxic optic neuropathy, and traumatic optic neuropathy. Signs that should prompt consideration of glaucoma mimickers include normal or low IOP, optic nerve head pallor, a normal cup-to-disc ratio, visual field defects that respect the vertical meridian (but not the horizontal one), dyschromatopsia, the presence of a RAPD, and sudden onset of vision loss. Obtaining a thorough medical history and asking about systemic symptoms can help narrow the differential. In challenging cases, neuroimaging can be performed to rule out other potential causes. Additionally, monitoring disease progression over time and the response to glaucoma medications can further support the diagnosis."
Glaucoma	Bleb leak	"A 70-year-old male with advanced glaucoma is seen on post-operative day 1 after undergoing a trabeculectomy in his right eye. He complains of blurry vision but doesn’t have pain. He has been using his post-operative drops as prescribed, which include moxifloxacin and prednisolone.  On examination, his visual acuity was 20/100 in the right eye and 20/25 in the left eye. Intraocular pressure was 3 mm Hg in the right eye and 17 mm Hg in the left eye. Slit-lamp examination revealed a hazy cornea with folds, as well as a narrow anterior chamber with iridocorneal touch in the periphery. There was no corneal-lenticular touch. The bleb, located superiorly, was flat, and on the application of fluorescein, a slow leakage of aqueous humor was observed. The optic nerve had a cup-to-disc ratio of 0.8 in the right eye. No hypotony maculopathy was seen on fundus examination or optical coherence tomography. Small peripheral choroidal detachments, nasal and temporal, were present."	"The suspected diagnosis is a bleb leak. The presence of hypotony with a flat anterior chamber in the context of a positive Seidel test is highly suggestive of this diagnosis.  Trabeculectomy is a type of filtering glaucoma surgery in which an opening, or ostium, is created into the anterior chamber beneath a partial-thickness scleral flap. This new pathway allows aqueous humor to flow into the subconjunctival space, creating a bleb (conjunctival elevation). Early post-operative complications following this surgery can be divided into two main groups: high ocular pressure and low ocular pressure. These groups can then be subdivided into two categories: deep anterior chamber and narrow/flat anterior chamber. High IOP with a deep chamber indicates inadequate filtration due to obstruction of flow at the ostium. This can occur when the peripheral iris obstructs the ostium, which can be treated with laser or by mechanically pulling the iris away. Another common cause of ostium obstruction is a large blood clot. If the bleb is flat, the elevated pressure could be secondary to tight sutures.  High IOP with a narrow chamber suggests pupillary block, suprachoroidal hemorrhage, or aqueous misdirection. IOP-lowering medications are started, but the management revolves around addressing the underlying etiology. Low IOP with a deep chamber indicates overfiltration by the bleb. This is often managed with observation only unless the chamber narrows or hypotony maculopathy develops. A bleb revision might be required if hypotony persists. Low IOP with a narrow chamber suggests overfiltration or wound leak. A Seidel test is performed at the bleb site to evaluate wound integrity. If a leak is detected, as in this patient, the next step would be to place a large bandage contact lens to tamponade the leak. A pressure patch can also be applied. If the leak or hypotony persists, then a bleb revision is required. "
Glaucoma	Blebitis	"A 71-year-old male presented to the ophthalmology clinic with a two-day history of pain in his left eye. His symptoms came on suddenly and have been causing significant discomfort. He has a past history of primary open angle glaucoma for which he underwent a trabeculectomy 3-years ago. He also underwent cataract surgery at the same time for both eyes. His postoperative course was uneventful. His past medical history includes hypertension and diabetes that is controlled with medication. He has no history of ocular trauma.   On examination, there was eye redness noted OS. His corrected visual acuity was 20/20 OD and 20/80 OS. Extraocular movements were full and non-painful OU. Pupils were 3mm OU and there were normal pupil responses noted OU. On anterior segment examination, there was intense conjunctival hyperaemia, anterior chamber cells and hypopyon noted OS. The trabeculectomy bleb was not hyperaemic. After fluorescein instillation into the left eye there was green fluorescence noted around the bleb under cobalt blue lighting. On posterior segment examination, there were no findings noted OU. His intraocular pressures were 8mmHg OS and 14mmHg OD."	"The suspected diagnosis in this patient is blebitis. Blebitis is an infection of the bleb and can present any time after a patient has had a trabeculectomy. This patient has many findings of blebitis such as a non-hyperaemic or “white-on-red” bleb, anterior chamber reaction and evidence of a bleb leak on Seidel testing with an associated decrease in IOP. A bleb leak significantly increases the risk of bleb infection and is the main risk factor for blebitis. The use of anti-metabolites during trabeculectomy also increases the risk significantly but is not detailed in this case.   The most important differential diagnosis to exclude in this patient is bleb-associated endophthalmitis (BAE). BAE presents similarly to blebitis but is typically more severe (greater visual acuity loss and possible periorbital skin involvement) and causes vitreous involvement, so is unlikely in this case. Other conditions such as conjunctivitis, anterior uveitis and episcleritis are less likely due to involvement of the bleb.   For management of this patient’s condition, next the patient should have the bleb cultured or the anterior chamber tapped with initiation of antibiotic treatment. Bleb culture or anterior chamber tapping is important to identify the inciting organisms. To prevent some of the devastating complications of blebitis, antibiotic treatment should also be initiated. There is no universally agreed regimen, however options include fortified antibiotics such as vancomycin, cefazolin, tobramycin and fluoroquinolones. Topical steroids can also be considered following resolution of infection, to salvage flap patency and prevent fibrosis and scarring."
Refractive surgery	LASIK regression	"A 32-year-old female patient presented to the ophthalmology clinic complaining of progressive decrease in vision in both eyes. She underwent bilateral LASIK surgery for high myopia four years ago, where her preoperative refraction was -9.00 diopters in both eyes. Postoperatively, her vision improved significantly in both eyes, achieving 20/20 in each eye six months after the surgery. Over the past year, she has noticed a gradual decline in her vision in both eyes. On examination, her uncorrected visual acuity was 20/50 in the right eye and 20/60 in the left eye. With a refraction of -1.00 in the right eye, her visual acuity improved to 20/20, and with a refraction of -1.50 in the left eye, it improved to 20/20. Her anterior and posterior segment examinations were normal without signs of myopic degeneration. Corneal topography showed no signs of ectasia."	"The most likely diagnosis is LASIK regression secondary to epithelial cell migration and scarring. This leads to loss of effect from the myopic LASIK ablation. Based on the patient's vision needs, the patient may be considered for a re-enhancement. The primary differential diagnosis to consider is myopia progression from axial elongation. This condition involves an increase in the eye’s axial length, leading to the worsening of myopia over time. Myopia progression is more common in patients with a history of high myopia, as seen in this patient. Axial elongation can occur independently of LASIK regression and may be a contributing factor to the refractive changes observed. Post-LASIK ectasia must also be considered and would typically result in an abnormal topography with corneal steepening and astigmatism on refraction."
Refractive surgery	Post-LASIK ectasia	"A 39-year-old female patient presented to the ophthalmology clinic following a progressive decrease in vision OS. She underwent bilateral LASIK surgery for high myopia four years ago. The vision in her right eye improved immediately and has remained the same. However, the vision in her left eye did not improve as much as the right eye. She subsequently underwent enhancement surgery of her left eye six months after her original LASIK procedure but did not notice a significant improvement. Over the past two years, vision in her left eye has become progressively more blurry, and has not improved with multiple changes to her glasses prescription.   On examination, her uncorrected visual acuity was 20/200 OS and 20/20 OD. With a -6.00 + 4.00 x 173 lens, visual acuity in her left eye improved to 20/60. Confrontational visual fields and extraocular movements were full OU. Pupils were 4mm bilaterally and the pupil response was normal OU. On anterior segment examination, her right eye was normal and her left eye had an irregular corneal contour. Dilated fundus examination was normal OU. Intraocular pressures were 15 mmHg OD and 16 mmHg OS."	"The suspected diagnosis is post-LASIK ectasia. Post-LASIK ectasia is considered in patients with worsening myopia +/- increasing astigmatism, loss of best-corrected visual acuity and inferior corneal steepening following refractive surgery such as LASIK. There are many known risk factors for post-LASIK ectasia and two minor risk factors that are relevant in this patient’s history are high myopia and enhancement surgery.   This patient has a progressive decrease in vision in her left eye following LASIK that was not improved with multiple changes to her glasses prescription. On examination, she has a significantly decreased uncorrected visual acuity OS which improved with a toric contact lens to 20/60. She has increasing myopia with astigmatism and a loss of best-corrected visual acuity. In addition, the patient has an irregular corneal contour OS.   For confirmation of the characteristic corneal changes, the patient should undergo corneal topography. In post-LASIK ectasia there would most likely be marked inferior steepening of the cornea seen on topography that resembles keratoconus. Treatment for this condition is also the same as keratoconus. A rigid gas permeable contact lens for the left eye should be the first treatment offered to this patient."
Refractive surgery	Diffuse lamellar keratitis (DLK)	"A 32 year old man presents on post-operative day 3 status post LASIK surgery of both eyes. He is having slightly blurred vision of both eyes. There are no flashes, floaters, or eye pain.   On exam, visual acuity is 20/25 in each eye. Pupils are equal, round, and reactive to light without afferent pupillary defect. Intraocular pressure is 14 in the right eye and 13 in the left eye. Extraocular motility is full. Visual fields are full to confrontation.   Slit lamp examination demonstrates many diffuse very fine opaque infiltrates directly underneath the LASIK flap; this is seen in both eyes. Anterior chamber is deep and quiet in both eyes, and iris stroma is normal, round and reactive in both eyes."	"The patient has diffuse lamellar keratitis (DLK), a complication of LASIK. Characteristically, very fine infiltrates develop in the LASIK flap-corneal stroma interface, starting peripherally at postoperative day 1; these may progress centrally over the next couple of days. The infiltrates are thought to represent a localized inflammatory reaction in the area of surgical manipulation, but the pathophysiology remains uncertain. Treatment typically includes topical corticosteroids such as prednisolone acetate eyedrops, and the condition often resolves within 5-8 days after initiating therapy. It is important to initiate therapy when DLK is seen, as if it progresses, especially centrally, it can lead to permanent corneal stromal scarring."
Refractive surgery	Pressure-induced stromal keratitis (PISK)	"A 26-year-old man presents to the clinic one week after undergoing LASIK surgery to correct his myopia (-5.00 diopters in each eye). He initially experienced improved vision but now reports progressively worsening blurred vision in both eyes over the past three days. He denies pain, redness, or discharge. On examination, his uncorrected visual acuity has decreased from 20/20 on the first postoperative day to 20/60 in each eye. Slit-lamp examination reveals diffuse stromal haze without significant inflammation or flap dislocation. Intraocular pressure is elevated at 30 mmHg in both eyes. The corneal flaps appear well-apposed with no signs of infection."	"The most likely diagnosis is pressure-induced stromal keratitis (PISK) or interface fluid syndrome. The immediate next step is to lower the IOP to reduce corneal edema and prevent further visual deterioration. This involves reducing or discontinuing corticosteroid eye drops, which can elevate IOP in susceptible individuals, and initiating IOP-lowering medications such as topical beta-blockers or carbonic anhydrase inhibitors. The differential diagnosis includes diffuse lamellar keratitis, flap dislocation and infectious keratitis, all of which present with different signs. Diffuse lamellar keratitis will show early post-operative inflammation with cells under the LASIK flap and infectious keratitis will show a corneal infltrate. OCT can be useful is PISK to show interface fluid accumulation in the cap-bed interface and monitor the clinical course."
Refractive surgery	Flap dislocation	"A 28-year-old patient presents to your clinic for their post-operative day #1 visit after LASIK OU. The patient reports decreased vision in the left eye. Pre-operatively, the vision in both eyes was best corrected to 20/20 with a -7.00 sphere cycloplegic refraction OU. The best-corrected vision is 20/20 in the right eye and 20/40 in the left eye. On slit-lamp examination, the right eye is normal. In the left eye, you observe undulating stromal folds centrally, as well as a slightly wider gutter temporally compared to nasallly."	"The most likely diagnosis is LASIK flap slippage leading to macrostriae. Macrostriae arise when the flap slips or wrinkles after LASIK surgery, leading to full-thickness folds in the cornea with noticeable vision impairment. The main differential diagnosis is microstriae which are microscopic, superficial wrinkles involving Bowman's layer and that are generally asymptomatic. Given the vision decrease in this patient, the best treatment is to refloat the flap immediately This is achieved by lifting and irrigating under the flap with sterile balanced salt solution. The flap is then repeatedly stroked perpendicular to the direction of the folds."
\end{filecontents*}
\csvreader[
  head to column names,
  separator=tab,      
  respect all
]{supplementary_cases.tsv}{}{

  \stepcounter{rowctr}%
  \edef\thisAnchor{case-\therowctr}%

  \ifnum\pdfstrcmp{\Subspecialty}{\prevSubspecialty}=0\relax
  \else
    \par\medskip\noindent\textbf{\Subspecialty}\par
    \edef\prevSubspecialty{\Subspecialty}%
  \fi

  \noindent\hspace{1em}\hyperlink{\thisAnchor}{\Diagnosis}\dotfill \pageref{\thisAnchor}\par
}
\endgroup

\begingroup
\small
\RaggedRight
\setlength{\parskip}{0.5em}
\emergencystretch=3em

\newcounter{casectr}\setcounter{casectr}{0}

\csvreader[
  head to column names,
  separator=tab,
  respect all,
  before reading={\catcode`\"=9},
]{supplementary_cases.tsv}{}{

  \clearpage 
  \stepcounter{casectr}%
  \edef\thisAnchor{case-\thecasectr}%

  \hypertarget{\thisAnchor}{}\label{\thisAnchor}

  \begin{tcolorbox}[casecard, title={Case \thecasectr: \Diagnosis\ (\Subspecialty)}, colback=lightgray!25!white]
    \textbf{Vignette}\par
    \preservebreaks{\Vignette}

    \medskip
    \textbf{Reasoning}\par
    \preservebreaks{\Reasoning}
  \end{tcolorbox}
}
\endgroup

\end{refsection}

\end{document}